\RequirePackage{ifpdf}
\documentclass[a4paper,11pt]{article}
\usepackage{jheppub}

\usepackage{amsmath}
\usepackage{graphicx}

\usepackage{xcolor}
\usepackage{slashed}

\newcommand{\Sobs}{{\alpha, \beta; S}}
\newcommand{\Jobs}{{\alpha, \beta; J}}
\newcommand{\Fobs}{{\alpha, \beta; F}}
\newcommand{\Sobsang}{{1, 1-a; S}}
\newcommand{\Jobsang}{{1, 1-a; J}}

\def\df{{\rm d}}
\newcommand{\scetI}{{SCET$_{\rm I}$}}
\newcommand{\scetII}{{SCET$_{\rm II}$}}
\newcommand{\cusp}{{\rm cusp}}
\newcommand{\sub}{{\rm sub}}
\title{Disentangling observable dependence in \scetI{} and \scetII{}
  anomalous dimensions: angularities at two loops}

\author[a]{Christian W.~Bauer,}
\author[b]{Aneesh V.~Manohar,}
\author[c]{Pier Francesco Monni.}

\affiliation[a]{Ernest Orlando Lawrence Berkeley National Laboratory, University of California, Berkeley, CA 94720, USA}
\affiliation[b]{Department of Physics, University of California at San Diego, 9500 Gilman Drive, La Jolla, CA 92093-0319, USA}
\affiliation[c]{CERN, Theoretical Physics Department, CH-1211 Geneva 23, Switzerland}

\emailAdd {cwbauer@lbl.gov}
\emailAdd{amanohar@ucsd.edu}
\emailAdd{pier.monni@cern.ch}

\preprint{CERN-TH-2020-189}

\abstract{The resummation of radiative corrections to collider jet
  observables using soft collinear effective theory is encoded in
  differential renormalization group equations (RGEs), with anomalous
  dimensions depending on the observable under consideration.
  This observable dependence arises from the ultraviolet (UV) singular
  structure of real phase space integrals in the effective field
  theory.
  We show that the observable dependence of anomalous dimensions in
  \scetI{} problems can be disentangled by introducing a
  suitable UV regulator in real radiation integrals. 
  Resummation in the presence of the new regulator can be
  performed by solving a two-dimensional system of RGEs in the
  collinear and soft sectors, and resembles many features of 
  resummation in \scetII{} theories by means of the rapidity
  renormalization group. 
  We study the properties of \scetI{} with the additional regulator
  and explore the connection with the system of RGEs in \scetII{}
  theories, highlighting some universal patterns that can be exploited
  in perturbative calculations.
  As an application, we compute the two-loop soft and jet anomalous
  dimensions for a family of recoil-free angularities and give new
  analytic results. This allows us to study the relations between the
  \scetI{} and \scetII{} limits for these observables.
  We also discuss how the extra UV regulator can be exploited to
  calculate anomalous dimensions numerically, and the prospects for
  numerical resummation.
}

\keywords{}

\begin{document} 
\maketitle
\flushbottom

\section{Introduction}
\label{sec:introduction}

The resummation of radiative corrections in the framework of soft
collinear effective theory
(SCET)~\cite{Bauer:2000ew,Bauer:2000yr,Bauer:2001ct,Bauer:2001yt} is
achieved by integrating renormalization group
equations (RGEs) in the effective theory.
The anomalous dimensions governing such RGEs depend on the observable
under consideration.
In the resummation of jet collider observables, this observable
dependence is related to the presence of ultraviolet (UV) divergences
in real radiation integrals of the effective theory originating from
the expansion of the physical phase space using power counting
dictated by the SCET Lagrangian.
At the same time some elements of the anomalous dimensions, those
arising from virtual UV divergences are universal across observables
for a given physical process.
An interesting question is whether such observable dependent and
independent components can be understood and disentangled, hence
unveiling some common patterns and consistency relations that can be
exploited when performing perturbative calculations.

We limit ourselves to collider jet observables in \scetI{} and \scetII{}.
In \scetII{} problems, the UV singularities of the phase space
integrals can be handled with the rapidity renormalization
group~\cite{Chiu:2011qc,Chiu:2012ir}, which encodes the full
observable dependence in the rapidity anomalous dimension. An
alternative approach to this problem was originally formulated in
Refs.~\cite{Becher:2010tm,Becher:2011pf}.
In \scetI{} problems this separation does not occur as
both UV and IR divergences are regulated by pure dimensional
regularization.

To be concrete, we consider the toy example of generalized
angularities in electron-positron collisions (an analogous jet-based
observable was defined in Ref.~\cite{Larkoski:2014pca})
\begin{align}
\tau = \sum_i \left(\frac{k_{i,\perp}}{Q}\right)^\alpha e^{-\beta |\eta_i|}
\label{eq:angularity}
\,,\end{align}
where the transverse momentum $k_{i,\perp}$ and pseudorapidity
$\eta_i$ are taken with respect to the recoil-free winner-take-all
axis, and $Q$ is the center-of-mass energy of the collision.
Note that, if the sum runs over (massless) partons in the event, this
observable is only collinear safe for $\alpha=1$ (with
$\beta > -\alpha$ to ensure IR safety). This is the case of
conventional angularities~\cite{Berger:2003iw} for which one has
$\alpha=1$ and $\beta=1-a$; the case $a=0$ ($\beta=1$) corresponds to
a thrust-like angularity (denoted by thrust in the rest of the paper),
while $a=1$ ($\beta=0$) corresponds to a recoil-free version of jet
broadening.
For $\alpha\ne 1$, an alternative collinear safe version of the
observable~\eqref{eq:angularity} with the same scaling behavior as in
Eq.~\eqref{eq:scalesDef} can, for instance, be defined as in
Ref.~\cite{Dasgupta:2020fwr} by using Lund Jet
Plane~\cite{Dreyer:2018nbf} clusters rather partons. Alternatively,
one can adopt a track based definition as in
Ref.~\cite{Larkoski:2014pca}.
All explicit computations in this article will refer to the simple
case of Eq.~\eqref{eq:angularity} of $\alpha=1$ with the sum running
over massless partons. However, since many of the considerations made
in the paper only depend on the scaling~\eqref{eq:scalesDef}, we will
keep the dependence on $\alpha$ in the rest of the paper.
  While any explicit results given in the paper that refer to the
factorization theorem~\eqref{eq:factorization} only hold for
$\alpha=1$, by keeping the general dependence on $\alpha$ and $\beta$
we make our results easily extendable to other observables, albeit
with different factorization theorems.

In the limit $\tau\to 0$, the logarithms of $\tau$ can be resummed to
all orders in perturbation
theory~\cite{Berger:2003iw,Bauer:2008dt,Hornig:2009vb,Larkoski:2014uqa,Banfi:2018mcq,Bell:2018gce,Procura:2018zpn}
(see also Ref.~\cite{Kang:2018vgn} for groomed angularities at hadron
colliders).
In \scetI{} ($\beta\neq 0$), this resummation is accomplished by observing
that the problem contains three separate mass scales, a hard,
jet and soft scale
\begin{align}
\label{eq:scalesDef}
M_H &= Q\,, &  M_J &= Q \, \tau^{1/(\alpha+\beta)}\,,  &  M_S &= Q \, \tau^{1/\alpha}
\,,
\end{align} 
and the differential cross section can be expressed by means of the
following factorization theorem valid for the recoil-free case at
leading power~\cite{Larkoski:2014uqa}
\begin{align}
\label{eq:factorization}
\frac{1}{\sigma_{\rm Born}}\frac{\df\sigma}{\df \tau} = H(M_H,\mu)\int
  \df \tau_n\,\df \tau_{\bar n}\, \df \tau_s\, {\cal J}_n(M_J,\tau_n) \,{\cal
  J}_{\bar n}(M_J,\tau_{\bar n}) \,{\cal S}(M_S,\tau_s)
  \,\delta(\tau-\tau_s-\tau_n-\tau_{\bar n})\,,
\end{align}
where the soft and jet functions have the standard
definitions
\begin{align}
\label{eq:soft_and_jet}
{\cal J}_n(M_J,\tau_n) &= \frac{2\pi}{N_c}{\rm tr}\langle
  0|\frac{\slashed{\bar n}}{2}\chi_n\delta({\bar n}\cdot p -
  Q)\delta(\tau_n-\hat{\tau}_n) {\bar \chi}_n|0\rangle\,,\notag\\
{\cal J}_{\bar n}(M_J,\tau_{\bar n}) &= \frac{2\pi}{N_c}{\rm tr}\langle
  0|{\bar \chi}_{\bar n}\delta(n\cdot p -
  Q)\delta(\tau_{\bar n}-\hat{\tau}_{\bar n}) \frac{\slashed{n}}{2}
                                     \chi_{\bar n}|0\rangle\,,\notag\\
{\cal S}(M_S, \tau_s) &= \frac{1}{N_c}{\rm tr}\langle 0| S^\dagger_{\bar n}S_{
            n}\delta(\tau_s - \hat{\tau}_s)S^\dagger_{ n}S_{\bar n}|0\rangle\,,
\end{align}
and $\hat \tau_n$, $\hat \tau_{\bar n}$ and $\hat \tau_s$ are
operators that return the value of the generalized angularity in the
$n$-collinear, $\bar n$-collinear and soft sector, respectively.
The operators depend on the parameters $\alpha$ and $\beta$, and therefore
introduce dependence on these parameters into the jet and
soft function which we have not indicated explicitly.

Notice that for angularities defined with respect to the
winner-take-all axis the factorization theorem in
Eq.~\eqref{eq:factorization} is the same in both \scetI{} and
\scetII{} (with $\beta = 0$).
This allows us to study the transition between the two theories. 
For general observables (e.g. if one takes the thrust axis as a reference), the factorization theorem is different between the \scetI{} and \scetII{} case. The full structure of the \scetII{} results can not be obtained as the limit of the \scetI{} result in this case. 
The results of this paper regarding the structure of the anomalous dimensions in \scetI{} however still hold.

In Laplace space, the factorization theorem~\eqref{eq:factorization}
becomes a simple product between the hard function and the Laplace
transform of the soft ($\hat{\cal S}$) and jet
($\hat{\cal J}_{n,\bar n}$) functions, namely
\begin{align}
\label{eq:factorizationLaplace}
\hat{\sigma}[u] \equiv\int_0^\infty d\tau e^{-u\tau}\frac{1}{\sigma_{\rm Born}}\frac{\df\sigma}{\df \tau} = H(M_H) \hat{\cal J}_n(M_J[u]) \,\hat{\cal
  J}_{\bar n}(M_J[u]) \,\hat{\cal S}(M_S[u])\,,
\end{align}
where we have defined\footnote{In this paper, we will often use the symbol $F$ to denote $F=S, J$. This means that the equation is valid for both the soft and jet sectors, with all all quantities with subscripts $F$ being replaced by their soft and collinear values, respectively.}
\begin{equation}
\label{eq:laplace}
\hat{F}(M_F[u]) = \int_0^\infty\df \tau\,e^{-u\,\tau} F(\tau)\,,
\end{equation}
with $F={\cal S},\,{\cal J}_{n},\,{\cal J}_{\bar n}$.
The scales $M_F[u]$ in Laplace space are given by the replacement $\tau \to e^{-\gamma_E}/u$ in Eq.~\eqref{eq:scalesDef}.
The \scetII{} case ($\beta=0$) obeys the same
factorization theorem~\eqref{eq:factorization} although the soft and
jet functions do not depend on a single scale like in the \scetI{}
case.

In effective field theories such as  SMEFT, anomalous dimensions
are purely of short distance nature and do not depend on any long
distance parameters such as the Higgs vacuum expectation value or the observable being
measured.
Therefore, in SCET the observable dependence of the anomalous
dimensions might seem at first sight to be in contradiction with their
short-distance nature.
In other words, one might expect that UV divergences arise from
virtual corrections, whereas real radiation describes the propagation
of on-shell degrees of freedom which should only give rise to infrared
divergences, and not contribute to anomalous dimensions.
The reason this is not true in SCET is that in the effective theory
phase space constraints need to be multipole
expanded~\cite{Bauer:2000ew,Bauer:2000yr,Beneke:2002ph,Beneke:2002ni}
as dictated by the power counting.
Therefore, real particles can have energies that are arbitrarily
large, and are integrated over phase space regions which go to
infinity. This induces an observable dependence in the UV
singularities originating from real radiation integrals.
The above discussion hints at the fact that the observable dependence
of the anomalous dimensions can be identified by introducing an
additional UV regulator in the phase space integrals.
As we discuss in this paper, an exponential regulator analogous
to that proposed in Ref.~\cite{Li:2016axz} for \scetII{} problems can
be introduced to render the real contributions UV finite, while
keeping the structure of the virtual corrections unchanged.

As we show in this paper, introducing such an extra regulator in \scetI{} calculations has 
several advantages,
as it allows one to disentangle the observable dependence.
It introduces a
new  scale $\nu$, and resummation in \scetI{} problems
can be performed by solving a two-dimensional system of RGEs in the
soft and each of the collinear sectors. 
In particular, it can be shown that the $\mu$ anomalous dimension is
independent of the observable, while all observable dependence is
contained in the $\nu$ anomalous dimension.
Differential equations in $\mu$ and $\nu$ are analogous to the
rapidity RG equations~\cite{Chiu:2012ir} that are commonly used in
\scetII{} theories, where a rapidity regulator is required to render
the soft and collinear contributions separately finite.
Our approach highlights a number of similarities with the \scetII{} case,
and allows us to study the connections and differences between the
two theories. 
These lead to consistency conditions for the anomalous dimensions that
can be exploited in perturbative calculations.
Moreover, the introduction of the extra regulator allows one to make integration over the real radiation
suitable for a numerical calculations, as
discussed in the conclusions.
This has the advantage that
complicated observable-dependent integrals can be computed numerically
in four dimensions.  
This is also being exploited in the ongoing effort at obtaining a
numerical resummation framework that is systematically extendable to
higher perturbative accuracy in
SCET~\cite{Bauer:2018svx,Bauer:2019bsp}.

Introducing an extra UV regulator, however, also has some side
effects.
In standard \scetI{} regularized in dimensional regularization in the
regime $M_J > M_S$ (or equivalently $\beta>0$), collinear degrees of
freedom are integrated out below $M_J$.
The collinear jet function is the matching coefficient between SCET
with both soft and collinear degrees of freedom, and a low-energy soft
theory containing only Wilson lines interacting via soft degrees of
freedom.
The introduction of an extra UV regulator introduces a new scale
  into the soft and jet functions which seemingly breaks the above
  factorization picture. However, we show that the above issue
  can be handled by observing that the dependence on the new scale can
  be completely factorized within the soft and jet functions, which
  allows one to preserve the properties of standard \scetI{}
  theories.

This paper is organized as follows:
In Section~\ref{sec:RGEStructure} we briefly summarize the structure
of the RGEs in \scetI{} and \scetII{} theories. This section also
serves to define the notation and conventions used throughout the
paper.
Section~\ref{sec:RapRegulator} discusses the effect of an extra
UV regulator and the conditions it needs to satisfy to regularize the
real phase space integrals.
Section~\ref{sec:RGRelationship} discusses the similarities and
differences between \scetI{} and \scetII{} RGEs in the presence of
the extra UV regulator.
In Section~\ref{sec:ObsDep} we show how these considerations allow one  to
isolate the observable dependence in the anomalous dimensions and how
the dependence on the new UV regularization scale $\nu$ can be
factorized separately within the soft and collinear sectors making its
cancellation manifest.
In Section~\ref{sec:ExplicitCalc} we explicitly calculate the
anomalous dimensions at one- and two-loop order for the recoil-free
angularities introduced above, and relate our findings to existing
results in the literature.
Our conclusions and
outlook are given in Section~\ref{sec:Conclusions}.

\section{Resummation of radiative corrections in \scetI\ and \scetII}
\label{sec:RGEStructure}
In this section we briefly summarize the
resummation of leading power logarithmic corrections in \scetI{} and
\scetII{} theories, and present some of the results from a different point of view.
The section also serves to define the notation we will use throughout the
paper.

Before we start, we want to make a brief comment about our notation of scale dependence in the various objects appearing in the factorization theorems of \scetI{} and \scetII{}. 
The SCET objects depend on a single characteristic scale for both the rapidity and renormalization scales, and for each function $F$ we denote the characteristic scales corresponding to $\mu$ by $M_F$ and those for $\nu$ by $N_F$. 
So for example, in \scetI{} the ingredients of the factorization
theorem $F = H, J, S$ depend on the renormalization scale $\mu$ and
the characteristic scales $M_F$ through the ratio of these two scales.
Similarly in \scetII{}, the jet and soft functions depend on the
renormalization scale $\mu$, the rapidity scale $\nu$, as well as the
characteristic scales $M_F$, $N_F$ through the ratios $\mu / M_F$ and
$\nu / N_F$.
In order to simplify the notation, we will omit the dependence on the characteristic scales in the rest of this paper, unless this dependence is important for clarity of the discussion. 
This means that we will use
\begin{align}
F(\mu) & \equiv F(M_F; \mu)\,, &
F(\mu, \nu) & \equiv F(M_F, N_F; \mu, \nu) 
\,,
\end{align}
and similarly for anomalous dimensions
\begin{align}
\gamma_F(\mu) & \equiv \gamma_F(M_F; \mu) \,, &
\gamma_F(\mu, \nu) & \equiv \gamma_F(M_F, N_F; \mu, \nu) 
\,.\end{align}

\subsection{Resummation in \scetI}
\label{sec:SCETIResummation}
Resummation of large logarithms in \scetI{} is accomplished by using a
sequence of effective field theories, each of which has a single
characteristic scale, and with the scales being widely
separated from one another~\cite{Bauer:2000ew}.
The first step is to match QCD onto \scetI{} by writing the QCD currents in terms of operators containing \scetI{} fields, combined with short distance Wilson coefficients.
For many applications of interest, the current in the full theory is
conserved and hence $\mu$-independent, and we will assume this here for simplicity. 
This allows one to write the matching in position space onto \scetI{} as 
\begin{align}
\label{2.1}
J^{\rm QCD}_{\rm bare}(x) = J^{\rm QCD} _{\rm ren}(x)  = C_{{\rm bare}} \, J^{\rm SCET}_{{\rm bare}}(x) = C_{{\rm
               ren}}(\mu) \, J^{\rm SCET}_{{\rm ren}}(x;\mu)
\,.\end{align}
In this article we specialize to the case in which $ J^{\rm SCET}$
contains a single operator. The considerations below can be easily
generalized to the case of multiple operators for which the evolution
between two scales can be expressed in terms of $\mu$-ordered matrix
exponentials.

The factorization theorem holds for the
differential cross section, not the amplitude, and we therefore consider the
quantity ($M_H=Q$)
\begin{equation}
\int \! \df^4x \, e^{i q\cdot x}\langle 0| J^{\rm QCD}(x)J^{\rm QCD\,
  \dagger}(0) |0\rangle = H_{\rm bare}(M_H) \, O_{{\rm bare}}  = H_{\rm ren}(M_H,\mu) \, O_{{\rm ren}} (\mu)\,,
\end{equation}
where we defined the matrix element of the squared SCET current as
\begin{equation}
\label{eq:operatorME}
O \equiv \int \! \df^4x \, e^{i q\cdot x}\langle 0|  J^{\rm SCET}(x) J^{\rm
  SCET,\dagger}(0) |0\rangle 
\,,
\end{equation}
and introduced the hard function 
\begin{align}
H(M_H) \equiv |C(M_H)|^2
\,.
\end{align}
The bare and renormalized coefficients and matrix element of the
operator in \scetI{} are related by
\begin{align}
H_{\rm bare}(M_H) &= Z_O^{-1}(\mu) \, H_{\rm ren}(M_H;\mu)\,, & 
O_{{\rm bare}} &= Z_O(\mu) \, O_{{\rm ren}}(\mu)
\,.\end{align}
The $\mu$ dependence of the renormalized matching coefficient is
obtained from the $\mu$ independence of the bare matching coefficient,
\begin{align}
\frac{\df}{\df \ln \mu} H_{\rm bare}(M_H)= 0
\,,
\end{align}
from which follows the RG equation
\begin{align}
\label{eq:anomDimDef}
\gamma_H(\mu) \equiv \frac{\df}{\df \ln \mu} \ln H_{\rm ren}(\mu)  = \frac{\df}{\df \ln \mu} \ln Z_O(\mu)
\,.\end{align}
We will suppress the subscript ${\it ren}$ in the remaining
discussion, unless there is a possibility of confusion. We have also
dropped the $M_H$ dependence in $\gamma_H$, as mentioned at the beginning of this section.

The anomalous dimension has been proven to have the all-order
form~\cite{Manohar:2003vb,Bauer:2003pi,Chiu:2009mg}
\begin{align}
\gamma_H(\mu) = -4 \Gamma_{\rm cusp}[\alpha_s(\mu)] \ln \frac{\mu}{M_H}  + \widehat \gamma_H[\alpha_s(\mu)]
\,,
\label{5}
\end{align}
and contains only a single logarithm of $\mu$ to all orders.
The coefficient of the $\log \mu$ term is the cusp anomalous dimension, and the non-log term is denoted by $\widehat \gamma_H$.
Equation~\eqref{eq:anomDimDef} can be integrated to obtain $H_{\rm ren}$ giving the well-known result
\begin{align}
H(\mu_2)  = H(\mu_1)  \, U_H(\mu_1, \mu_2)\,, \qquad U_H(\mu_1, \mu_2) = \exp\left[ \int_{\mu_1}^{\mu_2} \frac{\df \mu'}{\mu'} \gamma_H(\mu')\right]
\,.
\label{7}
\end{align}
Given Eq.~\eqref{7}, one can write
\begin{align}
H(\mu) \, O_{{\rm ren}}(\mu) = H(\mu_H)  \,U_H(\mu_H, \mu_O) \,  O_{{\rm ren}}(\mu_O) 
\label{8}
\,.\end{align}

The matching coefficient has no large logarithms at the scale $\mu_H
\sim M_H$.
If one could find a scale $\mu_O$ at which the matrix element of the
operator is free of large logarithms one could sum all large
logarithms in the required product of $H$ and $O$ using the right hand
side of Eq.~\eqref{8}.
However, the matrix elements of \scetI{} operators still contain
multiple scales, and it is not possible to identify a single scale
$\mu_O$ at which they have no large logarithms.

One can further factorize $O_{{\rm ren}}$ into a convolution of soft
and jet functions, each of which depends on a single scale.
As long as $\beta >0$, the two scales satisfy $M_S \ll M_J$ for $\tau \ll 1$, and the two scales can be disentangled by another matching step. 
In particular, at the scale $\mu_J \sim M_J$ one can match SCET onto a
soft theory containing only Wilson lines interacting with soft degrees
of freedom.\footnote{For work towards a formulation of SCET without
  the separation of collinear and soft modes, see
  Refs.~\cite{Goerke:2017ioi,Inglis-Whalen:2020rpi}.}
This low energy effective theory reproduces exactly the soft function,
and the matching coefficient onto this theory is given by the two jet
functions.
After this matching step, one continues running in the soft theory. 
In the case of recoil-free angularities Eq.~\eqref{eq:angularity} in
$e^+e^-\to 2$ jets, described by the
factorization formula Eq.~\eqref{eq:factorization}, the soft and jet functions are
defined in Eq.~\eqref{eq:soft_and_jet}.
By means of a Laplace transform, the factorization formula for
$\beta\neq 0$ becomes a simple product and the soft and jet functions
satisfy RGEs similar to Eq.~\eqref{eq:anomDimDef}, i.e.
\begin{align}
\frac{\df\ln \hat{\cal S}(\mu)}{\df\ln\mu} &= \gamma_\Sobs(\mu)\,, &  \frac{\df\ln \hat{\cal J}_{n,\bar{n}}(\mu)}{\df\ln\mu} &= \gamma_\Jobs(\mu)\,,
\end{align}
with
\begin{align}
\gamma_\Sobs(\mu) & =  -4 \frac{\alpha}{\beta} \Gamma_{\rm cusp}[\alpha_s(\mu)] \ln \frac{\mu}{M_S} + \widehat\gamma^{\rm SCET_I}_\Sobs[\alpha_s(\mu)] \,,
\nonumber\\
\gamma_\Jobs(\mu) & = 2\frac{\alpha+\beta}{\beta} \Gamma_{\rm cusp}[\alpha_s(\mu)] \ln \frac{\mu}{M_J} + \widehat\gamma^{\rm SCET_I}_\Jobs[\alpha_s(\mu)]
\label{2.15}
\,.\end{align}
The non logarithmic terms $\hat \gamma$ of the anomalous dimensions
above will be given in Section~\ref{sec:SCETIAnomDim} (see also
Ref.~\cite{Hornig:2009vb}).
We have also added a superscript \scetI{} since we introduce many
closely related anomalous dimensions later in the paper. The cusp and
non-cusp terms depend on the angularity parameters
$\alpha,\beta$.\footnote{We remind the reader that we specifically
  refer to the choice $\alpha=1$, although in the expressions that
  follow the $\alpha$ dependence is kept explicit as the conclusions
  made here can be extended to observables other than conventional
  angularities.}

The above RGEs can be solved starting from initial conditions at
$\mu_S\sim M_S$ and $\mu_J\sim M_J$, at which the soft and jet
functions are free of large logarithms of $\mu$.
The factorization
theorem Eq.~\eqref{eq:factorizationLaplace} including the scale dependence
of the renormalized soft, jet and hard functions becomes
\begin{align}
\label{SCETIFact}
\hat\sigma[u]=H(\mu_H)  \, U_H^2(\mu_H, \mu)  \,
  \hat{\cal J}_n(\mu_J)  \,\hat{\cal J}_{\bar n}(\mu_J)  \,  U_J^2(\mu_J, \mu) \, \hat{\cal S}(\mu_S) \,  U_S(\mu_S, \mu)
\,,\end{align}
on evolving to a common scale $\mu$.
In Eq.~\eqref{SCETIFact}, we have as usual suppressed the dependence
on $M_F$ and
\begin{align}
\label{2.17}
\ln U_S(\mu_S, \mu) &= \int_{\mu_S}^\mu \frac{\df\mu'}{\mu'} \left[ -4  \,\frac{\alpha}{\beta}  \,\Gamma_{\rm cusp}[\alpha_s(\mu)] \ln \frac{\mu}{M_S} + \widehat\gamma^{\rm SCET_I}_\Sobs[\alpha_s(\mu)]\right] \,,\nonumber\\
\ln U_J(\mu_J, \mu) &= \int_{\mu_J}^\mu \frac{\df\mu'}{\mu'} \left[ 2 \,\frac{\alpha+\beta}{\beta}  \,\Gamma_{\rm cusp}[\alpha_s(\mu)] \ln \frac{\mu}{M_J} + \widehat\gamma^{\rm SCET_I}_\Jobs[\alpha_s(\mu)]\right]
\,,
\end{align}
are the evolution factors in the soft and collinear sectors.
%

\subsection{Resummation in \scetII{}}
\label{sec:SCETIIResummation}
In \scetII{}, matching QCD onto the effective theory proceeds in the
same way as in \scetI{}, and Eq.~\eqref{2.1} through Eq.~\eqref{8}
still hold.
As in \scetI{}, these equations could be used to resum all large logarithms if one identifies two (initial) scales $\mu_H$ and $\mu_O$ at which the Wilson coefficient and the matrix element of the operator have no large logarithms.
This was not possible in \scetI{} because two separate scales are
still present in the effective theory, which were disentangled by
defining jet and soft functions, each of which depended on a single scale.
Unlike \scetI{}, in \scetII{} the jet and soft functions actually live at the same scale, and one might naively think that at that common scale $\mu_O$ the perturbative expression of $O_{{\rm ren}}(\mu_O)$ contains no large logarithms.
However, one can show that to all orders in $\alpha_s$ a single
logarithm of the hard scale $\mu / M_H$ survives in the combination of
the soft and jet
functions~\cite{Beneke,Chiu:2007yn,Chiu:2007dg,Chiu:2008vv,Chiu:2009mg,Becher:2010tm},
as a consequence of the presence of rapidity divergences in the
calculation of radiative corrections to the soft and jet functions.
The introduction of an additional (rapidity) regulator, associated
with a new scale $\nu$, allows one to define separately the soft and
jet functions and compute the coefficient of this residual single
logarithm~\cite{Chiu:2009mg,Becher:2010tm,Becher:2011pf,Becher:2011xn}.

A related approach is the so-called rapidity renormalization group~\cite{Chiu:2011qc,Chiu:2012ir},
where one derives a coupled system of two RGEs in the scales $\mu$ and
$\nu$, whose solution can be exploited to sum all sources of large
logarithms.
Consider the example of the factorization theorem in
Eq.~\eqref{eq:factorization} for $\beta=0$. The Laplace transform
$\hat{O}_{\rm bare}$ of the operator matrix element $O_{\rm bare}$ in
Eq.~\eqref{eq:operatorME} can be written as
\begin{align}
\label{eq:hardCoefficientSplitII}
\hat{O}_{\rm bare}  &= \hat{\cal S}_{\rm bare}(\nu)  \, \hat{\cal
                      J}_{n,\rm bare}(\nu) \, \hat{\cal J}_{\bar{n},\rm bare}(\nu)
\,.
\end{align}
One can subtract the $1/\epsilon$ divergences by defining
\begin{align}
\label{eq:scetIIRenorm}
\hat{\cal S}_{\rm bare}(\nu) &= Z_S(\mu, \nu)\, \hat{\cal S}_{\sub}(\mu, \nu)\,, & \hat{\cal J}_{\rm bare}(\nu) &= Z_J(\mu, \nu)\, \hat{\cal J}_{\sub}(\mu, \nu)
\,,
\end{align}
so that $\hat{\cal S}_{\sub}(\mu, \nu)$ and $\hat{\cal J}_{\sub}(\mu, \nu)$ are finite.
Specific rapidity regularization schemes (for
example~\cite{Becher:2011dz,Li:2016axz}) regulate only the real
radiation integrals but not the virtual corrections.
Of course, a consistent scheme requires using the same regulators in
the real and virtual corrections in order not to break unitarity (see also the discussion in Ref.~\cite{Becher:2011dz}).
The breaking of unitarity is reflected in an apparent IR unsafety of the soft and jet
functions, which implies that some of the $1/\epsilon$ divergences are
of IR nature. 
However, this issue can be overcome by noticing
that these spurious divergence cancel in the computation of physical
quantities, that is in the combination of soft and jet functions that
appear in the factorization theorem. 
Therefore, schemes of this type
can still be used for practical computations and one can still
define
  \begin{align}
 \label{2.21}
 \hat{O}_{\rm ren}(\mu) &= \hat{\cal S}_{\sub}(\mu, \nu)\, \hat{\cal J}_{n,
 \rm sub}(\mu, \nu)\, \hat{\cal J}_{\bar{n},\rm sub}(\mu, \nu), &    \hat{O}_{\rm bare}   &= Z_{O, \rm ren}(\mu)   \hat{O}_{\rm ren}(\mu) 
 \,,
 \end{align}
 with
 \begin{align}
 \label{2.21a}
 Z_{O, \rm ren}(\mu) = Z_{S}(\mu, \nu)\, Z^2_{J}(\mu, \nu)
 \,.\end{align}
We have deliberately denoted the {\it renormalized}
  soft and jet functions with the subscript {\it sub} to emphasize
  that  in some regularization schemes  the
  definition Eq.~\eqref{eq:scetIIRenorm} is not a renormalization in the
  strict sense.
  For the same reason, in the derivation of the RGEs that follows, we
  do not explicitly use the fact that the $1/\epsilon$ divergences are
  of UV origin\footnote{This means we don't assume that the
    $1/\epsilon$ divergences cannot depend on infrared scales, or
    cannot have observable dependence.}.  In this sense, the use of
  the rapidity renormalization group is to be interpreted only as a
  computational tool.

One can now derive the differential evolution equations in the
renormalization scale $\mu$ as
\begin{align}
\label{eq:muRGE_withLambda}
\frac{\df}{\df \ln \mu} \ln F_{\sub}(\mu, \nu) = -\frac{\df}{\df \ln \mu} \ln Z_{F}(\mu, \nu) \equiv \gamma^{(\mu)}_F(\mu, \nu) 
\,,
\end{align}
with $F=\hat{\cal S},\,\hat{\cal J}_{n},\,\hat{\cal J}_{\bar n}$. The
$\nu$ dependence in the $\mu$-anomalous dimensions cancels in the
combination
\begin{align}
\gamma^{(\mu)}_O(\mu) = \gamma^{(\mu)}_S(\mu, \nu) + 2 \gamma^{(\mu)}_J(\mu, \nu)=-\gamma^{(\mu)}_H(\mu)
\label{eq:2.22}
\,,\end{align}
since $H_{\rm ren}(\mu_H)$ does not depend on $\nu$.

Consistency arguments can be used to derive an all order expression
for the form of the $\mu$-anomalous dimensions.
First, using arguments analogous to those in
Refs.~\cite{Manohar:2003vb,Bauer:2003pi,Chiu:2009mg},
Eq.~\eqref{eq:2.22} implies that the soft and collinear
$\mu$-anomalous dimensions can depend at most on a single logarithm of
the rapidity regularization scale $\nu$.
Second, since the $\nu$ dependence cancels between the soft and jet functions, it is determined by the simultaneous soft and collinear limit, and is therefore proportional to the cusp anomalous dimension. 
Third, the rapidity regulator regulates the entire UV divergence in
the simultaneous soft and collinear limit, so that the jet anomalous
dimension does not contain an explicit $\ln \mu$.
We use these three conditions together with the fact that the $\mu$ and $\nu$ dependence enters in ratios $\mu / M_F$ and $\nu / N_F$ and that the canonical scales satisfy
\begin{align}
\label{nuFDef}
\nu_J \sim N_J \equiv M_H\,, \qquad \nu_S \sim N_S \equiv M_S\,.
\end{align}
One finds\footnote{Note that the anomalous dimensions $\widehat\gamma_F[\alpha_s(\mu)]$ is not the same as the \scetI{} anomalous dimension $\widehat\gamma^{\rm SCET_I}_{\Fobs}[\alpha_s(\mu)]$ discussed in Eq.~\eqref{2.15}} 
\begin{align}
\label{gamma_SCETII}
\gamma^{(\mu)}_S(\mu, \nu)  &= 4\Gamma_\cusp[\alpha_s(\mu)] \ln \frac{\mu}{\nu} + \widehat\gamma_S[\alpha_s(\mu)]\,,
\nonumber\\
\gamma^{(\mu)}_J(\mu, \nu)  &= 2\Gamma_\cusp[\alpha_s(\mu)] \ln \frac{\nu}{N_J} + \widehat\gamma_J[\alpha_s(\mu)]
\,.
\end{align}
The observable dependence in \scetII{} anomalous dimensions arises
from real diagrams in the large rapidity region.
Since those divergences are regulated by the rapidity regulator, the $\mu$ anomalous dimension is observable independent.

The solution to
Eq.~\eqref{eq:muRGE_withLambda}
\begin{align}
\label{eq:muRGESol}
F_{\sub}(\mu, \nu) &= F_{\sub}(\mu_F, \nu) \, \exp\left[ \int_{\mu_F}^{\mu} \frac{\df \mu'}{\mu'} \gamma^{(\mu)}_F(\mu', \nu)\right] 
\nonumber\\
& \equiv F_{\sub}(\mu_F, \nu) \, U_F(\mu_F, \mu; \nu) \,,
\end{align}
is not sufficient to perform the resummation since the initial condition
$F_{\sub}(\mu_F, \nu)$ still contains large logarithms of the ratio
$\nu/N_F$. Therefore a second differential equation in the rapidity
scale $\nu$ is necessary.
The nature of the scale $\nu$ is quite different from that of the
renormalization scale $\mu$.
Unlike for the scale $\mu$, the dependence on $\nu$ cancels only between
the soft and the zero-bin subtracted~\cite{Manohar:2006nz} collinear sectors, since $ \hat{\cal S}_{\rm bare}(\nu) $ and $\hat{\cal J}_{\rm bare}(\nu)$ depend on $\nu$,
so that
\begin{align}
\label{eq:WrongLambdaDeriv}
\frac{\df}{\df \ln \nu} \ln F_{\sub}(\mu, \nu)  \neq - \frac{\df}{\df \ln \nu} \ln Z_F(\mu, \nu)
\,.
\end{align}
However, a differential equation describing the change in $\nu$ can be obtained by  taking the derivative of Eq.~\eqref{eq:muRGESol} with respect to $\nu$. 
This yields
\begin{align}
\label{eq:RapRGE}
\frac{\df}{\df \ln \nu} F_{\sub}(\mu, \nu) & = \left[\frac{\df}{\df \ln \nu}F_{\sub}(\mu_F, \nu) \right] U_F(\mu_F, \mu; \nu) + F_{\sub}(\mu_F, \nu) \left[\frac{\df}{\df \ln \nu}U_F(\mu_F, \mu; \nu)\right]
\nonumber\\
&= F_{\sub}(\mu, \nu) \left[\frac{\df}{\df \ln \nu} \ln F_{\sub}(\mu_F, \nu) + \int_{\mu_F}^{\mu} \frac{\df \mu'}{\mu'} \frac{\df}{\df \ln \nu}\gamma^{(\mu)}_F(\mu', \nu) \right]
\nonumber\\
& = F_{\sub}(\mu, \nu) \left[\frac{\df}{\df \ln \nu} \ln F_{\sub}(\mu_F, \nu) - 2 a_F \int_{\mu_F}^{\mu} \frac{\df \mu'}{\mu'} \Gamma_\cusp[\alpha_s(\mu)] \right]
\,,
\end{align}
with
\begin{align}
a_S = 2\,, \qquad a_J = -1
\,,
\label{2.29}
\end{align}
and where we have used Eq.~\eqref{gamma_SCETII} in the last line of \eqref{eq:RapRGE}. 

One can obtain more constraints on the $\nu$ dependence, following
again an argument similar to that in
Refs.~\cite{Manohar:2003vb,Bauer:2003pi,Chiu:2009mg}.
The combination of soft and jet functions in the factorization theorem, Eq.~\eqref{2.21}, is independent of $\nu$.
The last term in square brackets vanishes when the soft and jet
contributions in the factorization theorem are combined, by
Eq.~\eqref{eq:2.22}.
This gives a constraint on the first term in square bracket of Eq.~\eqref{eq:RapRGE},
\begin{align}
\label{JSnuindep}
\frac{\df}{\df \ln \nu} \ln \hat{\mathcal S}_{\sub} (\mu_S, \nu) +
  2\frac{\df}{\df \ln \nu} \ln \hat{\mathcal J}_{\sub} (\mu_J, \nu) &=0
\,. \end{align}
Since any dependence on $\ln \nu$ of $\df \ln \mathcal{S}_{\sub} (\mu_S, \nu) / \df \ln \nu$ and $\df \ln \mathcal{J}_{\sub} (\mu_J, \nu) / \df \ln\nu$ is through the ratios $\nu / N_S$ and $\nu/N_J$, respectively, these derivatives can in fact not depend on $\nu$ at all. Combining this with \eqref{JSnuindep} implies
\begin{align}\label{2.28}
\frac{\df}{\df \ln \nu} \ln F_{\sub}(\mu, \nu) \equiv  \gamma_\Fobs^{(\nu)}(\mu)
\,,\end{align}
with
\begin{align}
\label{2.30}
\gamma^{(\nu)}_\Fobs(\mu) = a_F \left[ \gamma^{(\nu)}_{\alpha, \beta}(\mu_F) - 2 \int_{\mu_F}^{\mu} \frac{\df \mu'}{\mu'} \Gamma_\cusp[\alpha_s(\mu')]\right] 
\,.
\end{align}

An important observation is that the derivatives in $\mu$ and $\nu$ commute
\begin{align}
\label{2.32}
\left[ \frac{\df}{\df \ln \nu},  \frac{\df}{\df \ln \mu}\right] \ln F_{\sub}(\mu, \nu) = 0
\,,\end{align}
since
\begin{align}
\frac{\df}{\df \ln \mu} \gamma^{(\nu)}_\Fobs(\mu) &= \frac{\df}{\df \ln \nu} \gamma^{(\mu)}_\Fobs(\mu,\nu)
\,,\end{align}
and therefore one can resum all logarithms of $\mu$ and $\nu$ by
solving the system of differential equations in
Eq.~\eqref{eq:muRGE_withLambda},~\eqref{eq:RapRGE}
\begin{align}
\label{2.33}
F_{\sub}(\mu, \nu) = F_{\sub}(\mu_F, \nu_F) \, U_F(\mu_F, \nu_F, \mu, \nu)
\,,
\end{align}
where $F_{\sub}(\mu_F, \nu_F)$ is now free of large logarithms.
To obtain the evolution kernel $U_F$, one performs the integration along
the path $(\mu_F, \nu_F) \to (\mu, \nu_F) \to (\mu, \nu)$,\footnote{Due to Eq.~\eqref{2.32}, one can perform the integration along any path in $\mu$ and $\nu$, and the path chosen here is just a convenient choice} obtaining
\begin{align}
\label{eq:combinedEvolution1}
U_F(\mu_F, \nu_F, \mu, \nu) = U_F^{(\mu)}(\mu_F, \mu; \nu_F) \, U_F^{(\nu)}(\nu_F, \nu; \mu) 
\,,
\end{align}
with
\begin{subequations}
\label{eq:SCETIIRapidityEvolution}
\begin{align}
\label{eq:SCETIIRapidityEvolutiona}
\ln U_S^{(\mu)}(\mu_S, \mu; \nu_S) &= \int_{\mu_S}^\mu \frac{\df\mu'}{\mu'} \left[ 4\Gamma_\cusp[\alpha_s(\mu')] \ln  \frac{\mu'}{\nu_S} + \widehat\gamma_S[\alpha_s(\mu')] \right] \,,
\\
\label{eq:SCETIIRapidityEvolutionb}
\ln U_J^{(\mu)}(\mu_J, \mu; \nu_J) &= \int_{\mu_J}^\mu \frac{\df\mu'}{\mu'} \widehat\gamma_J[\alpha_s(\mu^\prime)] \,,
\\
\label{eq:SCETIIRapidityEvolutionc}
\ln U_S^{(\nu)}(\nu_S, \nu; \mu)  &=  \int_{\nu_S}^\nu\frac{\df\nu'}{\nu'} \left[  - 4 \int_{\mu_S}^\mu\frac{\df\mu'}{\mu'} \Gamma_\cusp[\alpha_s(\mu')]  + 2\gamma^{(\nu)}_{\alpha, \beta}[\alpha_s(\mu_S) ] \right]  \, \nonumber \\
&= \left[- 4 \int_{\mu_S}^\mu\frac{\df\mu'}{\mu'} \Gamma_\cusp[\alpha_s(\mu')]   +  2\gamma^{(\nu)}_{\alpha, \beta}[\alpha_s(\mu_S) ] \right]  \ln \frac{\nu}{\nu_S} \,,
\\
\label{eq:SCETIIRapidityEvolutiond}
\ln U_J^{(\nu)}(\nu_J, \nu; \mu)  &= \int_{\nu_J}^\nu \frac{\df\nu'}{\nu'}  \left[  2 \int_{\mu_J}^\mu \frac{\df\mu'}{\mu'}\Gamma_\cusp[\alpha_s(\mu')] - \gamma^{(\nu)}_{\alpha, \beta}[\alpha_s(\mu_J) ] \right]  \, \nonumber \\
&=    \left[ 2 \int_{\mu_J}^\mu \frac{\df\mu'}{\mu'}\Gamma_\cusp[\alpha_s(\mu')] -  \gamma^{(\nu)}_{\alpha, \beta}[\alpha_s(\mu_J) ] \right]  \ln \frac{\nu}{\nu_J}
\,.
\end{align}
\end{subequations}
$ U_J^{(\mu)}(\mu_J, \mu; \nu_J) $ has no cusp piece, and the $\mu$
dependence in the jet function is therefore single logarithmic.
Note that in \scetII{} $\mu_J \sim\mu_S \sim M_S = M_J$, and hence the
argument of $\gamma^{(\nu)}$ is evaluated at the $\nu$-independent
scale $\mu_F$. This ensures that the net effect of the $\nu$
dependence in the combination of soft and jet functions is only
single logarithmic.
Given these evolution equations, the factorization
theorem Eq.~\eqref{eq:factorizationLaplace} can be written as
\begin{align}
\label{SCETIIFact}
\hat\sigma[u]=H(\mu_H)  U_H^2(\mu_H, \mu)  \,
  \hat{\cal J}_n(\mu_J,\nu_J) \hat{\cal J}_{\bar n}(\mu_J,\nu_J)  \,  U_J(\mu_J, \nu_J, \mu, \nu)  \, \hat{\cal S}(\mu_S,\nu_S) \,  U_S(\mu_S, \nu_S, \mu, \nu)
\,,\end{align}
where we have again not shown explicitly the dependence on $M_F$. 

We conclude this section by pointing out that the fact that the combination of the functions $\hat{\cal S}(\mu, \nu) \hat{\cal J}_n (\mu, \nu) \hat{\cal J}_{\bar n}(\mu, \nu) $ has to be independent of the rapidity scale $\nu$ can be used to derive a tight constraint on the functional form of the functions $\hat{\cal S}(\mu, \nu)$ and $\hat{\cal J}(\mu, \nu)$. 
Obviously one needs to have
\begin{align}
\frac{\df \ln \hat{\cal S}}{\df \ln \nu}(M_S, N_S; \mu, \nu) = -
  \frac{\df \ln \hat{\cal J}_n}{\df \ln \nu}(M_J, N_J; \mu,
  \nu)-\frac{\df \ln \hat{\cal J}_{\bar n}}{\df \ln \nu}(M_J, N_J; \mu, \nu)\,.
\end{align}
Using that the dependence is only through the ratios $\mu / M_F$ and
$\nu / N_F$, and that $M_S = M_J$ (but $N_S\neq N_J$) in \scetII{},
one finds that the derivatives can not depend on the ratios
$\nu / N_F$ and therefore
\begin{align}
\frac{\df \ln \hat{\cal S}}{\df \ln \nu}(M_S, N_S; \mu, \nu) &= \frac{\df \ln \hat{\cal S}}{\df \ln \nu}(\mu / M_S) \equiv c(\mu / M_S)\,,
\nonumber\\
\frac{\df \ln \hat{\cal J}_{\bar n}}{\df \ln \nu}(M_J, N_J; \mu,
  \nu) &=\frac{\df \ln \hat{\cal J}_{n}}{\df \ln \nu}(M_J, N_J; \mu, \nu) = \frac{\df \ln \hat{\cal J}_n}{\df \ln \nu}(\mu / M_J)= -\frac{1}{2} c(\mu / M_J)\,,
\end{align}
where it is crucial that $M_S = M_J$. This means that the soft and jet functions have the general form
\begin{align}
\ln \hat{\cal S}(\mu, \nu) &= \ln \widetilde{\cal S}(\mu / M_S) +
  \int_{N_S}^{\nu} \frac{\df \nu'}{\nu'} \,  c(\mu /
  M_S)= \ln \widetilde{\cal S}(\mu / M_S) +
  \,  c(\mu /
  M_S)\ln\frac{\nu}{N_S}\,,
\nonumber\\
\ln \hat{\cal J}_n(\mu, \nu) &= \ln \widetilde{\cal J}_n(\mu /
  M_J) - \frac{1}{2} \int_{N_J}^{\nu} \frac{\df \nu'}{\nu'} \,
  c(\mu / M_J)= \ln \widetilde{\cal J}_n(\mu / M_J) - \frac{1}{2} \,  c(\mu / M_J)\ln\frac{\nu}{N_J}
\,.
\end{align}
All functions in the above two equations depend on $\alpha_s(\mu)$,
which can be equivalently rewritten in terms of
$\alpha_s(M_S)=\alpha_s(M_J)$ and a different functional dependence on
$\mu/M_S$ or $\mu/M_J$.
One can easily see that the solution to the RGEs given in
Eqs.~\eqref{eq:SCETIIRapidityEvolution} satisfies this constraint.

\section{Choice of UV regulator in real radiation
  integrals}
\label{sec:RapRegulator}

In this section we discuss the criteria for choosing a regulator for
real phase space integrals.
As already discussed in the introduction,
these integrals become UV divergent in the effective theory after the
integration measure and physical phase space constraints have been
multipole expanded. 
This can be easily seen by considering the angularity
Eq.~\eqref{eq:angularity} that for a single parton state can be
expressed as
\begin{align}
  \tau_{\alpha,\beta}(k^+,k^-) = \frac{\left[{\rm min}(k^+, k^-)\right]^\frac{\alpha+\beta}{2}\left[{\rm max}(k^+, k^-)\right]^\frac{\alpha-\beta}{2}}{Q}
  \,,\end{align}
which, at the one-loop level, gives rise to the following schematic phase
space integral\footnote{We assume, without loss of generality, that $k^- >
  k^+$, and we impose the on-shell condition $k_\perp^2=k^+k^-$.}
\begin{equation}
\label{eq:integral}
\int_0^\infty \frac{d k_\perp}{k_\perp^{1+2\epsilon}} \int_0^\infty \frac{d k^-}{k^-}\Theta\left(k^--\frac{k_\perp^2}{k^-}\right)\delta\left(\tau-\tau_{\alpha,\beta}\left(\frac{k_\perp^2}{k^-},k^-\right)\right)\,.
\end{equation}
The $k_\perp$ integral is regulated by standard dimensional regularization
both in the IR and UV limits.
In the \scetI{} case ($\beta\neq 0$), this
is sufficient also to regulate the integral over the light cone
component $k^-$ due to the constraint imposed by the observable $\tau_{\alpha,\beta}$
that relates $k_\perp$ and $k^-$. 
As is well known, this is not the case
in \scetII{} ($\beta=0$), and one has an additional rapidity divergence from
the limit in which one of the light cone components of $k$ tends to infinity.

In order to cope with these divergences, common rapidity
regularization schemes in
\scetII{}~\cite{Ji:2004wu,Chiu:2009yx,Becher:2011dz,Collins:2011zzd,Chiu:2012ir,Echevarria:2015byo,Li:2016axz,Chay:2020jzn}
proceed by introducing a new regulator in the $k^-$
integral~\eqref{eq:integral}, which effectively acts to damp the
integral above a certain scale $ k^-, k^+ \sim \nu$.
This regulates the divergence of the integral over the light cone
variables $k^\pm$, while the value of $k_\perp$ is instead fixed by
the observable's measurement function.
In general, one does not want the
rapidity regulator to affect the infrared divergences of the phase
space integral, and this is easily avoided by taking the limit in the
regulator {\it before} one takes the $\epsilon\to 0$ limit. 
This
ensures that the infrared limit is regulated by dimensional
regularization, and the infrared structure of QCD is reproduced on
combining the soft and collinear sectors.
In problems involving the resummation of jet observables, such as the
one discussed in this article, one often regulates only real radiation
integrals while leaving the virtual integrals untouched by the
regularization procedure. As discussed in the context of \scetII{}
theories (cf. Section~\ref{sec:SCETIIResummation}), some care is
needed to ensure that the dependence on the rapidity regulator in the
real radiation cancels in physical quantities.
As a result, all  UV divergences associated with real radiation
in \scetII{} are captured by the rapidity regulator. 
This makes the anomalous dimension governing the $\mu$ RGEs of the
soft and jet functions observable independent, while the rapidity
anomalous dimension governing the $\nu$ RGE is observable dependent.

We wish to achieve the same separation for the \scetI{}
anomalous dimension into an observable independent and an observable
dependent component, as for \scetII.
In \scetI, the observable dependent contribution will arise from the
large momentum region of the real radiation phase space, but
separating them from other singularities is a little more subtle than
in the \scetII{} case.
In analogy with \scetII{}, we consider the introduction of an extra UV
regulator (we refrain from calling it a rapidity regulator as no
rapidity divergences are present in \scetI{} theories).
The important property required for the extra UV regulator is that it
should not modify the IR structure of the effective theory, and that
it cancels between soft and jet functions, leaving the hard function
unaffected.
This ensures that the IR structure 
continues to reproduce that of QCD, which removes the condition that the $\epsilon \to 0$ has to be taken last.
Contrary to what happens in \scetII{}, dimensional regularization is sufficient to regulate all UV divergences in \scetI.
Therefore, separating out the observable dependence in the \scetI{}
case crucially requires taking the $\epsilon\to 0$ limit first,
otherwise the procedure would naively collapse to standard dimensional
regularization.

A second important condition is that the introduction of the extra
regulator must lead to a consistent system of RGEs to perform the
resummation. 
In particular, this implies that there needs to be an
integration path in the $\{\mu,\nu\}$ plane that allows one to resum
all large logarithms. 
If
\begin{equation}
\left[\frac{d}{d\ln\mu}, \frac{d}{d\ln\nu}\right] \ln F= 0\,,
\label{eq:criterion2}
\end{equation}
where $F=\hat{\cal S}, \hat{\cal J}_n, \hat{\cal J}_{\bar n}$ are the terms in the factorization
theorem Eq.~\eqref{eq:factorization}, then the integration is path independent and any path can be used to integrate the RGEs.

It is natural to expect that a subset of the regularization schemes
currently used for rapidity regularization satisfy the two criteria
above and thus can be adopted for this task.
In particular, the condition stemming from the first criterion
requires that the limit $\epsilon\to 0$ and the limit in the rapidity
regulator commute in \scetII{} problems.
For instance, the analytic regulator proposed in
Ref.~\cite{Chiu:2012ir} does not satisfy this requirement and
therefore cannot be adopted for our purposes.
However, the exponential regulator of
Ref.~\cite{Li:2016axz} satisfies both criteria given
above. This procedure amounts to replacing the integration measure for
each real particle as follows
\begin{equation}
\label{eq:expreg}
\df^dk \,\delta(k^2)\theta(k^0)\to \df^dk \,\delta(k^2)\theta(k^0)\,e^{-\frac{k^++k^-}{\nu}e^{-\gamma_E}}\,,
\end{equation}
which regulates the integral when its energy (or equivalently  either
of its light cone components) becomes larger than a regularization
scale $\nu$.\footnote{This also regulates the $k_\perp$ integral in the UV due
  to the on-shellness condition $k^+k^-=k_\perp^2$.}
In coordinate space,  this procedure 
amounts to shifting the light cone coordinates $x^{\pm}$ by an
imaginary amount $i e^{-\gamma_E}/(2\nu)$, hence regularizing the
$x^{\pm}\to 0$ UV singularity. At the same time, the
prescription Eq.~\eqref{eq:expreg} does not affect the IR limit of
phase space integrals, which is dealt with in standard dimensional
regularization. 

\begin{figure}
\begin{center}
\includegraphics[width=8cm]{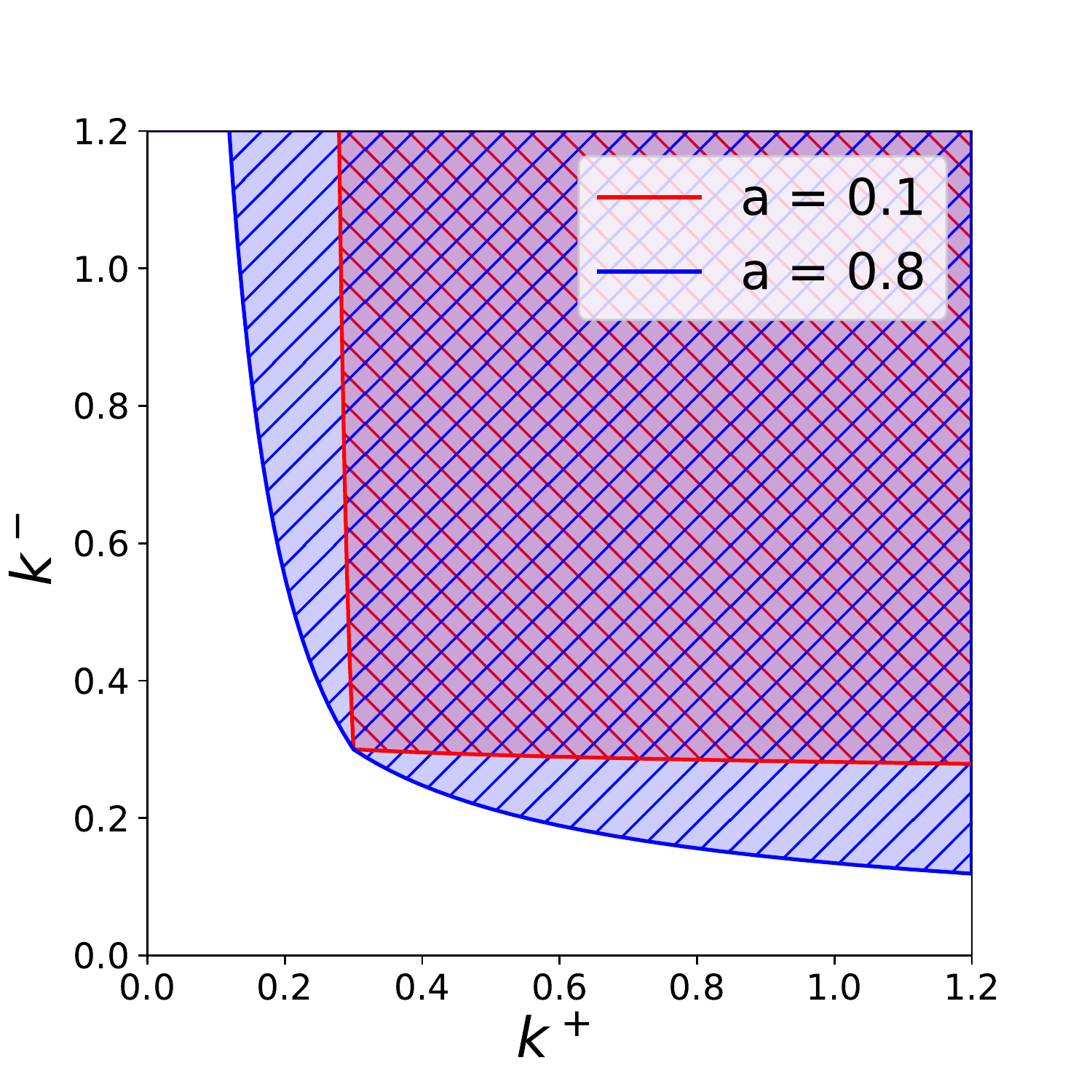}\includegraphics[width=8cm]{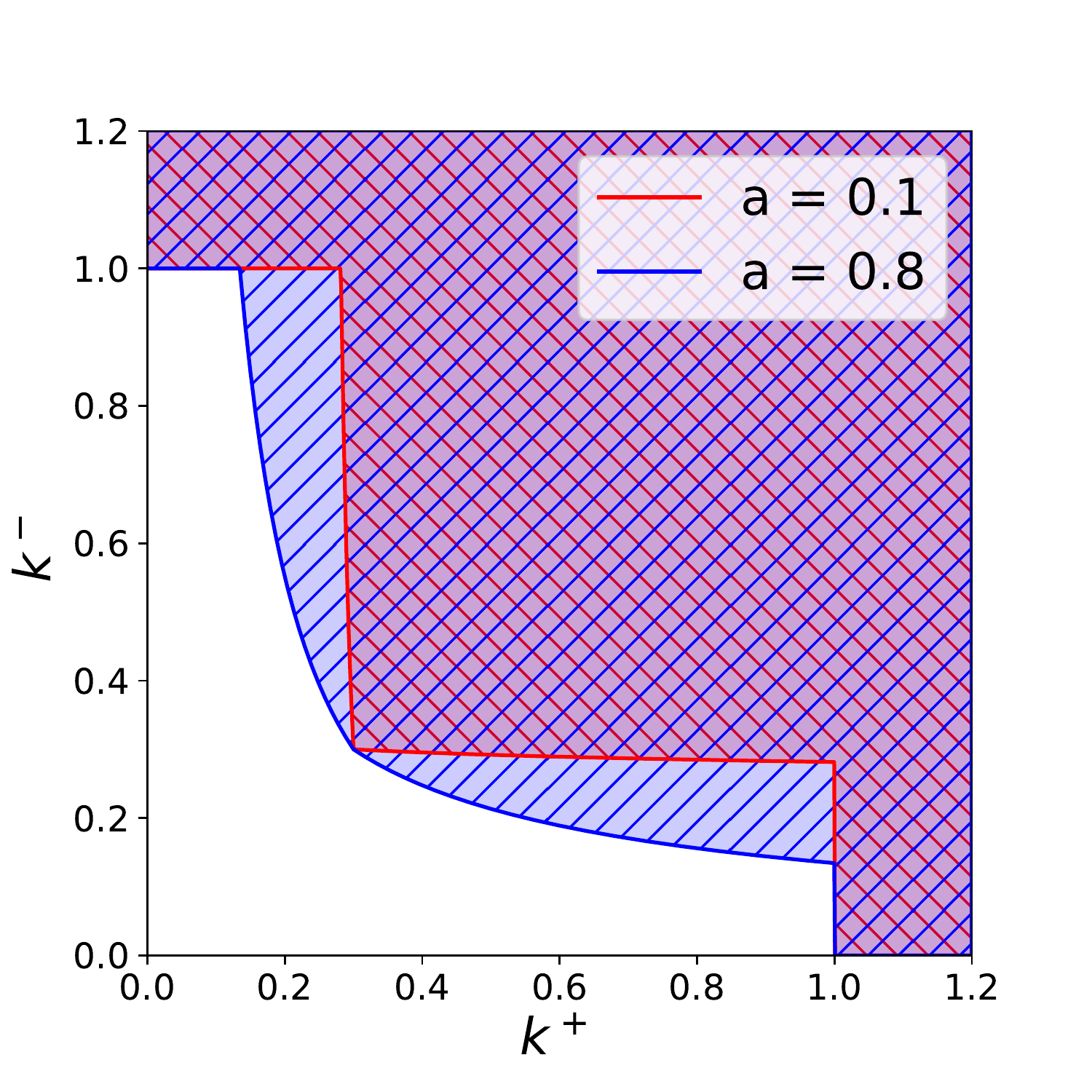}
\end{center}
\caption{\label{fig:phase}Phase space diagram in the $\{k^+, k^-\}$
  plane. The phase space region where there is a virtual contribution
  but no real radiation is shaded, $a = 0.1$ (red) and
  $a=0.8$ (blue). On the left we show the phase space region without
  the extra UV regulator, and on the right the region with the extra
  UV regulator which has been drawn as a hard cutoff $k^{+}\lesssim\nu$,
  $k^{-}\lesssim\nu$ for illustrative purposes.}
\end{figure}

One can understand the effect of the extra UV regulator by
looking at the phase space for the one-loop soft function in the
$\{k^+,k^-\}$ plane for the cumulative distribution, shown in
Fig.~\ref{fig:phase} for a conventional angularity
$\alpha=1$ and
$\beta=1-a$. The virtual graphs are integrated over all $k^\pm$,
whereas the real radiation graphs are constrained to have
$\tau_{\alpha,\beta}(k^+,k^-)<\tau_s$. The IR singularities cancel
between real and virtual graphs, and we have shaded the region where
there is only a virtual contribution.
The fact that without the extra UV regulator (shown on the left) UV divergences are observable dependent can be understood quite easily from this phase space diagram. 
The difference between two observables (in the figure represented by
the two values $a = 0.1$ and $a = 0.8$) is computed by integrating
over the region shaded blue but not red, which extends all the way to infinity. 
Thus, the $1/\epsilon$ divergences, and therefore the anomalous dimensions are observable dependent.
In the presence of the extra UV regulator (shown on the right) there are no $1/\epsilon$ divergences in the difference between two observables, from which one can expect that the $\mu$ anomalous dimensions are observable independent. 
The dependence on the extra UV regulator $\nu$, on the other hand, does depend on the observable.

\section{Structure of the RGEs and relationship between \scetI{} and
  \scetII{}}
\label{sec:RGRelationship}
In this section we will derive and discuss the system of RGE  in \scetI{} in the presence of the extra UV regulator.
We begin by briefly discussing the case of the one loop soft function, where one can explicitly
see some of the features introduced to all perturbative orders later.
\subsection{An example: the one loop soft function for $a=0$}
\label{sec:softOneLoopthrust}

Consider the soft
function Eq.~\eqref{eq:soft_and_jet} in the case of thrust ($\alpha=\beta=1$)
supplemented with the exponential regulator prescription Eq.~\eqref{eq:expreg}. At one loop
in the $\overline{\rm MS}$ scheme,
\begin{align}
\hat{\cal S}_{\rm bare}(\mu,\nu) = 1+2 \, \frac{\alpha_s(\mu)}{\pi}\, C_F \,
  e^{-\gamma_E\frac{Q\tau}{\nu}}\left(\frac{\mu^2}{Q\nu}\right)^{\epsilon}\tau^{-1-\epsilon}\frac{
  \Gamma\left(-\epsilon,\frac{e^{-\gamma_E}Q\tau}{\nu}\right)}{\Gamma(1-\epsilon)}\,.
\end{align}
Taking the Laplace transform~\eqref{eq:laplace} of the soft function
and performing the Laurent expansion for $\epsilon\to 0$ followed by that
for $\nu\to \infty$ gives (with $u_0=e^{-\gamma_E}$) 
\begin{align}
\label{eq:soft_thrust}
\hat{\cal S}_{\rm bare}(\mu,\nu) = 1+\frac{\alpha_s(\mu)}{\pi} C_F
  \left(\frac{1}{\epsilon^2}+\frac{2}{\epsilon}\,{\ln\frac{\mu}{\nu}}+\ln^2\frac{\mu}{\nu}-\ln^2\frac{\mu u}{Q
  u_0}+2\ln\frac{\mu}{\nu}\ln\frac{\mu u}{Q
  u_0}-\frac{\pi^2}{12}\right)\,.
\end{align}
We now renormalize the $\epsilon$ singularities using the standard
procedure outlined in Section~\ref{sec:RGEStructure} and obtain the
renormalized soft function
\begin{align}
\hat{\cal S}_{\rm ren}(\mu,\nu) = 1+\frac{\alpha_s(\mu)}{\pi} C_F
  \left(\ln^2\frac{\mu}{\nu}-\ln^2\frac{\mu u}{Q
  u_0}+2\ln\frac{\mu}{\nu}\ln\frac{\mu u}{Q
  u_0}-\frac{\pi^2}{12}\right)\,.
\end{align}
It is instructive to first analyze the system of RGEs governing
  the evolution of the soft function in the $\{\mu,\nu\}$ plane in a
  fixed coupling approximation, obtaining
\begin{align}
\label{eq:RGE_thrust}
\frac{\df\ln \hat{\cal S}_{\rm ren}}{\df\ln\mu} &=4 C_F\frac{\alpha_s(\mu)}{\pi}\ln\frac{\mu}{\nu}\,,\notag\\
\frac{\df\ln \hat{\cal S}_{\rm ren}}{\df\ln\nu} &= - 2 C_F\frac{\alpha_s(\mu)}{\pi} \left[ \ln\frac{\mu}{\nu} +  \ln\frac{\mu u}{Q u_0} \right]
\,, \end{align}
where the two differential equations are coupled due to the
contribution to the single pole proportional to $\ln\nu$. Using
Eq.~\eqref{eq:scalesDef} with $\tau \to u_0/u$ in Laplace space, we
can write the second of these as
\begin{align}
\label{eq:RGE_thrust1}
\frac{\df\ln \hat{\cal S}_{\rm ren}}{\df\ln\nu} &= - 2 C_F\frac{\alpha_s(\mu)}{\pi} \ln \frac{\mu^2}{M_S \nu} = - 4 C_F\frac{\alpha_s(\mu)}{\pi} \ln \frac{\mu}{\mu(\nu)} 
\,,
\end{align}
where $\mu(\nu) = M_S \sqrt{\nu/N_S} = \sqrt{M_S \nu}$ (since
$N_S=M_S$) is a new scale that appears in the soft and jet functions,
which will be discussed in more detail in the sections
below. Including the running coupling effects, the RGEs become
\begin{align}
\label{eq:RGE_thrust2}
\frac{\df\ln \hat{\cal S}_{\rm ren}}{\df\ln\mu} &=4 C_F\frac{\alpha_s(\mu)}{\pi}\ln\frac{\mu}{\nu}\,,\notag\\
\frac{\df\ln \hat{\cal S}_{\rm ren}}{\df\ln\nu} &= 
                                   -\int_{\mu(\nu)}^\mu\frac{d\mu^\prime}{\mu^\prime}
                                   4 C_F\frac{\alpha_s(\mu^\prime)}{\pi}\,.
\end{align}

\subsection{Evolution equations of \scetI{} with a UV regulator for
  real radiation}
As illustrated in the one-loop example of the previous section, in the
presence of an additional UV regulator for the real radiation, the soft
and jet functions depend both on $\mu$ and $\nu$, and much of the
discussion will proceed along very similar lines to what was discussed
for the case of \scetII\ in Section~\ref{sec:SCETIIResummation}, with
a few crucial differences.
At the canonical scales
\begin{align}
\mu_S =  M_S \,, \qquad \mu_J = M_J \,, \qquad \nu_S = N_S = M_S\,, \qquad \nu_J = N_J = M_H\,,
\end{align}
with the $M_F$ defined in Eq.~\eqref{eq:scalesDef}, the soft and jet
functions contain no large logarithms.
As in \scetII{}, the $\mu$ dependence of the soft and jet functions is
obtained by requiring that the bare functions are independent of the
scale $\mu$, leading to Eq.~\eqref{eq:muRGE_withLambda}, repeated here
for convenience
\begin{align}
\label{4.2}
\frac{\df}{\df \ln \mu} \ln F_{\sub}(\mu, \nu) = -\frac{\df}{\df \ln \mu} \ln Z_{F}(\mu, \nu) = \gamma^{(\mu)}_F(\mu, \nu)
\,,
\end{align}
with $F=\hat{\cal S},\,\hat{\cal J}_{n},\,\hat{\cal J}_{\bar n}$.
The form of the anomalous dimensions $\gamma^{(\mu)}_F(\mu, \nu) $ are
also the same as in \scetII{},
\begin{align}
\label{gamma_SCETIIlike}
\gamma^{(\mu)}_S(\mu, \nu)  &= 4\Gamma_\cusp[\alpha_s(\mu)] \ln \frac{\mu}{\nu} + \widehat\gamma_S[\alpha_s(\mu)]\,,
\nonumber\\
\gamma^{(\mu)}_J(\mu, \nu)  &= 2\Gamma_\cusp[\alpha_s(\mu)] \ln \frac{\nu}{N_J} + \widehat\gamma_J[\alpha_s(\mu)]
\,.
\end{align}
However, extra care must be taken because, contrary to
\scetII, one has
\begin{align}
\mu_S \sim M_S \neq M_J \sim \mu_J
\,.\end{align}
In particular, the arguments given at the end of
Section~\ref{sec:SCETIIResummation} leading to the general form for
the soft and jet functions need to be revisited.
Following similar arguments as in \scetII{} one can show that the most general form has to be 
\begin{align}
\ln \hat{\cal S}(\mu, \nu) &= \ln \widetilde{\cal S}(\mu / M_S) + \int_{N_S}^{\nu} \frac{\df \nu'}{\nu'} \,  c(\mu / \mu_S(\nu))\,,
\nonumber\\
\ln \hat{\cal J}(\mu, \nu) &= \ln \widetilde{\cal J}(\mu / M_J) - \frac{1}{2} \int_{N_J}^{\nu} \frac{\df \nu'}{\nu'} \,  c(\mu / \mu_J(\nu))
\,.
\end{align}
and we introduced the new scales $\mu_S(\nu)$ and $\mu_J(\nu)$ such
that $\mu_S(\nu) = \mu_J(\nu)\equiv\mu(\nu)$, in order for the $\nu$ dependence to
cancel in the combination of the soft and jet functions.
Each function in the above equation also depends on $\alpha_s(\mu)$
which can in turn be re-expressed in terms of $\alpha_s(\mu(\nu))$ in all
quantities (upon changing the functional dependence on $\mu/\mu(\nu)$).
Since the dependence on $\mu$ and $\nu$ is always through the ratio
with $M_F$ and $N_F$, by dimensional analysis the most general form
for $\mu_F(\nu)$ is
\begin{align}
\mu_F(\nu) = M_F \, f(\nu / N_F)
\,,
\end{align}
and one needs to find a function $f$ for which $\mu_S(\nu) = \mu_J(\nu)$.

In order for the function $F(M_F, N_F)$ to be free of large logarithms
one requires $\mu_F(N_F) = M_F$ and therefore $f(1) = 1$. 
The functional form for the function $f$ can be found by evaluating $\mu_F(\nu)$ for $\nu = N_J = M_H$. This gives
\begin{align}
\mu_S(\nu=M_H) = M_S \,  f(M_H / M_S) \overset{!}{=} \mu_J(\nu=M_H)= M_J f(M_H / M_H) = M_J
\,.
\end{align}
Using Eq.~\eqref{eq:scalesDef}, this immediately implies $f(z) = z^{\beta/(\alpha+\beta)}$,  therefore
\begin{align}
\label{4.16}
\mu_F(\nu) = M_F \left( \frac{\nu}{N_F} \right)^{\frac{\beta}{\alpha+\beta}}
\,,
\end{align}
and $\mu_F(\nu)$ is in fact independent of $F$
\begin{align}
\mu(\nu) \equiv \mu_S(\nu) = \mu_J(\nu) = \nu^{\frac{\beta}{\alpha+\beta}} Q^{\frac{\alpha}{\alpha+\beta}} \tau^{\frac{1}{\alpha+\beta}}
\,.
\end{align}
From this discussion one sees that the results are very similar to the \scetII{} case, with the only difference being that the derivatives with respect to $\nu$ are functions of $\mu / \mu_F(\nu)$, rather than $\mu / M_F$.

Given this, one can now write the solution to the differential
equation Eq.~\eqref{4.2} as
\begin{align}
\label{4.7}
F_{\sub}(\mu, \nu) &= F_{\sub}(\mu_F(\nu), \nu) \exp\left[ \int_{\mu_F(\nu)}^{\mu} \frac{\df \mu'}{\mu'} \gamma^{(\mu)}_F(\mu', \nu)\right] 
\nonumber\\
& \equiv F_{\sub}(\mu_F(\nu), \nu)U_F(\mu_F(\nu), \mu; \nu) \,,
\end{align}
but as in \scetII{} it is not sufficient to perform the resummation. A second differential
equation in the new regularization scale $\nu$ can however be derived as in
Sec.~\ref{sec:SCETIIResummation} by taking the $\nu$ derivative of the
resummed result $F_{\sub}(\mu, \nu)$ 
\begin{align}
\frac{\df}{\df \ln \nu} F_{\sub}(\mu, \nu) &= \left[ \frac{\partial \ln \mu_F(\nu)}{\partial \ln \nu} \gamma_F^{(\mu)}(\mu_F(\nu), \nu) F_{\sub}(\mu, \nu)+ U_F(\mu_F(\nu), \mu; \nu)\frac{\partial}{\partial \ln \nu}F_{\sub}(\mu_F(\nu), \nu) \right] 
\nonumber\\
& \qquad  +  \left[ -\frac{\partial \ln \mu_F(\nu)}{\partial \ln \nu}\gamma_F^{(\mu)}(\mu_F(\nu), \nu) + \int_{\mu_F(\nu)}^{\mu} \frac{\df \mu'}{\mu'} \frac{\partial}{\partial \ln \nu} \, \gamma^{(\mu)}_F(\mu', \nu) \right] F_{\sub}(\mu , \nu)
\nonumber\\
&= F_{\sub}(\mu, \nu) \left[\frac{\partial}{\partial \ln \nu} \ln F_{\sub}(\mu_F(\nu), \nu) +\int_{\mu_F(\nu)}^{\mu} \frac{\df \mu'}{\mu'} \frac{\partial}{\partial \ln \nu} \, \gamma^{(\mu)}_F(\mu', \nu)\right]
\,.
\label{4.19}
\end{align}

One can define in analogy with Eq.~\eqref{2.28} 
\begin{align}
\label{eq:gammanuDef}
\frac{\partial}{\partial \ln \nu} \ln F_{\sub}(\mu_F(\nu), \nu) \equiv  a_F \, \gamma^{(\nu)}_{\alpha, \beta}[\alpha_s(\mu_F(\nu))]
\,,\end{align}
with $a_F$ given in Eq.~\eqref{2.29}.
This is possible because $\mu_F(\nu)$ is the same function for the
soft and jet sector, and because the $\nu$ dependence cancels between
the two sectors.
We therefore obtain from Eq.~\eqref{4.19}
\begin{align}
\label{4.13}
\frac{\df}{\df \ln \nu} \ln F_{\sub}(\mu, \nu) & = a_F \, \left[ \gamma^{(\nu)}_{\alpha, \beta}[\alpha_s(\mu_F(\nu))] -2 \int_{\mu_F(\nu)}^{\mu} \frac{\df \mu'}{\mu'}  \, \Gamma_\cusp[\alpha_s(\mu')] \right]
\nonumber\\
&\equiv \gamma^{(\nu)}_\Fobs(\mu, \nu)
\,,\end{align}
analogous to Eq.~\eqref{2.30}.

With the dependence of $F_{\sub}(\mu(\nu), \nu)$ on $\mu$ and $\nu$, given in Eqs.~\eqref{4.2} and \eqref{4.13}, respectively, one can sum all logarithms by performing the integration along the path $(\mu_F, \nu_F) \to (\mu, \nu_F) \to (\mu, \nu)$, just as was done for \scetII.
This gives
\begin{align}
\label{eq:combinedEvolution}
U_F(\mu_F, \nu_F; \mu, \nu) = U_F^{(\mu)}(\mu_F, \mu; \nu_F) U_F^{(\nu)}(\nu_F, \nu; \mu) \,,
\end{align}
with
\begin{subequations}
\label{eq:SCETIRapidityEvolution}
\begin{align}
\label{eq:SCETIRapidityEvolutiona}
\ln U_S^{(\mu)}(\mu_S, \mu; \nu_S) &= \int_{\mu_S}^\mu \frac{\df\mu'}{\mu'} \left[ 4\Gamma_\cusp[\alpha_s(\mu')] \ln  \frac{\mu'}{\nu_S} + \widehat\gamma_S[\alpha_s(\mu')] \right]\,,
\\
\label{eq:SCETIRapidityEvolutionb}
\ln U_J^{(\mu)}(\mu_J, \mu; \nu_J) &= \int_{\mu_J}^\mu \frac{\df\mu'}{\mu'} \widehat\gamma_J[\alpha_s(\mu)]\,,
\\
\label{eq:SCETIRapidityEvolutionc}
\ln U_S^{(\nu)}(\nu_S, \nu; \mu)  &= \int_{\nu_S}^\nu\frac{\df\nu'}{\nu'} \left[ - 4 \int_{\mu(\nu')}^\mu\frac{\df\mu'}{\mu'}  \Gamma_\cusp[\alpha_s(\mu')] + 2 \gamma^{(\nu)}_{\alpha, \beta}[\alpha_s(\mu(\nu')) ]\right]\,,
\\
\label{eq:SCETIRapidityEvolutiond}
\ln U_J^{(\nu)}(\nu_J, \nu; \mu)  &= \int_{\nu_J}^\nu \frac{\df\nu'}{\nu'}  \left[2  \int_{\mu(\nu')}^\mu \frac{\df\mu'}{\mu'} \Gamma_\cusp[\alpha_s(\mu')- \gamma^{(\nu)}_{\alpha, \beta}[\alpha_s(\mu(\nu')) ]\right]
\,.
\end{align}
\end{subequations}
$U_J^{(\mu)}(\mu_J, \mu; \nu_J)$ has no term proportional to the cusp
anomalous dimension.
The factorization theorem Eq.~\eqref{eq:factorizationLaplace} in the
presence of the extra UV regulator takes the same form as
\scetII{} Eq.~\eqref{SCETIIFact}, namely
\begin{align}
\label{eq:SCETIFactnu}
\hat\sigma[u]=H(\mu_H)  U_H^2(\mu_H, \mu)  \,
  \hat{\cal J}_n(\mu_J,\nu_J) \hat{\cal J}_{\bar n}(\mu_J,\nu_J)  \,  U_J(\mu_S, \nu_S, \mu, \nu)  \, \hat{\cal S}(\mu_S,\nu_S) \,  U_S(\mu_S, \nu_S, \mu, \nu)
\,,\end{align}
and as before we have not explicitly shown  the dependence on the scales $M_H, M_J, M_S$. 

It is illustrative to compare these equations with the \scetII{}
equations obtained in Sec.~\ref{sec:SCETIIResummation}.
The crucial difference is that the scales $\mu_F$ that appeared in the
functions $U_F^{(\nu)}$ are now replaced by the scale $\mu(\nu)$ that
is common to the soft and  zero-bin subtracted jet functions. 
As
already mentioned, this is a consequence of the fact that the
dependence on the new unphysical regularization scale $\nu$ must
cancel in their combination.
In the following section we will discuss the implications of
introducing an extra UV regulator in the effective theory.

\section{Implications of the $\nu$ regulator on \scetI{}}
\label{sec:ObsDep}

We have shown that with the introduction of a suitably defined UV
regulator into \scetI{} theories, resummation of large logarithms can
be achieved via a system of RGEs that involves two types of anomalous
dimensions, namely the $\mu$ anomalous dimensions
\begin{align}
\gamma^{(\mu)}_S(\mu, \nu)  &= 4\Gamma_\cusp[\alpha_s(\mu)] \ln \frac{\mu}{\nu} + \widehat\gamma_S[\alpha_s(\mu)]\,,
\nonumber\\
\gamma^{(\mu)}_J(\mu, \nu)  &= 2\Gamma_\cusp[\alpha_s(\mu)] \ln \frac{\nu}{N_J} + \widehat\gamma_J[\alpha_s(\mu)]
\,,
\end{align}
and the $\nu$ anomalous dimension
\begin{align}
\label{eq:5.2}
\gamma^{(\nu)}_\Fobs(\mu, \nu) = a_F \, \left[ \gamma^{(\nu)}_{\alpha, \beta}[\alpha_s(\mu(\nu))] -2 \int_{\mu(\nu)}^{\mu} \frac{\df \mu'}{\mu'} \, \Gamma_\cusp[\alpha_s(\mu')] \right]
\,.
\end{align}
This has a number of implications that we would like to comment upon
below and in more detail in the subsections that follow.

\paragraph{Observable (in)dependence of the anomalous dimensions, and
  implications for multi-leg processes.} As was already discussed, the
extra regulator handles all UV divergences in the real contributions to
operator matrix elements, while not affecting the virtual corrections.
This implies that the $\mu$ anomalous dimension is related to the virtual diagrams, and therefore independent of the observable, which only affects real phase space integrals. 
It is therefore identical to the $\mu$ anomalous dimension in
\scetII{} problems, that does not depend on the specific constraint on
the real radiation.
The term in $\gamma^{(\nu)}_\Fobs$ proportional to the cusp anomalous
dimension is also universal, and only depends on the observable
through the definition of the scale $\mu(\nu)$ given in Eq.~\eqref{4.16}.
All observable dependence is therefore contained in the non-cusp term
of the $\nu$ anomalous dimension Eq.~\eqref{eq:5.2}, which is
determined by the real contributions to the matrix elements of
operators.
As discussed, this $\nu$ dependence has to cancel between the jet and
soft functions, and as we will see shortly it can be seen to arise
from the zero-bin region~\cite{Manohar:2006nz} of phase space integrals, which are expanded
simultaneously around the soft and collinear limit, and are therefore
significantly easier to compute.
This observation is particularly useful when tackling the computation
for multi-leg processes, in which case the factorization theorem will
be of the (schematic) form
\begin{equation}
\sigma \sim \mathbf{H}\, {\cal \mathbf S}\, \prod_{i}{\cal J}_{n_i}\,,
\end{equation}
where now the hard and soft functions will be matrices in color
space.\footnote{Here we assume the absence of additional modes,
  e.g. Glauber in the corresponding SCET Lagrangian.}
In this case, the considerations above suggest that the observable
dependence of the anomalous dimensions is entirely of soft-collinear
origin and it can be extracted from a calculation of the zero-bin
subtraction~\cite{Manohar:2006nz} in the presence of the extra UV regulator, which is a
diagonal quantity in color space, and hence would not require the
explicit computation of the more complicated soft function. We leave
the exploration of this property to future work.

\paragraph{Connection to \scetI{}.}
While one can directly perform the resummation in \scetI{} in the
presence of the extra UV regulator, one can also connect the new system
of RGEs to the standard \scetI{} case regularized in dimensional
regularization. This results in an interesting connection between the
anomalous dimensions obtained with and without the extra UV regulator.
As will be shown in detail in Section~\ref{eq:refactorization}, and
explicitly at 2-loop order in Section~\ref{sec:ExplicitCalc}, this can be used to
carry out the direct derivation of \scetI{} anomalous dimensions
starting from a computation in the presence of the extra UV regulator. This
property can become quite useful for perturbative calculations.

\paragraph{Numerical resummation.} Another important feature discussed in more detail in the next section is that when we perform the
explicit 2-loop calculation of the soft function, the observable
dependence can be computed by dividing the real phase space integrals
into different separate contributions.
One can for instance define a first contribution where a simplified
version of the observable is used rather than the full observable.
This simplified observable only depends on the singular scaling of the original
observable and is therefore universal and gives rise to much
simpler integrals.
The observable dependence in this contribution can be determined
relatively straightforwardly.
The second contribution, that we call non-inclusive, computes the
difference of the full observable to the simplified version.
While this contribution depends on the full details of the observable,
the difference is infrared and collinear finite, hence allowing one to
perform the computation numerically in a rather efficient fashion.
This observation is the basis of the numerical resummation technique
developed recently in the framework of SCET~\cite{Bauer:2018svx,Bauer:2019bsp}.

\subsection{Zero-bin subtraction and refactorization: relation with
  standard \scetI{}}
\label{eq:refactorization}

We now relate the system of RGEs given by
Eqs.~\eqref{4.2},~\eqref{4.13} to the standard \scetI{} case in pure
dimensional regularization.
The solutions to the equations in
Eqs.~\eqref{eq:SCETIRapidityEvolution} sum the logarithmic
corrections in a form that is similar for both \scetI{} and \scetII{}.
To compare with the usual \scetI{} form, we invert the order of the $\nu$ and $\mu$
integration in the $\nu$ evolution
equations Eq.~\eqref{eq:SCETIRapidityEvolutionc}
and Eq.~\eqref{eq:SCETIRapidityEvolutiond}.
From now on we will always assume $\mu_S=M_S$, $\mu_J=M_J$ and
$\mu_H=M_H$, as well as $\nu_J = M_H$ and $\nu_S=M_S$, and use the
notation interchangeably.
Using Eq.~\eqref{4.16} one can write
\begin{align}
\label{reverseOrderOfIntegration}
\int_{\nu_F}^\nu \frac{\df\nu'}{\nu'} \int_{\mu(\nu^\prime)}^\mu \frac{\df\mu'}{\mu'}  = 
\left\{ 
\begin{array}{ll}
\int_{\mu_F}^{\mu} \frac{\df\mu'}{\mu'}  \int_{\nu_F}^{\nu(\mu')} \frac{\df\nu'}{\nu'} + \int_{\mu(\nu)}^\mu \frac{\df\mu'}{\mu'}  \int_{\nu(\mu')}^{\nu} \frac{\df\nu'}{\nu'}& \beta \neq 0\\
\int_{\mu_F}^\mu \frac{\df\mu'}{\mu'}   \int_{\nu_F}^\nu \frac{\df\nu'}{\nu'} & \beta = 0
\end{array}
\right.
\,,
\end{align}
with $\nu(\mu)$ being the inverse of $\mu(\nu)$
\begin{align}
\label{eq:nuofmu}
\nu(\mu) = \nu_F \left(\frac{\mu}{\mu_F}\right)^{\frac{\alpha+\beta}{\beta}}
\,.\end{align}
Note that we had to distinguish between $\beta = 0$ (\scetII{}) and
$\beta \neq 0$ (\scetI{}), since for $\beta = 0$ the quantity
$\nu(\mu)$ is clearly not defined and $\mu(\nu)$ is independent of
$\nu$.

For $\beta \neq 0$ we can rewrite the solution to the $\nu$ evolution as
\begin{align}
\label{5.6}
\ln U_S^{(\nu)}(\nu_S, \nu; \mu)  &= 
- 4\frac{\alpha+\beta}{\beta}  \int_{\mu_S}^\mu\frac{\df\mu'}{\mu'}  \Gamma_\cusp[\alpha_s(\mu')] \ln \frac{\mu'}{\mu_S} 
\\
& \qquad + 4\frac{\alpha+\beta}{\beta}  \int_{\mu(\nu)}^\mu\frac{\df\mu'}{\mu'}  \Gamma_\cusp[\alpha_s(\mu')] \ln \frac{\mu'}{\mu(\nu)}
+ 2   \int_{\nu_S}^\nu \frac{\df\nu'}{\nu'} \gamma^{(\nu)}_{\alpha, \beta}[\alpha_s(\mu(\nu')) ]\,,
\nonumber\\
\ln U_J^{(\nu)}(\nu_S, \nu; \mu)  &= 
2 \frac{\alpha+\beta}{\beta}  \int_{\mu_J}^\mu\frac{\df\mu'}{\mu'}  \Gamma_\cusp[\alpha_s(\mu')] \ln \frac{\mu'}{\mu_J} 
\nonumber\\
& \qquad -2 \frac{\alpha+\beta}{\beta}  \int_{\mu(\nu)}^\mu\frac{\df\mu'}{\mu'}  \Gamma_\cusp[\alpha_s(\mu')]  \ln \frac{\mu'}{\mu(\nu)}
- \int_{\nu_J}^\nu \frac{\df\nu'}{\nu'} \gamma^{(\nu)}_{\alpha, \beta}[\alpha_s(\mu(\nu')) ]
\,,\nonumber
\end{align}
where we have used
\begin{align}
\ln \frac{\nu(\mu)}{\nu_F} = \frac{\alpha+\beta}{\beta} \ln \frac{\mu}{\mu_F}\,, \qquad \ln \frac{\nu}{\nu(\mu)} = -\frac{\alpha+\beta}{\beta} \ln \frac{\mu}{\mu(\nu)}
\,,
\end{align}
which follow directly from Eqs.~\eqref{4.16} and~\eqref{eq:nuofmu}.
Using this in Eq.~\eqref{eq:combinedEvolution} one finds
\begin{align}
\label{4.21}
\ln U_S(\mu_S, \nu_S; \mu, \nu) & = \int_{\mu_S}^\mu \frac{\df\mu'}{\mu'} \left[ - 4 \, \frac{\alpha}{\beta} \, \Gamma_\cusp[\alpha_s(\mu')]\ln \frac{\mu'}{\mu_S}  +  \widehat \gamma_S[\alpha_s(\mu')] \right] 
\\
& \quad +4  \frac{\alpha+\beta}{\beta} \int_{\mu(\nu)}^\mu\frac{\df\mu'}{\mu'}  \, \Gamma_\cusp[\alpha_s(\mu')] \ln  \frac{\mu'}{\mu(\nu)}
+ 2 \int_{\nu_S}^\nu \frac{\df\nu'}{\nu'}
\gamma^{(\nu)}_{\alpha, \beta}[\alpha_s(\mu(\nu')) ]\,,
\nonumber\\
\ln U_J(\mu_J, \nu_J; \mu, \nu) & = \int_{\mu_J}^\mu \frac{\df\mu'}{\mu'} \left[ 2\, \frac{\alpha+\beta}{\beta}\, \Gamma_\cusp[\alpha_s(\mu')] \ln \frac{\mu'}{\mu_J}  + \widehat \gamma_J[\alpha_s(\mu')] \right]
\nonumber\\
& \quad -2 \frac{\alpha+\beta}{\beta} \int_{\mu(\nu)}^\mu\frac{\df\mu'}{\mu'}  \Gamma_\cusp[\alpha_s(\mu')] \ln  \frac{\mu'}{\mu(\nu)}
-\int_{\nu_J}^\nu \frac{\df\nu'}{\nu'}
\gamma^{(\nu)}_{\alpha, \beta}[\alpha_s(\mu(\nu')) ]\,.
\nonumber
\end{align}

The first line in each of the two parts of Eq.~\eqref{4.21} starts
resembling the usual \scetI{} evolution with an integration over
$\df \ln \!\mu$ between the scales $\mu_F$ and $\mu$, and an anomalous dimension that depends on $\alpha$ and $\beta$.
The second line in each equation on the other hand does not have this form.
Using Eq.~\eqref{4.16}, however, one can change the integration variable  from $\nu$ to $\mu(\nu)$ 
\begin{align}
\label{tmp5.9}
\int_{\nu_F}^\nu \frac{\df \nu}{\nu}  f(\nu) & = \frac{\alpha+\beta}{\beta} \int_{\mu_F}^{\mu(\nu)}\frac{\df \mu(\nu)}{\mu(\nu)} f(\nu(\mu))
\,.\end{align}
This leads to 
\begin{align}
\label{eq:factorizedSudakov}
\ln U_S(\mu_S, \nu_S; \mu, \nu) & = \int_{\mu_S}^\mu \frac{\df\mu'}{\mu'} \left[ -4 \, \frac{\alpha}{\beta} \, \Gamma_\cusp[\alpha_s(\mu')] \ln \frac{\mu'}{\mu_S} + \widetilde \gamma_\Sobs[\alpha_s(\mu')] \right] + 2 R_{\alpha, \beta}(\mu(\nu);\mu)\,,
\nonumber\\
\ln U_J(\mu_J, \nu_J, \mu, \nu) & =  \int_{\mu_J}^\mu \frac{\df\mu'}{\mu'} \left[  2\, \frac{\alpha+\beta}{\beta} \, \Gamma_\cusp[\alpha_s(\mu')]\ln \frac{\mu'}{\mu_J} + \widetilde \gamma_\Jobs[\alpha_s(\mu')]  \right] - R_{\alpha, \beta}(\mu(\nu);\mu)
\,,
\end{align}
where we have used \eqref{eq:nuofmu} and defined
\begin{align}
\label{eq:intermediateAnomDim}
\widetilde \gamma_\Sobs[\alpha_s(\mu)] &= \widehat\gamma_S[\alpha_s(\mu)] + 2 \,\frac{\alpha+\beta}{\beta}\,\gamma^{(\nu)}_{\alpha, \beta}[\alpha_s(\mu) ]\,,
\nonumber\\
\widetilde \gamma_\Jobs[\alpha_s(\mu)]  &= \widehat\gamma_J[\alpha_s(\mu)] - \frac{\alpha+\beta}{\beta}\,\gamma^{(\nu)}_{\alpha, \beta}[\alpha_s(\mu) ]
\,,
\end{align}
and
\begin{align}
R_{\alpha, \beta}(\mu(\nu);\mu) = \frac{\alpha+\beta}{\beta}\int_{\mu(\nu)}^{\mu} \frac{\df\mu'}{\mu'} \left[ 2\,  \Gamma_\cusp[\alpha_s(\mu')] \ln \frac{\mu'}{\mu(\nu)} - \gamma^{(\nu)}_{\alpha, \beta}[\alpha_s(\mu') ] \right]\,.
\end{align}
Eq.~\eqref{eq:factorizedSudakov} indicates that the evolution operator
of each of the soft and jet functions in the \scetI{} case ($\beta\neq 0$)
can be factorized into the product in Laplace space of a term that only depends on the ratio of scales $\mu / \mu_F$ and a term $R(\mu(\nu); \mu)$ that depends on the ratio $\mu / \mu(\nu)$ which cancels in the
physical cross section.

In order to complete this re-factorization, we need to consider the
factorization of the initial condition to the soft and jet functions
in~\eqref{eq:SCETIFactnu} which, as already explained, depends on two
canonical scales $\mu_F$,~$\nu_F$.
The full soft and jet function in the presence of the extra UV
regulator can be written in terms of those in standard \scetI{} as
\begin{align}
\hat{\cal S}(\mu_S,\nu_S;\mu,\nu) &= \hat{\cal S}_{\rm SCET_I} (\mu_S;\mu)  \Delta_\Sobs(\mu_S,\nu_S;\mu, \nu)\,,
\nonumber\\
\hat{\cal J}(\mu_J,\nu_J;\mu,\nu) &= \hat{\cal J}_{\rm SCET_I} (\mu_J;\mu)  \Delta_\Jobs(\mu_J,\nu_J;\mu, \nu)\,,
\end{align}
which defines the functions $ \Delta_\Fobs(\mu_F,\nu_F;\mu, \nu)$.
Given the importance of the scale dependence in the soft and functions
in what follows, we show them explicitly in these equations.
Using, as before, the fact that the $\nu$ dependence cancels in the
combination of jet and soft functions, leads to the important
relations~\footnote{This refactorization is similar in spirit to that
  performed in Section 4.2 of Refs.~\cite{Beneke:2019slt,Beneke:2020vnb}, albeit in a
  different physical context.}
\begin{align}
\label{DeltaDefX}
\Delta_\Sobs(\mu_S,\nu_S;\mu, \nu) = \Delta_{\alpha, \beta}^2(\mu(\nu);\mu) \,, \qquad \Delta_\Jobs(\mu_J,\nu_J;\mu, \nu) = \Delta_{\alpha, \beta}^{-1}(\mu(\nu);\mu)
\,,
\end{align}
and that the function $\Delta_{\alpha, \beta}$ has no large logarithms when evaluated at the scale $\mu = \mu(\nu)$. 
Given this general form, one can write the evolution kernel for the soft sector as (and similarly for the collinear sector)
\begin{align}
\label{5.14}
U_S(\mu_S,\nu_S,\mu, \nu) &= \frac{ \hat{\cal S}(\mu_S,\nu_S;\mu,\nu) }{ \hat{\cal S}(\mu_S,\nu_S;\mu_S,\nu_S)} 
\\
&= \frac{ \hat{\cal S}_{\rm SCET_I} (\mu_S;\mu) }{\hat{\cal S}_{\rm SCET_I} (\mu_S;\mu_S) }  \frac{ \Delta_{\alpha, \beta}^2(\mu(\nu);\mu) }{  \Delta_{\alpha, \beta}^2(\mu(\nu_S);\mu_S)} 
\nonumber\\
&= U^{\rm SCET_I}_S(\mu_S,\mu) \frac{ \Delta_{\alpha, \beta}^2(\mu(\nu);\mu) }{  \Delta_{\alpha, \beta}^2(\mu(\nu);\mu(\nu))}  \frac{ \Delta_{\alpha, \beta}^2(\mu(\nu);\mu(\nu)) }{  \Delta_{\alpha, \beta}^2(\mu;\mu)} \frac{ \Delta_{\alpha, \beta}^2(\mu;\mu) }{  \Delta_{\alpha, \beta}^2(\mu_S;\mu_S)} 
\nonumber\\
&= \left[ U^{\rm SCET_I}_S(\mu_S,\mu) \frac{ \Delta_{\alpha, \beta}^2(\mu;\mu) }{  \Delta_{\alpha, \beta}^2(\mu_S;\mu_S)}  \right] \left[ \frac{ \Delta_{\alpha, \beta}^2(\mu(\nu);\mu) }{  \Delta_{\alpha, \beta}^2(\mu(\nu);\mu(\nu))}  \frac{ \Delta_{\alpha, \beta}^2(\mu(\nu);\mu(\nu)) }{  \Delta_{\alpha, \beta}^2(\mu;\mu)}\right] 
\nonumber
\,,
\end{align}
where we have used that $\mu(\nu_F) = \mu_F$. Here the term in the first square bracket only depends on the scales $\mu$ and $\mu_S$, while the one in the second square bracket  depends on the scales $\mu$ and $\mu(\nu)$. 
Since $\Delta(\mu(\nu);\mu(\nu))$ has no large logarithms, we can write it as
\begin{equation}
\label{eq:deltaexpansion}
\Delta_{\alpha, \beta}[\alpha_s(\mu(\nu))] \equiv \Delta_{\alpha, \beta}(\mu(\nu);\mu(\nu)) = 1 + \sum_{n=1}\left( \frac{\alpha_s(\mu(\nu))}{4\pi}\right) ^n d_{\alpha, \beta}^{(n)}\,,
\end{equation}
where the coefficients $d_{\alpha, \beta}^{(n)}$ can be obtained by
taking the ratio between the \scetI{} initial conditions with and
without the $\nu$ regulator.
This means that we can write
\begin{align}
\frac{\Delta_{\alpha, \beta}(\mu_1;\mu_1)}{\Delta_{\alpha, \beta}(\mu_2;\mu_2)} = \exp\left\{\int_{\mu_2}^{\mu_1}\frac{\df\mu^\prime}{\mu^\prime}2\beta[\alpha_s(\mu')]\frac{\df
                            \ln\Delta_{\alpha, \beta}}{\df\alpha_s}\right\}\,,
\end{align}
where  ${\df \ln\Delta_{\alpha, \beta}} / {\df\alpha_s}$ is a function of $\alpha_s$. The QCD $\beta$ function is given by
\begin{equation}
\beta[\alpha_s(\mu)]=\frac{\df \alpha_s(\mu)}{\df\ln\mu^2}=-\alpha_s(\mu)\left( b_0\frac{\alpha_s(\mu)}{4\pi}+\dots\right)
\,,
\end{equation}
where $b_0=(11 C_A-2 n_F)/3$.
This allows us to write
\begin{align}
\ln U_S(\mu_S,\nu_S,\mu, \nu) &= \left[ \ln U^{\rm SCET_I}_S(\mu_S,\mu) + 2 \int_{\mu_S}^{\mu}\frac{\df\mu^\prime}{\mu^\prime}2\beta[\alpha_s(\mu')]\frac{\df
                            \ln\Delta_{\alpha, \beta}}{\df\alpha_s} \right]
\\
& \qquad + \left[ 2 \ln \frac{ \Delta_{\alpha, \beta}(\mu(\nu);\mu) }{  \Delta_{\alpha, \beta}(\mu(\nu);\mu(\nu))} - 2 \int_{\mu(\nu)}^{\mu}\frac{\df\mu^\prime}{\mu^\prime}2\beta[\alpha_s(\mu')]\frac{\df
                            \ln\Delta_{\alpha, \beta}}{\df\alpha_s} \right]
                            \,.\nonumber
\end{align}
Comparing this result to Eq.~\eqref{eq:factorizedSudakov}, and
equating the terms that involve the evolution between $\mu_S$ and
$\mu$ and the ones that involve the $\mu(\nu) \to \mu$ evolution, one
can read off
\begin{align}
\label{eq:DeltaDef}
\ln  U_{S}^{\rm SCET_I}(\mu_S; \mu)  & = \int_{\mu_S}^\mu \frac{\df\mu'}{\mu'} \left[ -4 \, \frac{\alpha}{\beta} \, \Gamma_\cusp[\alpha_s(\mu')] \ln \frac{\mu'}{\mu_S} + \widetilde \gamma_\Sobs[\alpha_s(\mu')] - 4\beta[\alpha_s(\mu') ]\frac{\df \ln\Delta_{\alpha, \beta}}{\df\alpha_s} \right]\,,
\nonumber\\
\ln \frac{ \Delta_{\alpha, \beta}(\mu(\nu);\mu) }{  \Delta_{\alpha, \beta}(\mu(\nu);\mu(\nu))} &= R_{\alpha, \beta}(\mu(\nu);\mu) + \int_{\mu(\nu)}^{\mu}\frac{\df\mu^\prime}{\mu^\prime}2\beta[\alpha_s(\mu') ]\frac{\df
                            \ln\Delta_{\alpha, \beta}}{\df\alpha_s}
                            \,.
\end{align}
From this, we obtain the anomalous dimension in standard \scetI{} (see
Eq.~\eqref{2.17}), and the non-cusp pieces
$\widehat \gamma_{\alpha, \beta; F}^{\rm SCET_I}$ are
\begin{align}
\label{eq:finalAnomDim}
\widehat \gamma_\Sobs^{\rm SCET_I}[\alpha_s(\mu)] &= \widetilde \gamma_\Sobs[\alpha_s(\mu)] - 4\beta[\alpha_s(\mu)]\frac{\df \ln\Delta_{\alpha, \beta}}{\df\alpha_s}
\nonumber\\
&= \widehat\gamma_S[\alpha_s(\mu)] + 2 \,\frac{\alpha+\beta}{\beta}\,\gamma^{(\nu)}_{\alpha, \beta}[\alpha_s(\mu) ] - 4\beta[\alpha_s(\mu)]\frac{\df \ln\Delta_{\alpha, \beta}}{\df\alpha_s}\,,
\nonumber\\
\widehat \gamma_\Jobs^{\rm SCET_I}[\alpha_s(\mu)] &= \widetilde \gamma_\Jobs[\alpha_s(\mu)] + 2\beta[\alpha_s(\mu) ]\frac{\df \ln\Delta_{\alpha, \beta}}{\df\alpha_s}
\nonumber\\
&= \widehat\gamma_J[\alpha_s(\mu)] - \,\frac{\alpha+\beta}{\beta}\,\gamma^{(\nu)}_{\alpha, \beta}[\alpha_s(\mu) ] + 2\beta[\alpha_s(\mu) ]\frac{\df \ln\Delta_{\alpha, \beta}}{\df\alpha_s}\,,
\end{align}
where we have used Eq.~\eqref{eq:intermediateAnomDim} and analogous arguments to obtain the jet anomalous dimension.%

Eqs.~\eqref{eq:finalAnomDim} relate the anomalous dimensions
calculated with the additional $\nu$ regulator to the standard \scetI{}
result calculated in dimensional regularization. Notice that due to
the presence of $\beta[\alpha_s(\mu)]$ in the last term, to compute an
anomalous dimension at $k$-th order one only requires the
initial condition for $\Delta_{\alpha, \beta}$ at  $(k-1)$-th order.
It is important to stress again that the above discussion relating the
system of RGEs in the presence of a UV regulator to the standard RGE
is clearly not allowed when $\beta=0$ (i.e. in \scetII{}), in which
case one is forced to keep a coupled system of evolution
equations.

The above discussion highlights an important point: as mentioned in
Section~\ref{sec:SCETIResummation}, \scetI{} is characterized by a
scale separation between the soft and the collinear sectors.
In particular, if $M_J\gg M_S$ the two
jet functions can be interpreted as a matching coefficient between
\scetI{} and the lower-energy purely soft theory, described by the
soft function. 
The introduction of the extra UV regulator, however,
introduces a new scale $\mu(\nu)$ that interpolates between the soft
scale $M_S$ and the collinear scale $M_J$ depending on the value of
the regularization scale $\nu$. 
Defining the soft theory as before, namely only containing Wilson lines regulated by dimensional regularization, the dependence on the extra UV regulator cancels in the matching coefficient. This is because the matching coefficient (defined by the difference of the theory above and below the matching scale) is not equal to the jet functions with the extra UV regulator. It also includes the $\Delta_\Sobs$ from the soft function in Eq.~\eqref{DeltaDefX}. This is never possible in \scetII{}, since the soft theory is not defined without a rapidity regulator.

It is interesting to understand if one can formulate an operator
definition of $\Delta_{\alpha, \beta}$ defined in this section.
The quantity $\Delta_{\alpha, \beta}$ has to cancel in the product (in
Laplace space for the observables considered here) between the soft
and jet functions, and hence it necessarily has to be entirely
determined by radiation that is simultaneously soft and collinear.
Contrary to the standard \scetI{} case, the introduction of the extra UV regulator implies the existence of a
non-vanishing zero-bin subtraction~\cite{Manohar:2006nz} that induces a cross-talk between the
soft and the jet functions.
It is then natural to identify this cross talk, parameterized by
$\Delta_{\alpha, \beta}$, with the eikonalized jet function that
enters the zero-bin subtraction calculated with the additional $\nu$
regulator.
This is an interesting observation as it implies that a calculation of
the zero-bin subtraction is sufficient to determine both the
coefficients $d_{\alpha, \beta}^{(n)}$ of
Eq.~\eqref{eq:deltaexpansion} and the anomalous dimension
$\gamma^{(\nu)}$ that provides the only observable-dependent
contribution to the anomalous dimension.
As a result, the structure of the anomalous dimension is entirely
constrained by consistency of the theory, and the observable
dependence is only encoded in a specific contribution arising from the
soft-collinear region.
In particular, since any result in the
soft-collinear region, as in the collinear region itself, depends on
only a single light-cone direction, it is diagonal in color
space, and for example, does not depend on
  ${\mathbf T}_{n_1} \cdot {\mathbf T}_{n_2}$, the dot product of color generators in two
  different directions.
 This significantly simplifies part of the calculation of the
 anomalous dimensions in \scetI{} problems.
 Notice that these constraints only apply to the anomalous dimensions
 and not to the constants (i.e. the initial conditions to the RGEs),
 which still require an explicit calculation.

\subsection{Dependence of $\gamma^{(\nu)}$ on the $\nu$-regularization
  scheme}
\label{sec:SchemeDep}

We now wish to discuss the dependence of the soft and jet
anomalous dimensions on the specific regularization scheme used to
single out the UV divergences in the real radiation, and contrast the results between \scetI{} and \scetII.
We first consider \scetI{}. 
In Eq.~\eqref{eq:finalAnomDim}, the left hand side
is obviously independent of the specific scheme used to regulate the UV
limit of the real radiation integrals. 
On the right hand side, the
anomalous dimensions
$\widehat\gamma_S[\alpha_s]$,~$\widehat\gamma_J[\alpha_s]$
are scheme independent by definition, and therefore one
obtains
\begin{align}
\label{eq:schemeinvariant}
 \,\frac{\alpha+\beta}{\beta}\,\gamma^{(\nu)}_{\alpha, \beta}[\alpha_s ]-2\beta[\alpha_s]\frac{\df
                            \ln\Delta_{\alpha, \beta}}{\df\alpha_s} \to {\rm \nu~scheme~invariant~in~SCET_I}\,.
\end{align}
The previous equation can be used to relate the $\gamma^{(\nu)}$
anomalous dimension calculated in different schemes. 
On the other hand, in \scetII{}
one has $\beta=0$ and therefore Eq.~\eqref{eq:schemeinvariant} is not
defined. 
Given that
$\widehat\gamma_S[\alpha_s]$,~$\widehat\gamma_J[\alpha_s]$ are independent of the UV regularization scheme for real
corrections, and the $\nu$ dependence must cancel in the product of
soft and jet functions, one trivially gets that
\begin{align}
\label{eq:schemeinvariantSCETII}
\gamma^{(\nu)}_{\alpha, \beta}[\alpha_s]\to {\rm \nu~scheme~invariant~in~SCET_{II}}\,,
\end{align}
consistent with the conclusion of Ref.~\cite{Li:2016axz}.
This marks another important difference between the two theories.
The properties~\eqref{eq:schemeinvariant}
and~\eqref{eq:schemeinvariantSCETII} are quite powerful and can be very
useful in practical calculations,
for example to carry out the calculation for the anomalous
dimensions numerically. As we will show in
Section~\ref{sec:SchemeDepCalc}, one can adopt a UV
regularization scheme that is suitable for numerical calculation, for
instance a cutoff on the light-cone momentum components, and then use
the equations obtained in this section to convert the result to a
scheme with better theoretical properties (e.g.\ boost invariance)
such as the exponential regulator used in this article.

\section{Soft and jet anomalous dimensions for angularities up to two loops}
\label{sec:ExplicitCalc}
In this section we perform a computation of the anomalous
dimensions up to two loops, which allows us to explicitly verify the
structure of the system of RGEs and the relations between anomalous
dimensions derived in Sections~\ref{sec:RGRelationship}
and~\ref{sec:ObsDep} in the case of recoil-free angularities in
$e^+e^-$. The relevant factorization theorem is given in
Eq.~\eqref{eq:factorization}, where we set $\alpha=1$, $\beta=1-a$.
We adopt the exponential regulator~\cite{Li:2016axz}, and we also
report results with an alternative regulator in
Section~\ref{sec:SchemeDepCalc}.
Throughout the section we use the following notation for the
perturbative expansion of the anomalous dimensions:
\begin{align}
\Gamma_\cusp[\alpha_s(\mu)] &= \sum_{n=0}
\left(\frac{\alpha_s(\mu)}{4\pi}\right)^{n+1} \Gamma_\cusp^{(n)}\,,\notag\\
\gamma_{\alpha,\beta}^{(\nu)}[\alpha_s(\mu)] &= \sum_{n=0}
\left(\frac{\alpha_s(\mu)}{4\pi}\right)^{n+1}\gamma_{\alpha,\beta}^{(\nu,\,n)}\,,\notag\\
\widehat\gamma_F[\alpha_s(\mu)]&= \sum_{n=0}
\left(\frac{\alpha_s(\mu)}{4\pi}\right)^{n+1}\widehat{\gamma}^{(n)}_F\,,\qquad
                                   F=\{S,J\}\,,\notag\\
  \gamma_H[\alpha_s(\mu)]&= \sum_{n=0}
\left(\frac{\alpha_s(\mu)}{4\pi}\right)^{n+1}\gamma^{(n)}_H\,,
\notag\\
\Delta_{\alpha, \beta}[\alpha_s(\mu)] &= 1 + \sum_{n=1}\left( \frac{\alpha_s(\mu)}{4\pi}\right) ^n d_{\alpha, \beta}^{(n)}\,.
\end{align}
\subsection{One loop result}
\begin{figure}
\centering
\includegraphics[width=3cm]{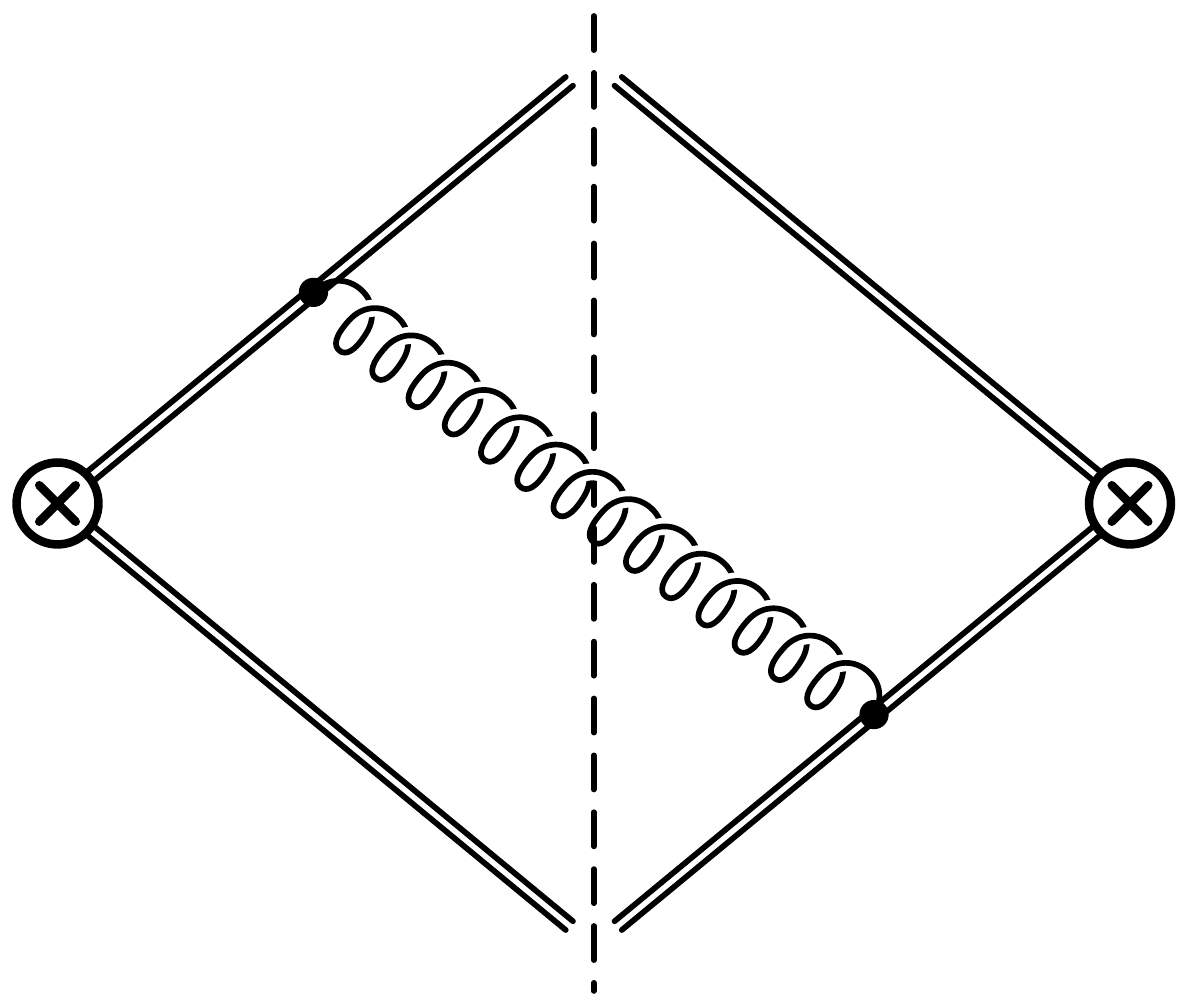}\hspace{1cm}
\includegraphics[width=3cm]{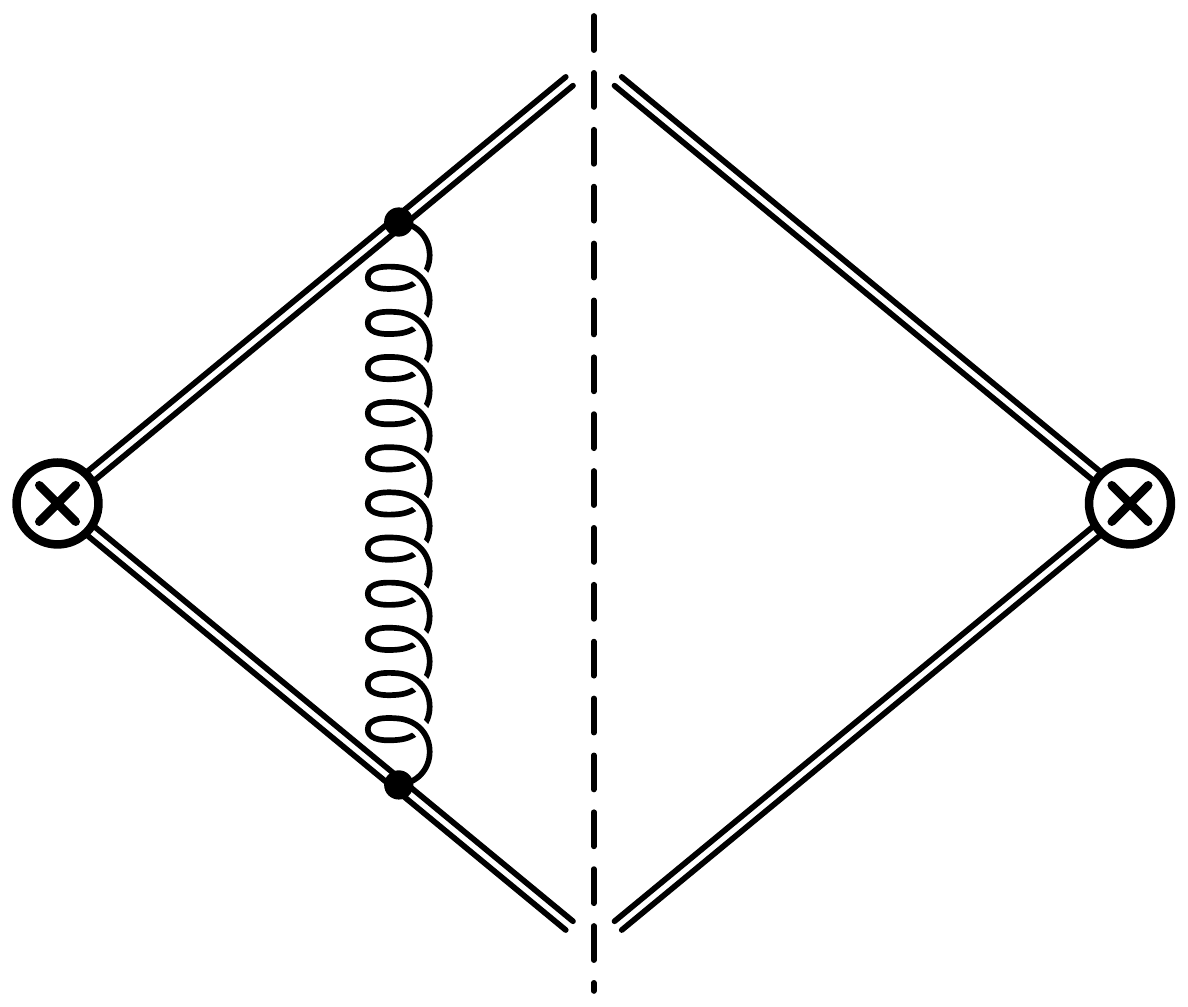}
\caption{Diagrams contributing to the one-loop soft function.}
\label{fig:soft1L}
\end{figure}

\begin{figure}
\centering
\includegraphics[width=4cm]{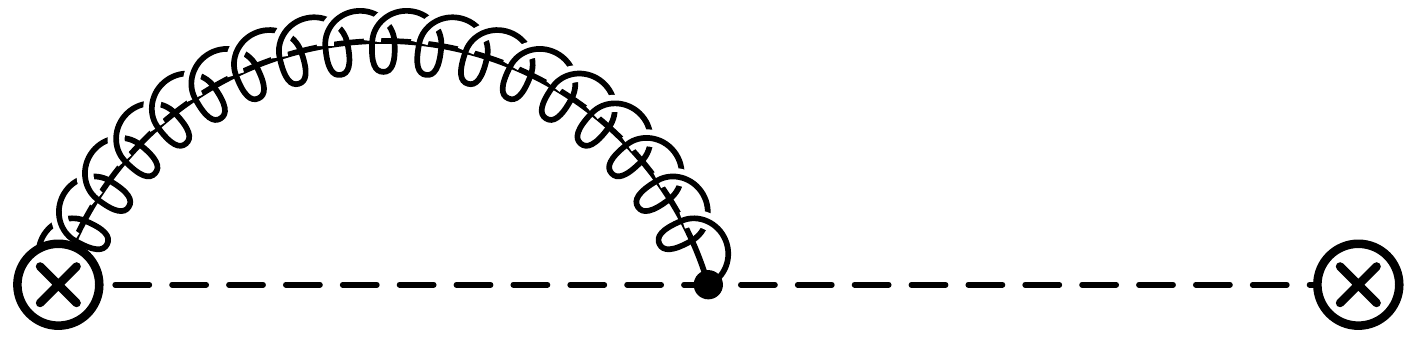}\hspace{1cm}
\includegraphics[width=4cm]{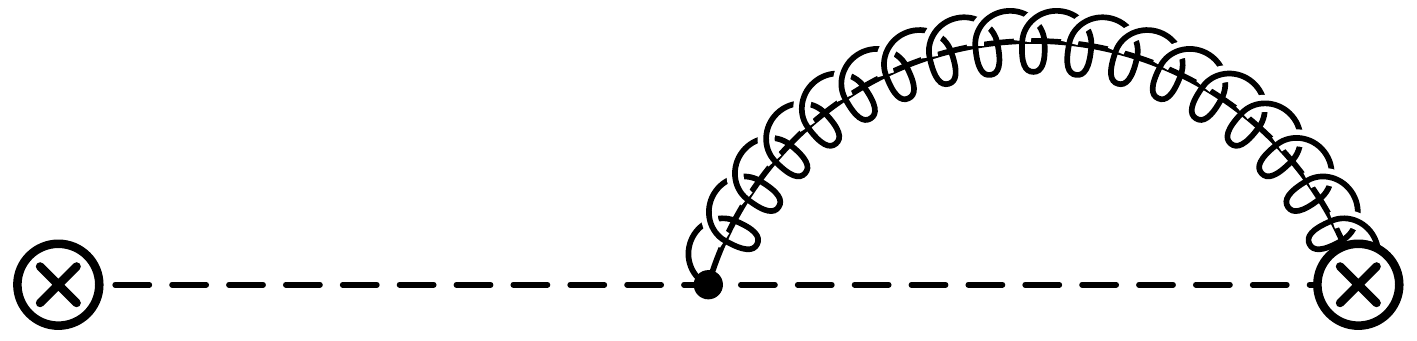}

\vspace{1cm}

\includegraphics[width=4cm]{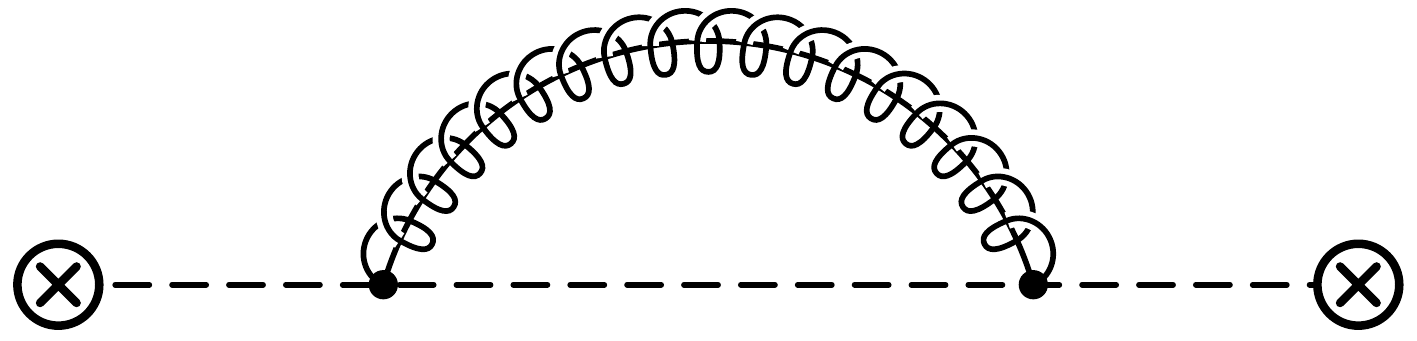} \hspace{1cm}
\includegraphics[width=4cm]{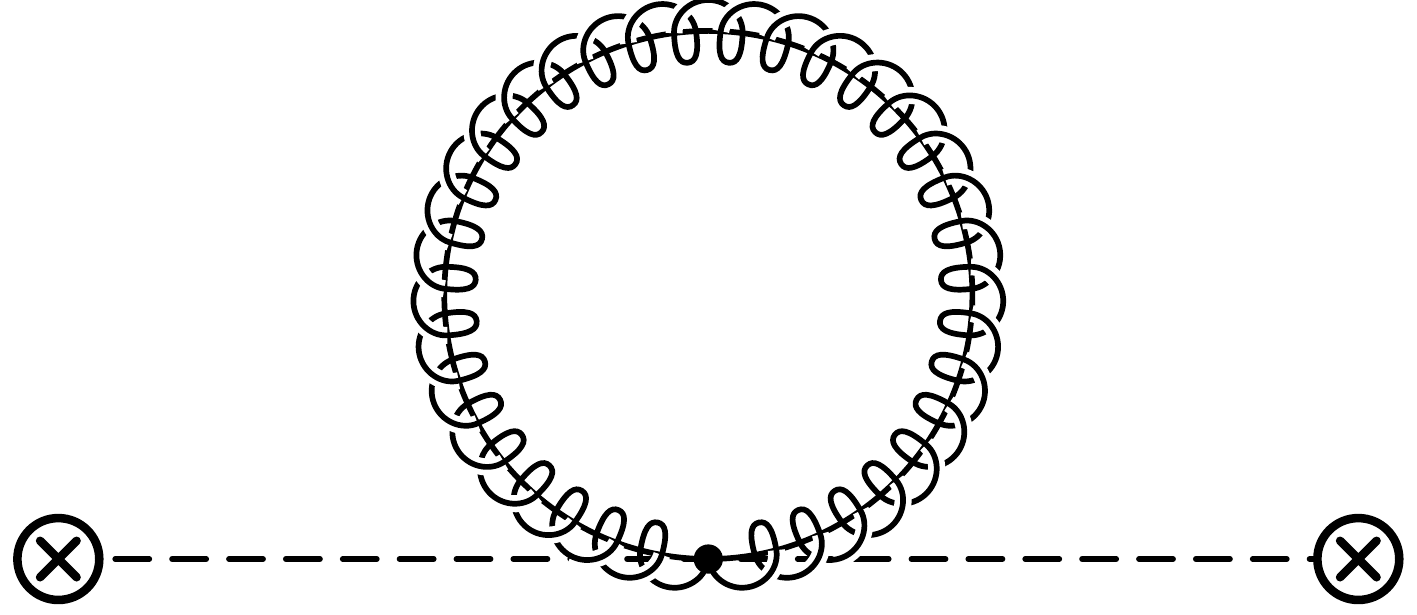} 
\caption{Diagrams contributing to the one-loop jet function. The jet function is given by the sum over all cuts. }
\label{fig:jet1L}
\end{figure}

The one-loop result is a generalization of that for thrust given in
Sec.~\ref{sec:softOneLoopthrust}. The soft function is given by the
diagrams given in Fig.~\ref{fig:soft1L} (and the corresponding
mirror conjugate ones).
For a generic angularity one obtains
in Laplace space 
\begin{align}
\label{eq:softoneloop}
\hat{\cal S}_{\rm bare}(\mu,\nu) = 1&+\frac{\alpha_s(\mu)}{\pi}
C_F\left[\frac{1}{\epsilon^2}+\frac{2}{\epsilon}\ln\frac{\mu}{\nu} \right.\notag\\
&\left.+  \frac{1}{2-a}\left(2(1-a)\ln^2\frac{\mu}{\nu}-2 \ln^2\frac{\mu u}{Q
      u_0}+4
    \ln\frac{\mu}{\nu}\ln\frac{\mu u}{Q u_0}-\frac{\pi^2}{12}(2+3 a)\right)\right]\,.
\end{align}
The soft anomalous dimensions are then extracted using Eq.~\eqref{4.2}
and~\eqref{eq:gammanuDef}, which at one loop give
\begin{align}
\gamma^{(\mu)}_S(\mu, \nu) =& \,4\frac{\alpha_s(\mu)}{\pi}
  C_F\ln\frac{\mu}{\nu}+{\cal O}(\alpha_s^2)\,,\notag\\
\gamma^{(\nu)}_\Sobsang(\mu, \nu) =&
                                     -4\frac{\alpha_s(\mu)}{\pi}C_F\ln\frac{\mu}{\mu(\nu)}+{\cal O}(\alpha_s^2)\,.
\end{align}
This verifies Eqs.~\eqref{gamma_SCETIIlike} and~\eqref{4.13} to
${\cal O}(\alpha_s)$ and gives the well known results
$\Gamma_\cusp^{(0)}=4 C_F$,
\begin{align}
\widehat{\gamma}^{(0)}_S[\alpha_s(\mu)] = 0\,,
\end{align}
and
\begin{align}
\label{gammanu1loop}
\qquad \gamma_{1, 1-a}^{(\nu,\,0)}[\alpha_s(\mu)]=0
\,. \end{align}
Similarly, the zero-bin subtracted one-loop jet function is given by
all possible cuts of the diagrams shown in
Fig.~\ref{fig:jet1L}.
After performing the zero-bin subtraction, which becomes non-trivial
in the presence of the extra UV regulator, we get
\begin{align}
\hat{\cal J}_{n, {\rm bare}}(\mu,\nu) = 1&+\frac{\alpha_s(\mu)}{\pi}
C_F\bigg\{\frac{3}{4
                                     \epsilon}+\frac{\ln\frac{\nu}{Q}}{\epsilon}\notag\\
& +\frac{1}{2-a}\left[\ln\left(\frac{\mu}{Q}\left(\frac{u}{u_0}\right)^{\frac{1}{2-a}}\right)\left(3-\frac{3}{2}a
                                     + 2
        (2-a)\ln\frac{\nu}{Q}\right)-(1-a)\ln^2\frac{\nu}{Q}\right.\notag\\
&\left.\,+\frac{1}{12}\left(60-39 a + \pi^2(6 a-8)-18 \ln 2\right)\right] \bigg\}\,,
\end{align}
from which we obtain the one loop anomalous dimensions of the jet function
\begin{align}
\gamma^{(\mu)}_J(\mu, \nu) &= \frac{\alpha_s(\mu)}{\pi}
  C_F\left(2\ln\frac{\nu}{Q}+\frac{3}{2}\right) +{\cal O}(\alpha_s^2), \quad\notag\\
\gamma^{(\nu)}_\Jobsang(\mu, \nu) &= \,2\frac{\alpha_s(\mu)}{\pi}C_F\ln\frac{\mu}{\mu(\nu)}+{\cal O}(\alpha_s^2)
\,.
\end{align}
One therefore confirms Eq.~\eqref{gammanu1loop} and obtains
\begin{align}
\widehat{\gamma}^{(0)}_J[\alpha_s(\mu)]  &= 6 \, C_F\,.
\end{align}
Note that at one-loop order the $\nu$ anomalous dimension is
independent of the observable since $\gamma_{1,
  1-a}^{(\nu,\,0)}=0$.
It is necessary to go to two loop order in order to analyze its
observable dependence.

\subsection{Two loop result}
\begin{figure}
\centering
\includegraphics[width=3cm]{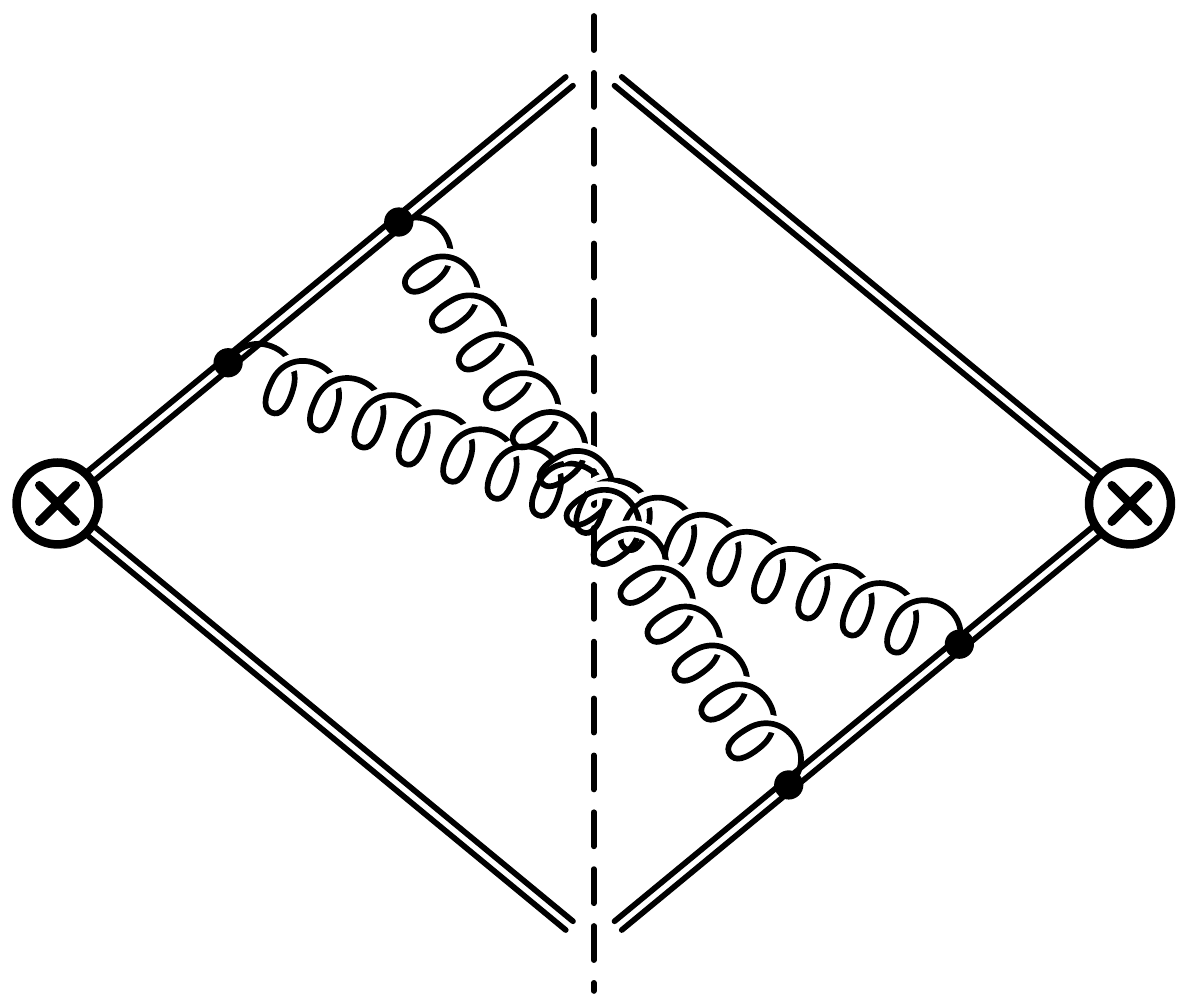}\hspace{0.5cm}
\includegraphics[width=3cm]{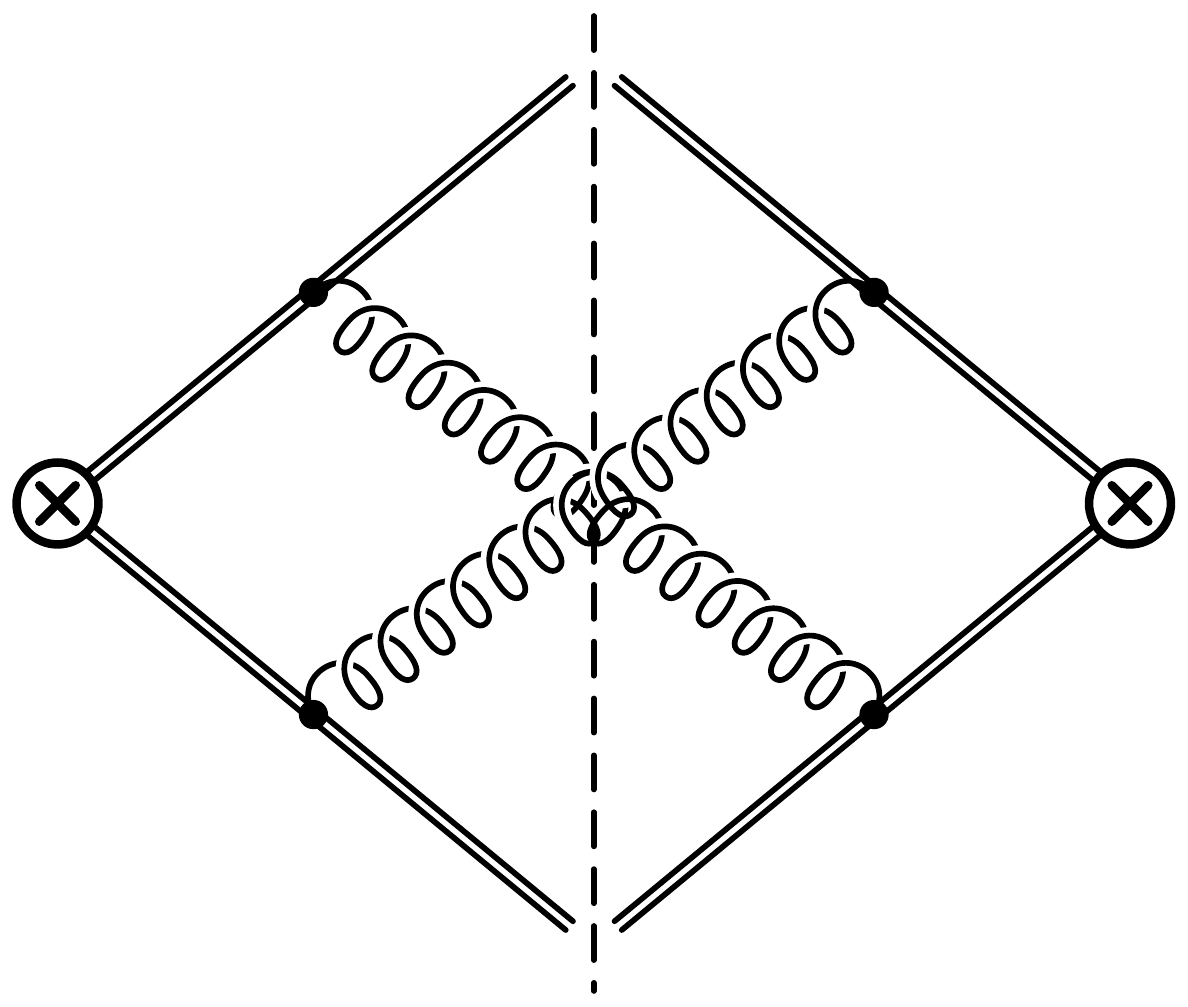}\hspace{0.5cm}
\includegraphics[width=3cm]{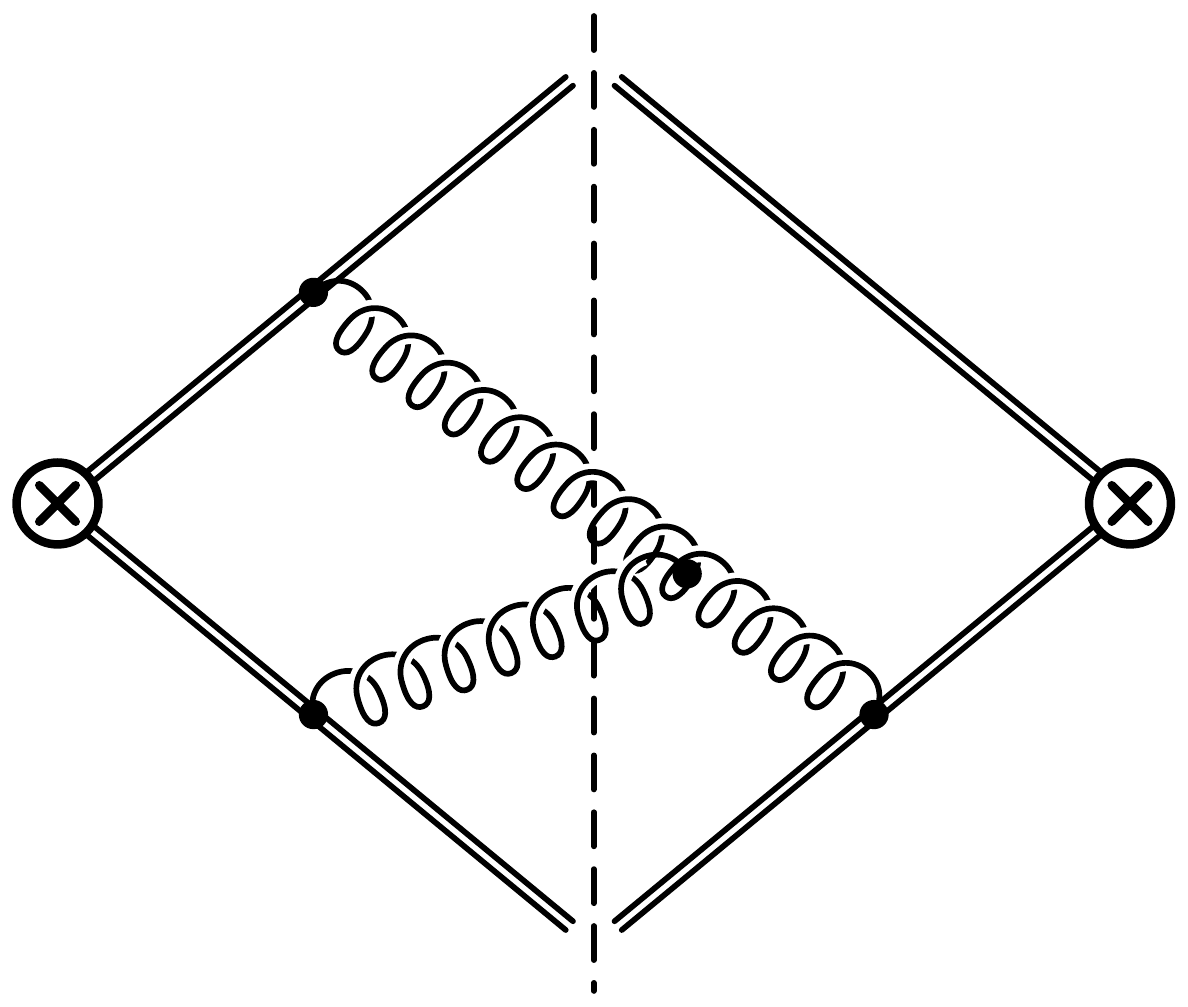}\hspace{0.5cm}
\includegraphics[width=3cm]{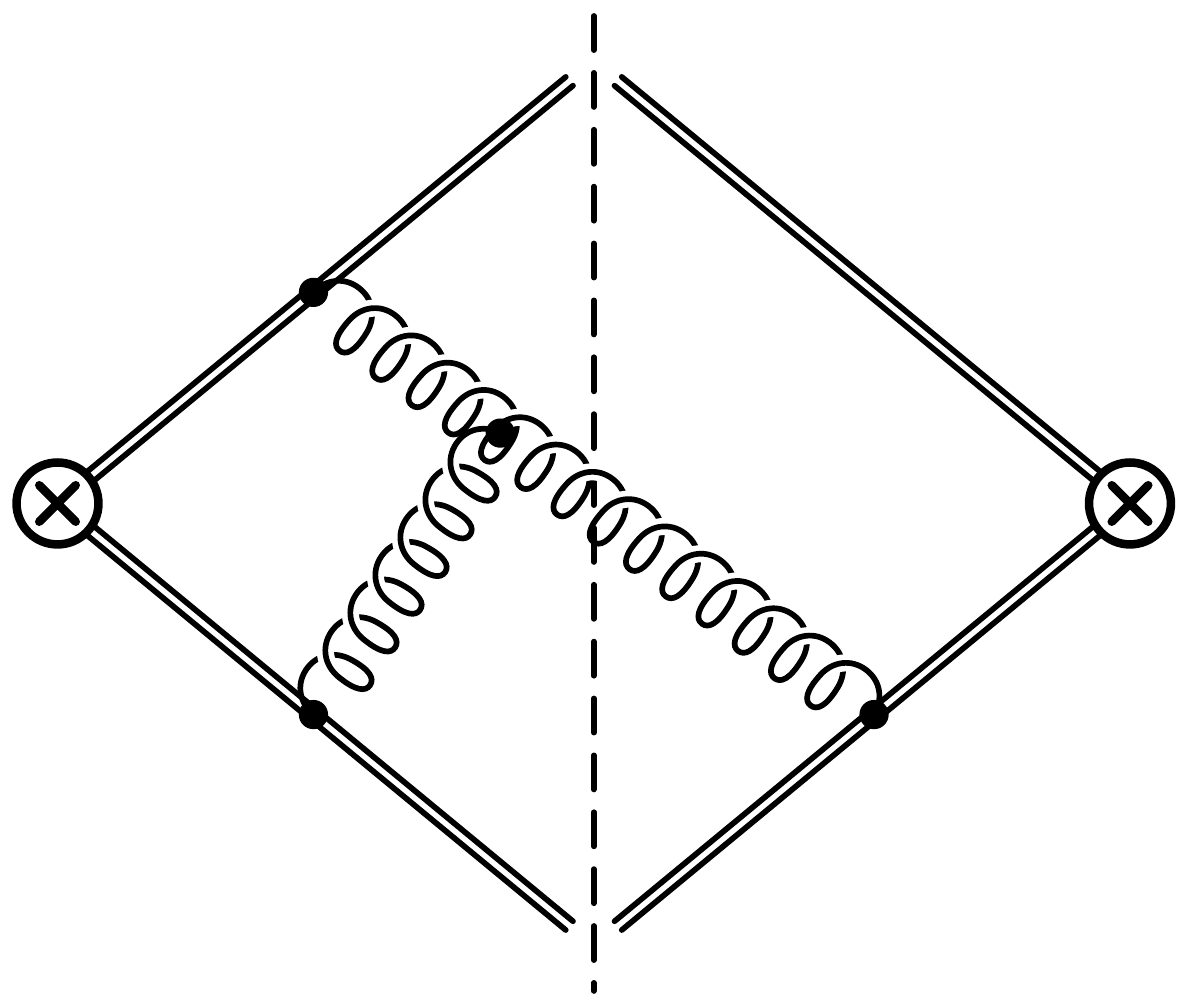}
\includegraphics[width=3cm]{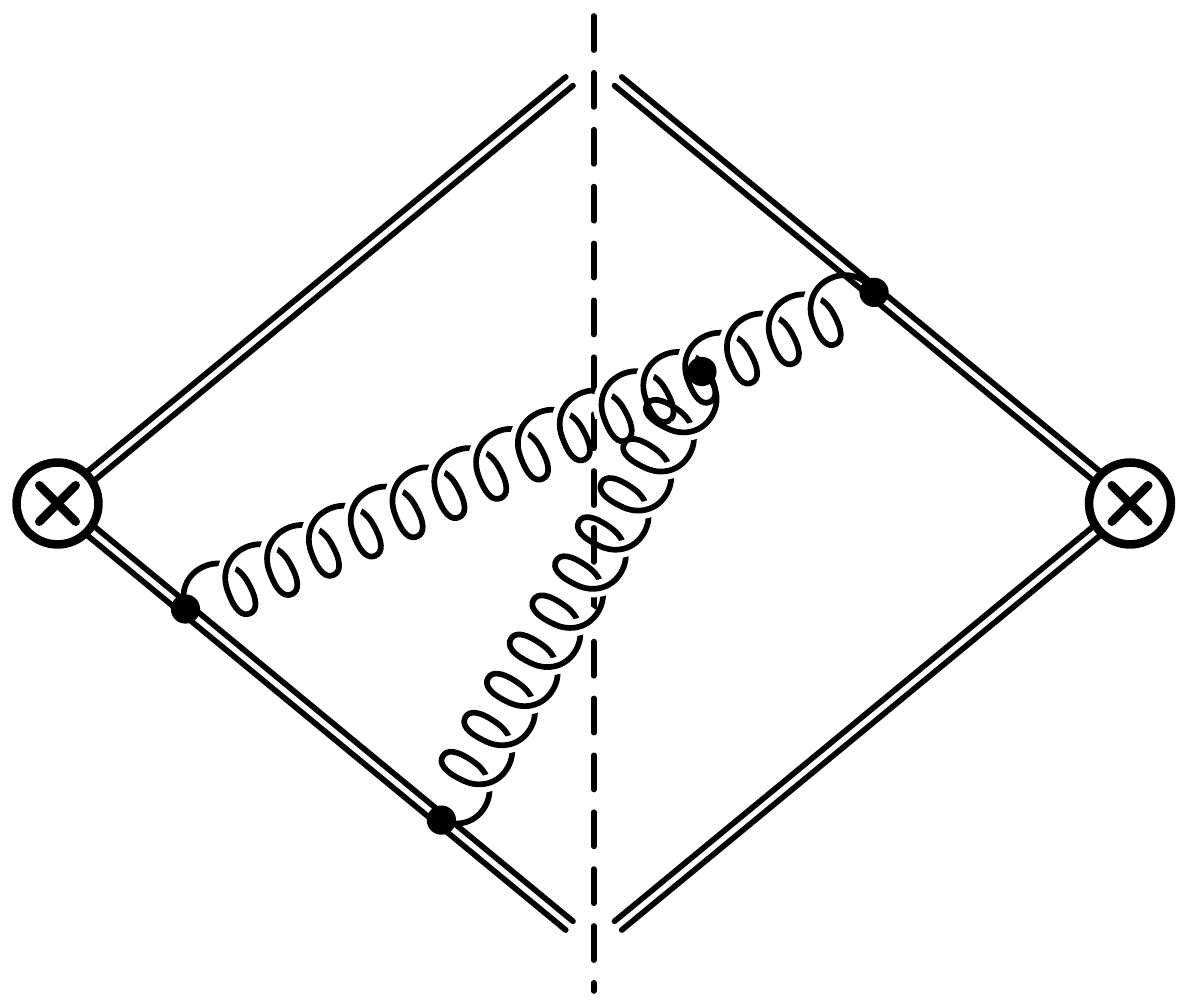}\hspace{0.5cm}
\includegraphics[width=3cm]{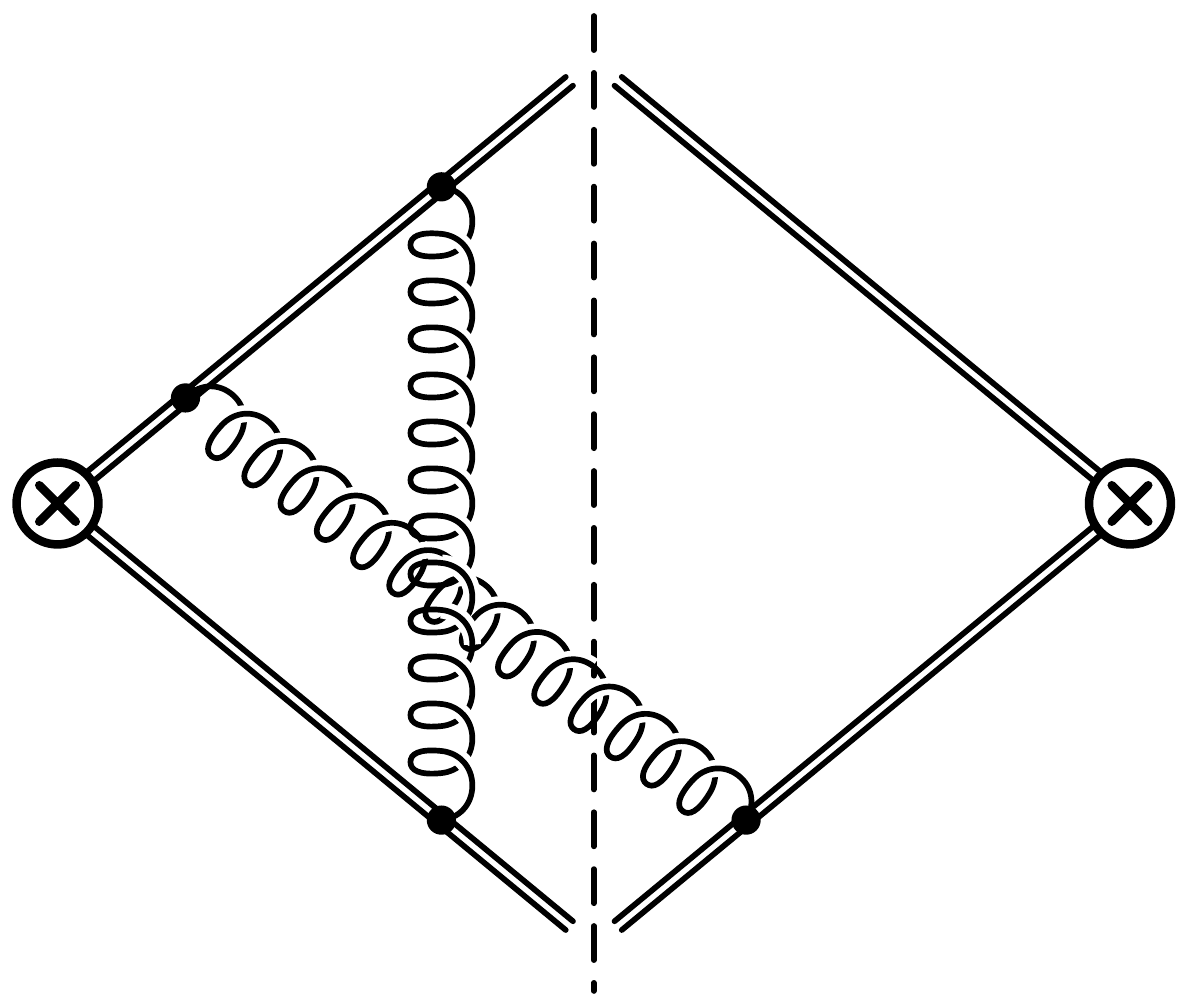}\hspace{0.5cm}
\includegraphics[width=3cm]{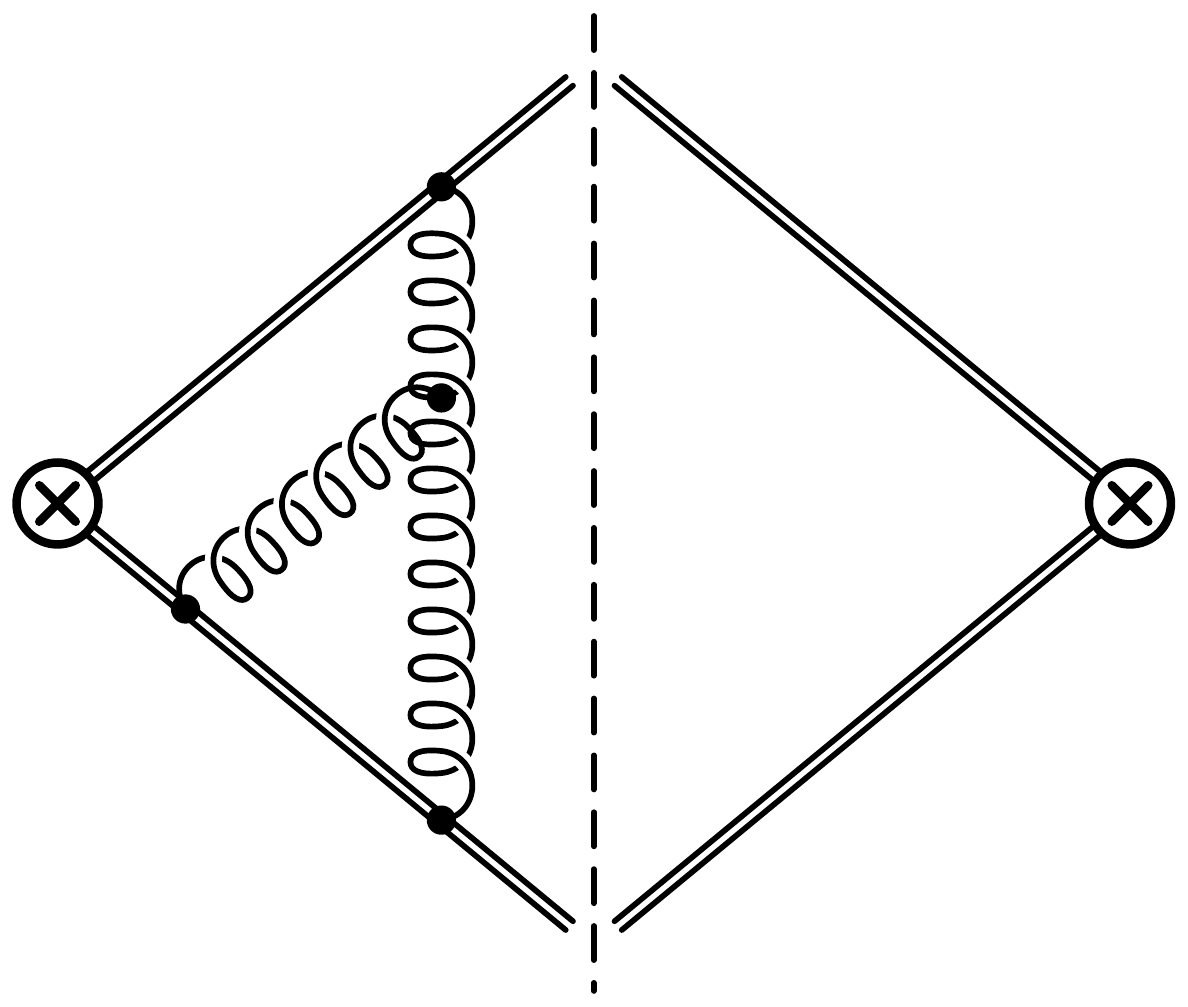}\hspace{0.5cm}
\includegraphics[width=3cm]{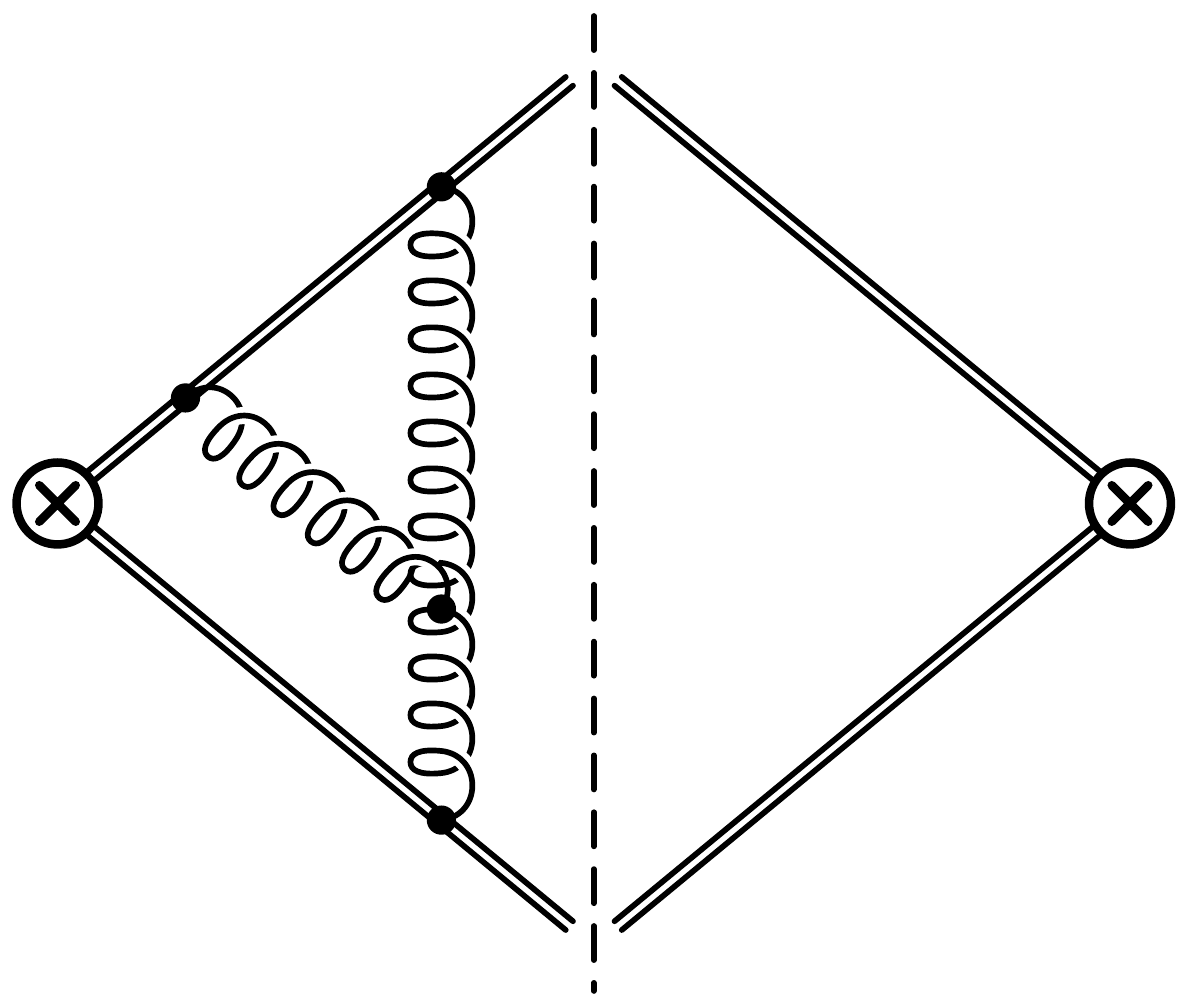}
\includegraphics[width=3cm]{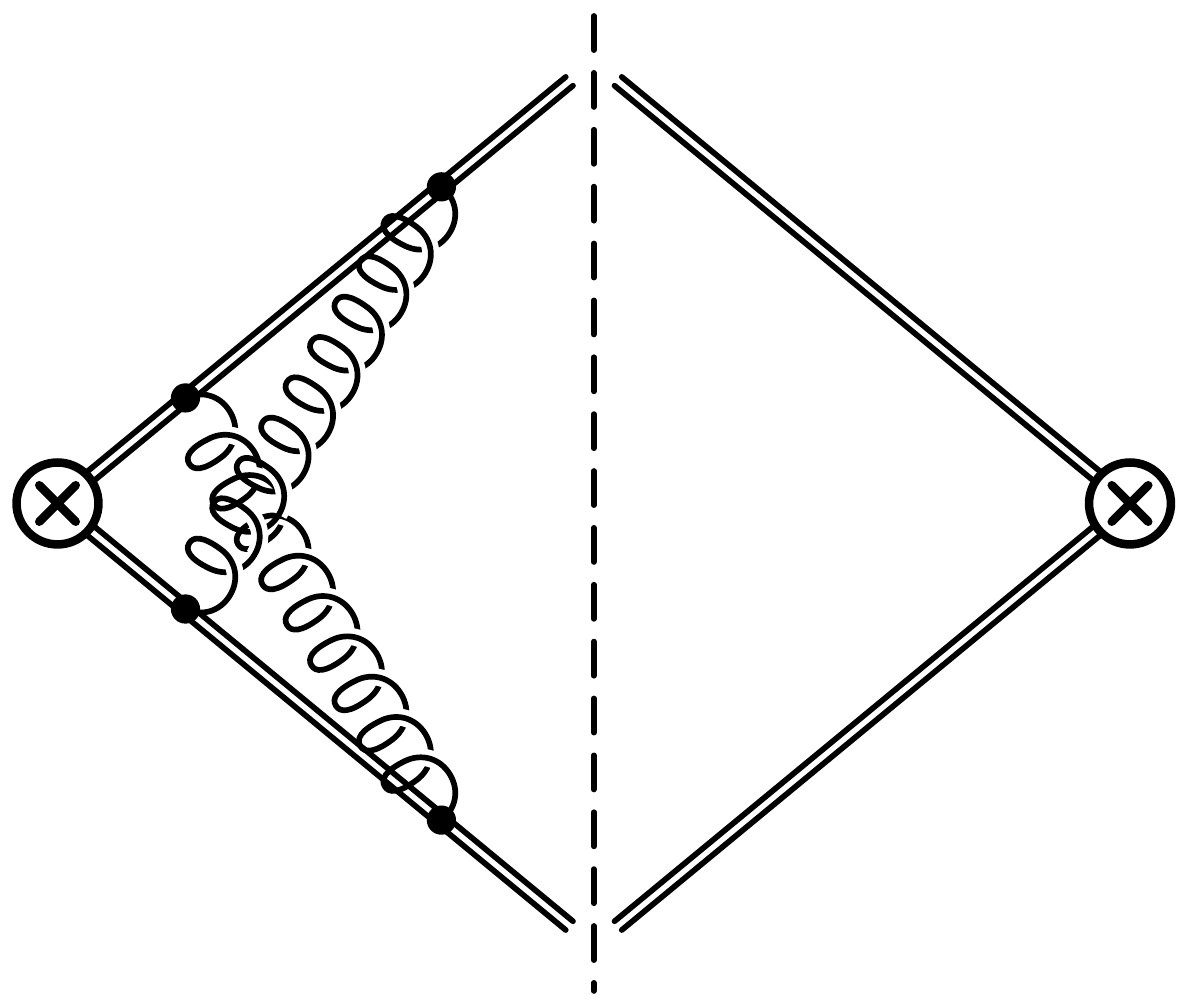}\hspace{0.5cm}
\includegraphics[width=3cm]{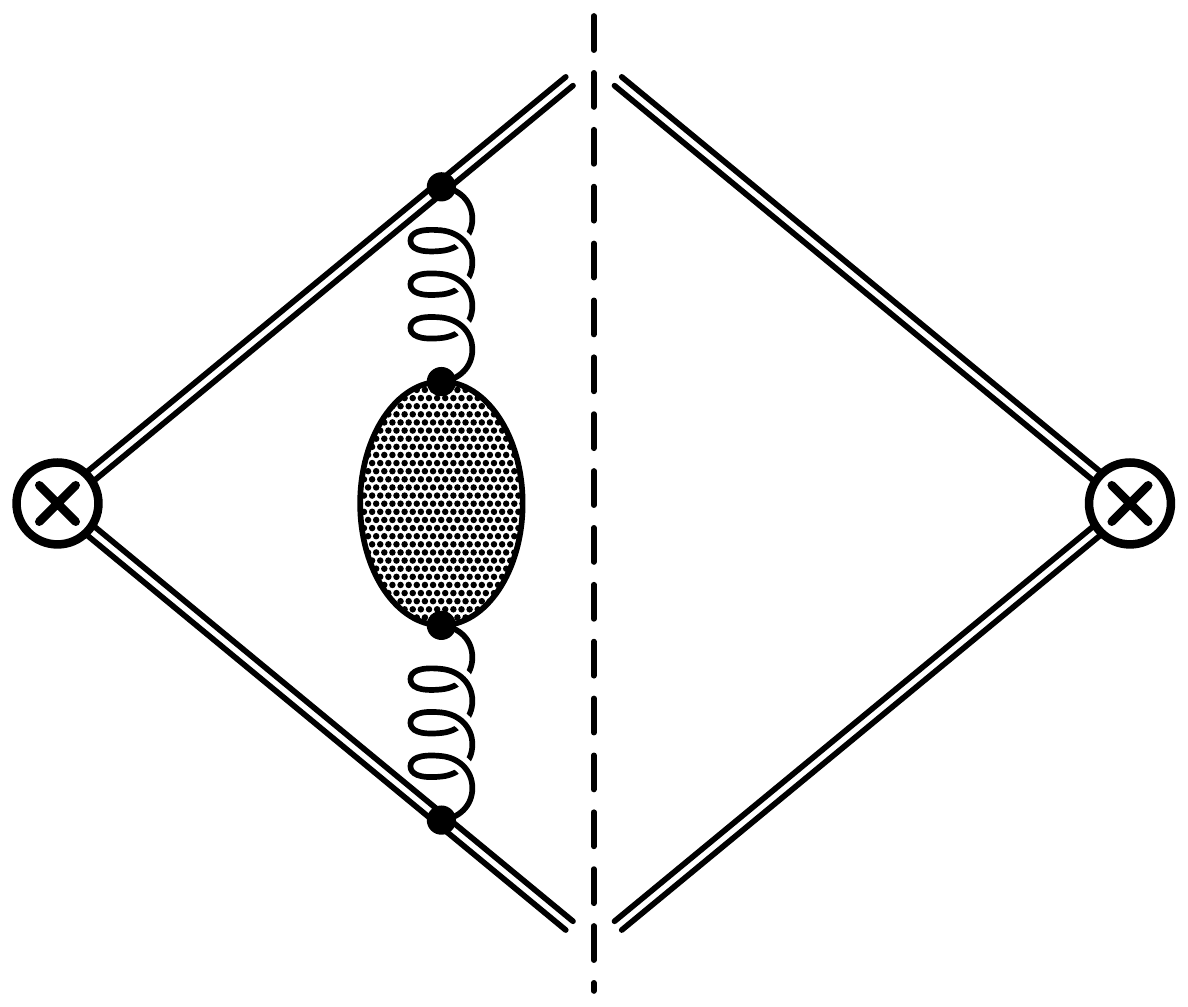}\hspace{0.5cm}
\includegraphics[width=3cm]{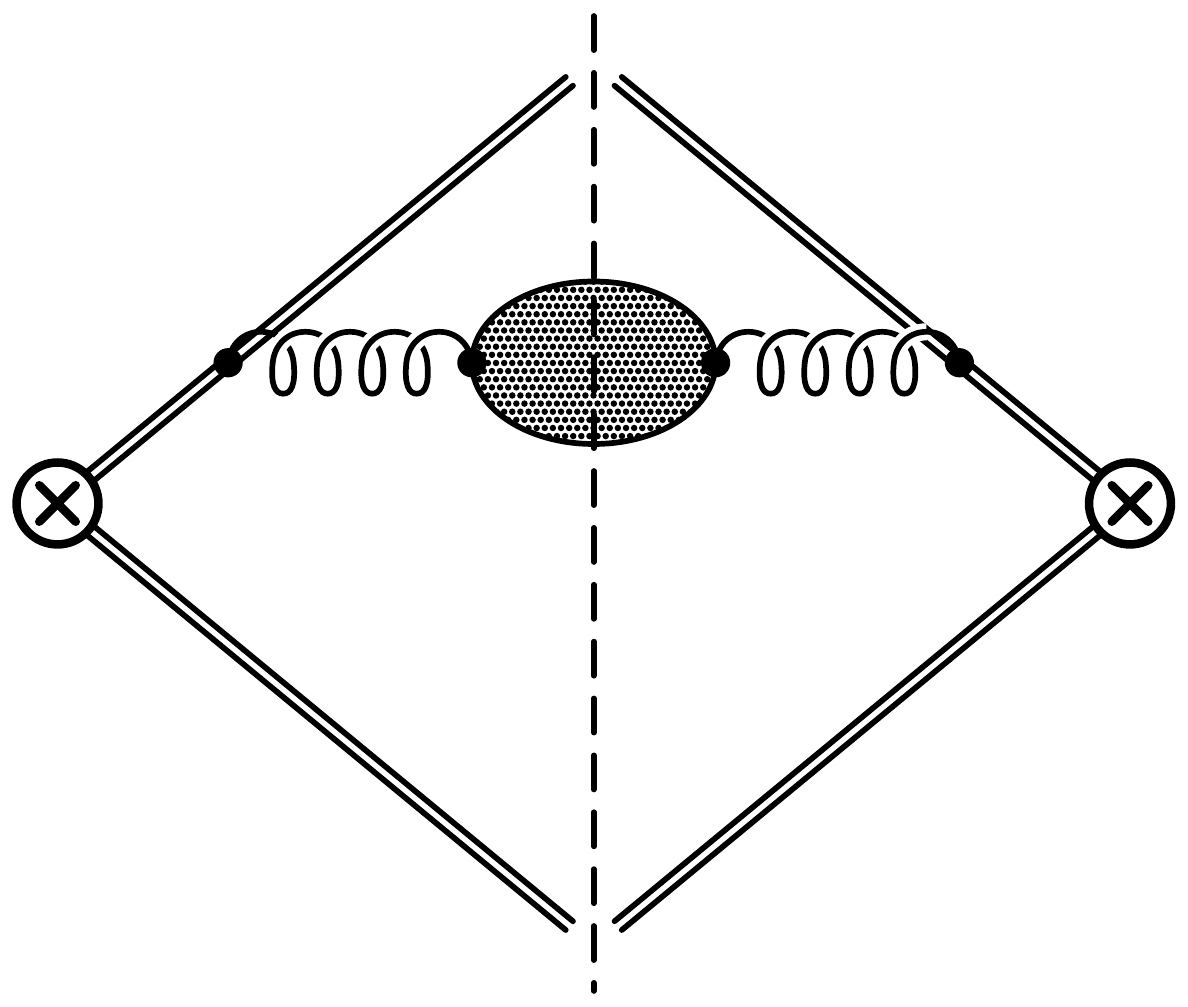}\hspace{0.5cm}
\includegraphics[width=3cm]{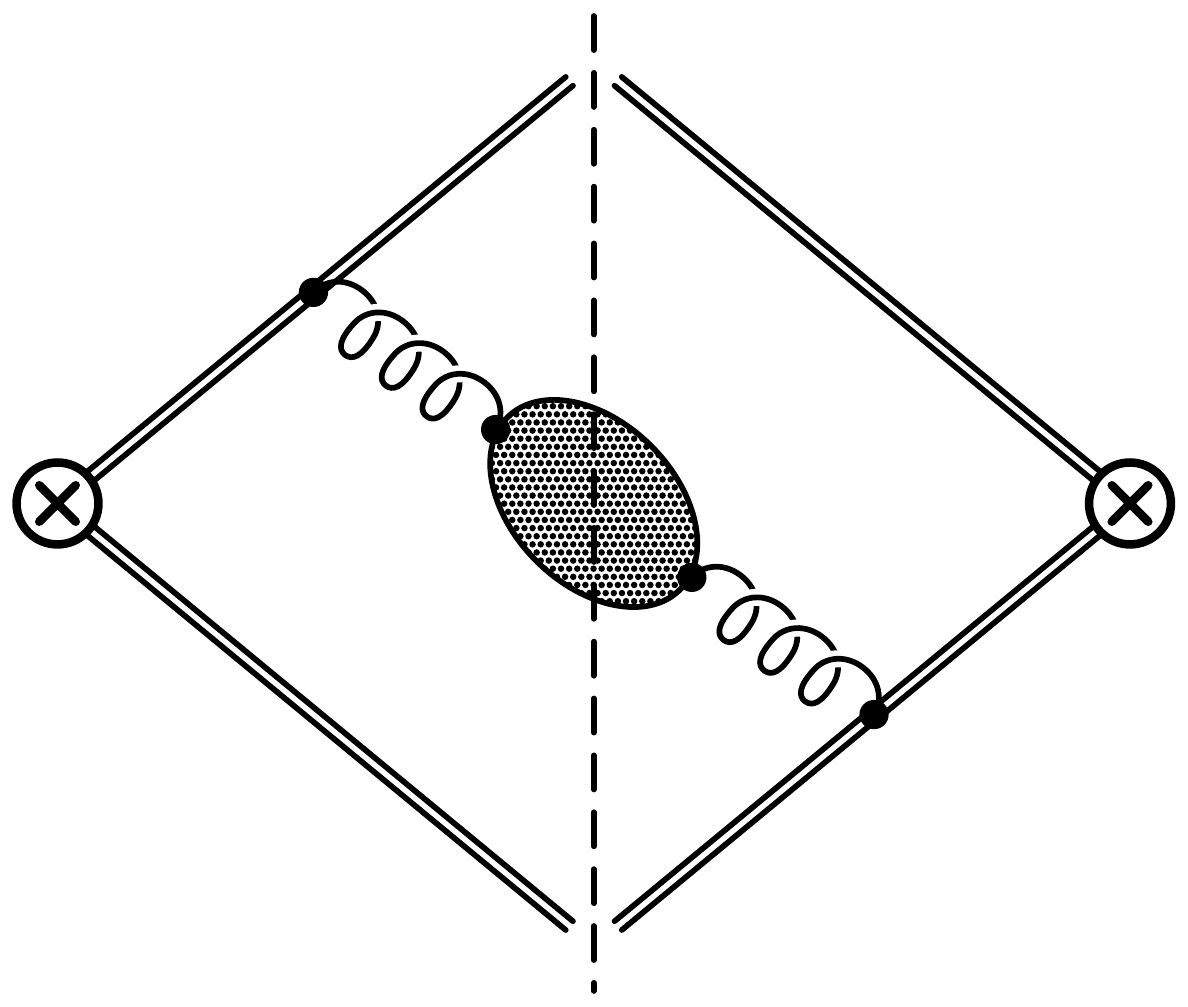}

\caption{Diagrams contributing to the non-Abelian part of the two-loop
  soft function. The gray blobs include the contribution of quarks,
  gluons and ghosts to the self energy of the gluon.}
\label{fig:soft2L}
\end{figure}

We now describe the main steps of the two loop calculation of the soft
anomalous dimensions, while the jet anomalous dimensions can be
derived via consistency relations by requiring the physical
distribution of the angularities to be both $\mu$ and $\nu$
independent. The relevant Feynman diagrams are given in
Fig.~\ref{fig:soft2L}, where the mirror conjugate graphs have been
omitted for simplicity.
We start from the standard Sudakov decomposition for the two soft
partons $k_a$ and $k_b$,
\begin{align}
k_a^\mu &= k_a^- \frac{n^\mu}{2}+k_a^+ \frac{\bar{n}^\mu}{2}+k^\mu_{a,t}\,,\notag\\
k_b^\mu &= k_b^- \frac{n^\mu}{2}+k_b^+ \frac{\bar{n}^\mu}{2}+k^\mu_{b,t}\,,
\end{align}
with $k_{i}^{+}=n\cdot k_{i}$, $k_{i}^{-}=\bar{n}\cdot k_{i}$, and
$n\cdot\bar{n}=2$. We also define $k_{i,\perp}^2=-k_{i,t}^2$ and
$k_{i}^{+}k_{i}^{-}=k_{i,\perp}^2$.
The double virtual diagrams give a scaleless contribution. For the
real-virtual corrections, the one-loop amplitude for the emission of a
soft gluon was derived in~\cite{Catani:2000pi}, and the phase space of
the real gluon is parameterized as
\begin{align}
\frac{\df^dk_a}{(2\pi)^{d-1}} \, \delta(k_a^2) \, \Theta(k_a^0)=\frac{1}{2}\df k_a^+ \, \df
  k_a^- \, \frac{\df^{2-2\epsilon}k_{a,t}}{(2\pi)^{3-2\epsilon}} \, \delta(k_a^+k_a^--k_{a,\perp}^2)\,.
\end{align}
For the double real correction, we introduce the variable $z$ as
\begin{align}
k_a^- = z k^-,\quad k_b^- = (1-z) k^-\,,
\end{align}
where $k^\pm$ are the light cone components of the
$k^\mu\equiv k_a^\mu+k_b^\mu$ momentum of invariant mass $m$ and now
\begin{equation}
k^+k^-=k_{\perp}^2+m^2\,.
\end{equation}
The $d$-dimensional phase space for the emission of $k_a$ and $k_b$
can then be parameterized as
\begin{align}
[\df k_{ab}]&\equiv\frac{\df^dk_a}{(2\pi)^{d-1}}\frac{\df^dk_b}{(2\pi)^{d-1}} \, \delta(k_a^2) \, \Theta(k_a^0)
 \,  \delta(k_b^2)\Theta(k_b^0) \notag\\
 &= \frac{1}{2}\df k^+\df
  k^-\frac{\df^{2-2\epsilon}k_{t}}{(2\pi)^{3-2\epsilon}}\frac{\df
  m^2}{m^{2\epsilon}}\delta(k^+k^--k_{\perp}^2-m^2) \frac{1}{(4 \pi)^2}\,\df z\,
  z^{-\epsilon}(1-z)^{-\epsilon}\frac{\df\Omega_{2-2\epsilon}}{(2\pi)^{1-2\epsilon}}\,,
\end{align}
with $\Omega_{2-2\epsilon}$ being the $(2-2\epsilon)$-dimensional
solid angle
\begin{equation}
\frac{\df\Omega_{2-2\epsilon}}{(2\pi)^{1-2\epsilon}} =
\frac{(4\pi)^\epsilon}{\sqrt{\pi}\Gamma\left(\frac12-\epsilon\right)}\df
\phi \sin^{-2\epsilon}\phi\,.
\end{equation}
The $C_F C_A$ and $C_F n_F$ contributions
to the double soft, tree-level squared amplitude (hereby denoted by
$M^2_{s,0}(k_a,k_b)$) can be found in several places and it is given in
the above parameterization in Appendix~\ref{app:doublesoft}. 
Due to the non-Abelian exponentiation
theorem~\cite{Gatheral:1983cz,Frenkel:1984pz}\footnote{Note that the exponential regulator preserves the structure predicted by the non-Abelian exponentiation theorem.}, we do not consider the
$C_F^2$ contribution from the radiation of two independent gluons off
the Wilson lines, as that is determined entirely by the leading order
calculation given in the previous subsection and hence does not
contribute to the two-loop soft anomalous dimension.

For a given angularity $\tau_a(k_a,k_b)$ evaluated on a double real
final state $\{k_a,k_b\}$, we then organize the calculation as
follows:
\begin{itemize}
\item 
We express the value of the angularity $\tau_a(k_a,k_b)$ in terms
of the above phase space variables as
\begin{align}
\label{eq:angularitydoublereal}
\tau_a(k_a,k_b) &=\,
  \frac{k_\perp}{Q}e^{-(1-a)|\eta|}\,(1+\mu^2)^{\frac{a-1}{2}}\,f_{a}(z,\mu,\phi)\,,\notag\\
f_{a}(z,\mu,\phi)&=\Bigg[z
  \bigg(1+2\sqrt{\frac{1-z}{z}}\mu\cos\phi+\frac{1-z}{z}\mu^2\bigg)^{1-\frac{a}{2}}\notag\\
& \qquad +(1-z)
\bigg(1-2\sqrt{\frac{z}{1-z}}\mu\cos\phi+\frac{z}{1-z}\mu^2\bigg) ^{1-\frac{a}{2}}\Bigg]\,,
\end{align}
where $\mu^2\equiv m^2/k_\perp^2$ and $\eta = \frac12 \ln \frac{k^-}{k^+}$.

\item We split the double-real correction into two terms as follows
\begin{align}
{\cal I}_{RR}&\equiv \frac{1}{2} \int [\df
 k_{ab}]\,e^{-(k^++k^-)\frac{e^{-\gamma_E}}{\nu}}\,M^2_{s,0}(k_a,k_b)\,\delta(\tau-\tau_a(k_a,k_b))
               ={\cal I}_{RR}^{\small \rm{(I)}} + {\cal I}_{RR}^{\small \rm{(NI)}}\,,
               \label{eq:6.16}
\end{align}
where the factor $1/2$
in Eq.~\eqref{eq:6.16} is a combinatorial factor in the case of
two gluons, and represents $T_F=1/2$ in the case of a gluon splitting
into two quarks (factored out from the squared amplitude given in
Appendix~\ref{app:doublesoft}).
We introduce a simplified observable $\tilde{\tau}_a(k)$ defined as
\begin{equation}
\tilde{\tau}_a(k) = \frac{k_\perp}{Q}e^{-(1-a)|\eta|}\,,
\end{equation}
with, as above, $\eta = \frac12 \ln \frac{k^-}{k^+}$.
The double real contribution can then be split as the sum of the
following two integrals
\begin{align}
\label{eq:doublereal}
{\cal I}_{RR}^{\small \rm{(I)}} &= \frac{1}{2} \int [\df
 k_{ab}]\,e^{-(k^++k^-)\frac{e^{-\gamma_E}}{\nu}}\,M^2_{s,0}(k_a,k_b)\,\delta(\tau-\tilde{\tau}_a(k))\,,
\\
{\cal I}_{RR}^{\small \rm{(NI)}}&=\frac{1}{2} \int [\df
                                  k_{ab}]\,e^{-(k^++k^-)\frac{e^{-\gamma_E}}{\nu}}\,M^2_{s,0}(k_a,k_b)\left[\delta(\tau-\tau_a(k_a,k_b))-\delta(\tau-\tilde{\tau}_a(k))\right]
                                  \notag\,,
\end{align}
The {\it inclusive} integral ${\cal I}_{RR}^{\small \rm{(I)}}$ is
defined by the first equation in~\eqref{eq:doublereal}, that is
replacing the angularity with the observable $\tilde{\tau}_a$.
The {\it non-inclusive} correction ${\cal I}_{RR}^{\small \rm{(NI)}}$,
defined by the second equation in~\eqref{eq:doublereal}, accounts for
the difference between the actual observable $\tau_a(k_a,k_b)$ and its
inclusive approximation $\tilde{\tau}_a(k)$. 
 
\end{itemize}

The reason for splitting the calculation into an inclusive and non-inclusive contribution is that,
as discussed in this paper,
${\cal I}_{RR}^{\small \rm{(NI)}}$ encodes the difference between two IRC safe observables which only depends on the extra UV regulator, but is finite in dimensional regularization and does not contribute
to the $\widehat\gamma_S[\alpha_s(\mu)]$ anomalous dimension.
Therefore this non-inclusive piece, which contains the complexity associated with the
observable definition, is defined solely in terms of double real
diagrams and can be evaluated directly in four dimensions
(numerically if necessary). 
Similar ideas were
proposed and exploited in
Refs.~\cite{Jouttenus:2011wh,Bauer:2011hj,Banfi:2014sua,Gangal:2016kuo}.
Conversely, the inclusive contribution is
considerably simpler and can be easily computed analytically for a
generic observable.
The
$\gamma_{1,1-a}^{(\nu)}[\alpha_s(\mu)]$ anomalous dimension governing the
$\nu$ RGE receives contributions from both integrals. 
We recall that,
in general, one should take the limit $\epsilon\to 0$ first, in order
to isolate the observable dependence with the $\nu$ exponential
regulator. 
An exception is given by the case $a=1$ (broadening-like angularity),
where the two limits $\epsilon\to 0$ and $\nu\to \infty$ commute with
the regulator adopted here.

\begin{figure}
 \includegraphics[width=0.5\linewidth]{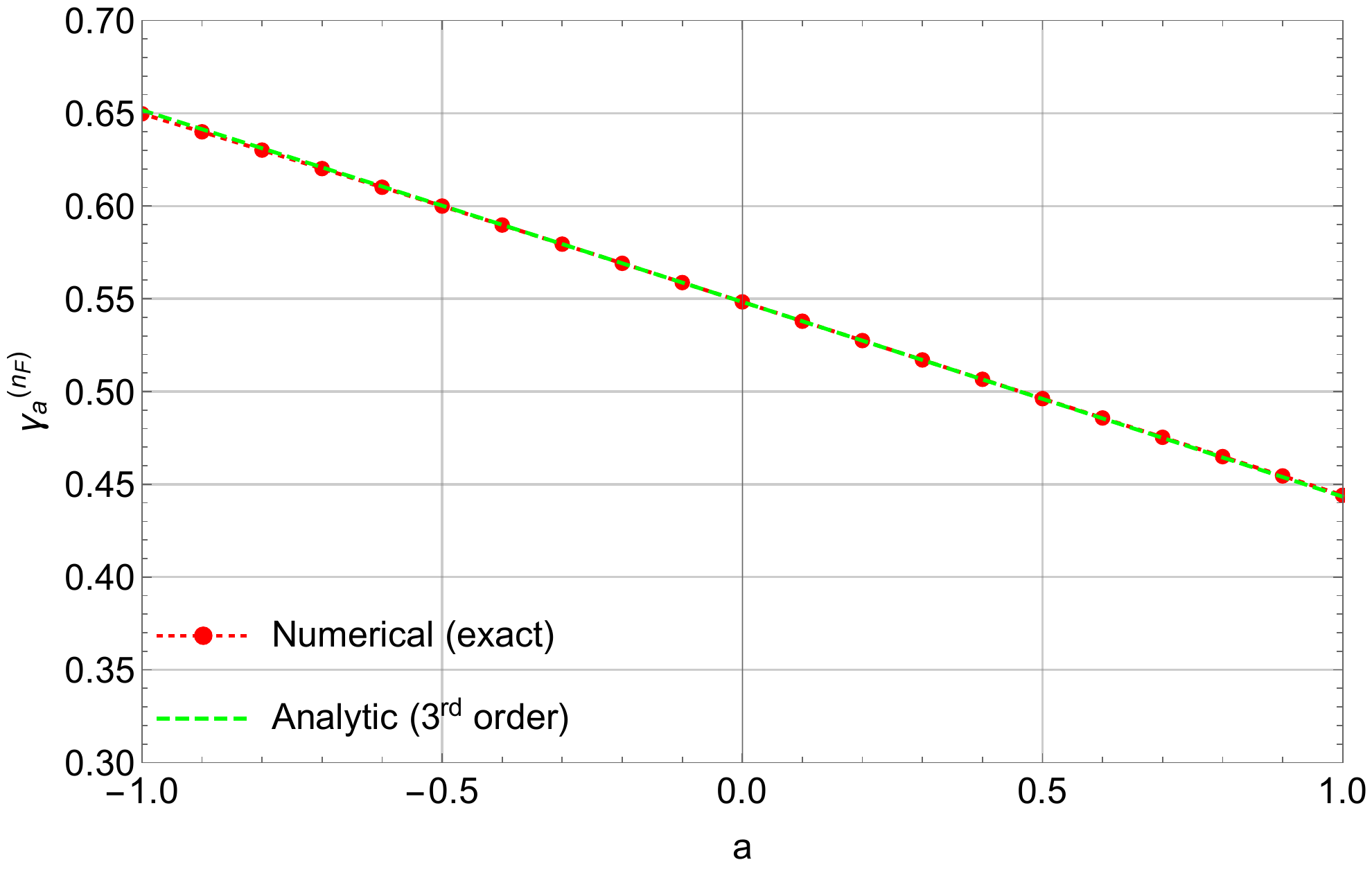}
\includegraphics[width=0.5\linewidth]{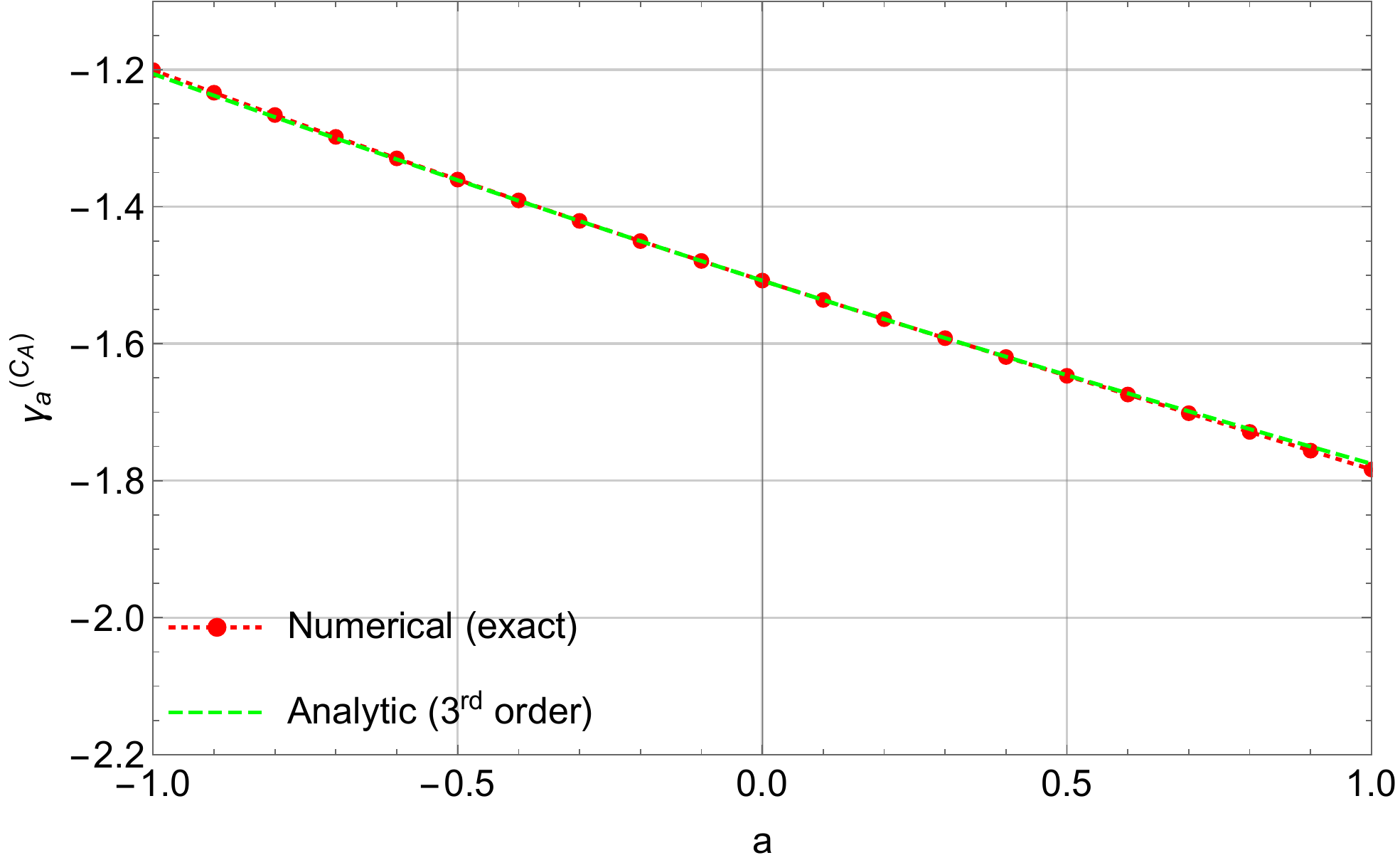}
\caption{The functions $\gamma_a^{(n_F)}$ (left) and
  $\gamma_a^{(C_A)}$ (right) as a function of $a$. The dashed (green)
  line shows the analytic result based on a third order Taylor
  expansion, while the points (red) represent the result of the
  numerical integration.}
 \label{fig:gammaNIcoeffs}
\end{figure}

Working in the $\overline{\rm MS}$ scheme, we obtain the following
two-loop anomalous dimensions 
\begin{align}
\label{eq:gamma2loops}
\widehat{\gamma}^{(1)}_S[\alpha_s(\mu)]  &= \,2 \,C_F
                           C_A\left(\frac{808}{27}-\frac{11}{9}\pi^2-28\zeta_3\right)-
                           2 \,C_F n_F
                           T_F\left(\frac{224}{27}-\frac{4}{9}\pi^2\right)\notag\\
& = \,2 \,C_F
                           C_A\left(\frac{808}{27}-28\zeta_3\right)-
                           2 \,C_F n_F
                           T_F\frac{224}{27} -
  \frac{2}{3}\pi^2 C_F \,b_0\,,\notag\\
\gamma_{1, 1-a}^{(\nu,\,1)}[\alpha_s(\mu)]  &= -\,C_F
                           C_A\left(\frac{808}{27}-28\zeta_3\right)+
                            \,C_F n_F T_F\frac{224}{27}
                                     -\frac{4}{3}\pi^2 C_F \frac{2-
                     a(2-a)}{(2-a)^2} \, b_0 \notag\\
& \quad- 16 C_F\left(C_A
                                     \gamma_a^{(C_A)}+T_F n_F
                                     \gamma_a^{(n_F)}\right)\,,
\end{align}
with $b_0= ({11 C_A-2 n_F})/{3}$.
In the computation we expanded the part of the exponential function in
the integrand relative to the smaller light cone component, and
neglect subleading power corrections that would not contribute to the
(leading power) anomalous dimensions.
The inclusive contribution was evaluated analytically using sector
decomposition with the help of {\tt HypExp}~\cite{Huber:2005yg}, and
cross checked numerically with {\tt pySecDec}~\cite{Borowka:2017idc}.
The quantities $\gamma_a^{(C_A)}$ and $\gamma_a^{(n_F)}$ arise from the
non-inclusive correction which can be calculated in four
dimensions. 
They are given by the finite integrals
\begin{align}
  \label{eq:gammaIntegrals}
  \gamma_a^{(C_A)} & = \frac{1}{2-a}\int_0^\infty\frac{d\mu^2}{\mu^2(1+\mu^2)} \int_0^1 dz
      \int_0^{2\pi}\frac{d\phi}{2\pi}\frac{1}{2!} \left(2
                   \mathcal{S}_{\rm s.o.}+\mathcal{H}_g\right)\ln f_{a}(z,\mu,\phi)\,, \\
  \gamma_a^{(n_F)} & = \frac{1}{2-a} \int_0^\infty\frac{d\mu^2}{\mu^2(1+\mu^2)} \int_0^1 dz
      \int_0^{2\pi}\frac{d\phi}{2\pi} \mathcal{H}_q\ln f_{a}(z,\mu,\phi)\,,
\end{align}
where the functions ${\cal S}_{\rm s.o.}$, ${\cal H}_g$ and
${\cal H}_q$ are given in Appendix~\ref{app:doublesoft}, and the
function $f_a$ is given in Eq.~\eqref{eq:angularitydoublereal}. 
A numerical computation shows that $\gamma_a^{(C_A)}$,
$\gamma_a^{(n_F)}$ (for $a<2$) exhibit an almost exactly linear
dependence on $a$, as displayed in Fig.~\ref{fig:gammaNIcoeffs}.
One can therefore expand these functions in a Taylor series around
$a=0$, and consider the first few terms as an analytic approximation
of the exact result. 
We evaluate the $\phi$ integrals by contour integration and, after
integrating over $\mu$, we carry out the final integration over $z$
either analytically or numerically with ${\cal O}(100)$ significant
digits, which allows us to reconstruct the analytic answer by means of
the PSLQ algorithm~\cite{10.2307/2585116}. We also perform a numerical
cross check using the {\tt Cuba} libraries~\cite{Hahn:2004fe}.
We give here the expansion to third order, which is sufficient to
reach a few-permille accuracy in the interesting range $a\in [-1,1]$
considered in our study:
\begin{align}
\label{eq:gammaIntegrals-taylor}
&  C_A\gamma_a^{(C_A)}+
  T_F  n_F\gamma_a^{(n_F)}=-\frac{\zeta_2}{4}b_0+\left[C_A\left(\frac{41}{96}-\frac{\zeta_2}{4}-\frac{\zeta_3}{4}\right)-\frac{10}{96}
  T_F n_F\right] a\notag\\
&+
  \left[C_A\left(\frac{57\,941}{537\,600}+\frac{277}{163\,840}\zeta_2-\frac{9}{32}\zeta_2\,\ln
  2 +\frac{121}{640}\zeta_3\right) + T_F n_F\left(-\frac{131}{1440}+\frac{3}{40}\zeta_3\right)\right] a^2  +{\cal O}(a^3)\,.
\end{align}

The numerical value (i.e. not based on a Taylor expansion) for
$\gamma_a^{(C_A)}$, $\gamma_a^{(n_F)}$ for $a<2$ is given in
Table~\ref{tab:gammaNIcoeffs} for several values of $a$.
The third order expansion of Eq.~\eqref{eq:gammaIntegrals-taylor}
provides an excellent approximation of the full result over the whole
$a$ range relevant for the theoretical considerations made here on the
transition between the \scetI{} and \scetII{} regimes, as can be seen
from the comparison in Fig.~\ref{fig:gammaNIcoeffs}.
\begin{table}
\parbox{.35\linewidth}{
\centering 
\begin{tabular}{|c|c|c|}
  \hline
  $a$ & $\gamma_a^{(n_F)}$ & $\gamma_a^{(C_A)}$ \\
  \hline
-1. & 0.650 & -1.201   \\
 \hline
-0.9 & 0.640 & -1.234   \\
 \hline
-0.8 & 0.630 & -1.266   \\
 \hline
-0.7 & 0.620 & -1.298   \\
 \hline
-0.6 & 0.610 & -1.330   \\
 \hline
-0.5 & 0.600 & -1.361   \\
 \hline
-0.4 & 0.590 & -1.391   \\
 \hline
-0.3 & 0.579 & -1.421   \\
 \hline
-0.2 & 0.569 & -1.450   \\
 \hline
-0.1 & 0.559 & -1.479   \\
 \hline\end{tabular}
}
\parbox{.3\linewidth}{
\centering
\begin{tabular}{|c|c|c|}
  \hline
  $a$ & $\gamma_a^{(n_F)}$ & $\gamma_a^{(C_A)}$ \\
  \hline
0. & 0.548 & -1.508   \\
 \hline
0.1 & 0.538 & -1.536   \\
 \hline
0.2 & 0.527 & -1.564   \\
 \hline
0.3 & 0.517 & -1.592   \\
 \hline
0.4 & 0.507 & -1.620  \\
 \hline
0.5 & 0.496 & -1.647   \\
 \hline
0.6 & 0.486 & -1.674   \\
 \hline
0.7 & 0.475 & -1.701  \\
 \hline
0.8 & 0.465 & -1.729   \\
 \hline
0.9 & 0.454 & -1.756   \\
 \hline
\end{tabular}
}
\hspace{0.2cm}
\parbox{.3\linewidth}{
\centering
\begin{tabular}{|c|c|c|}
  \hline
  $a$ & $\gamma_a^{(n_F)}$ & $\gamma_a^{(C_A)}$ \\
  \hline
1. & 0.444 & -1.784   \\
 \hline
1.1 & 0.434 & -1.811   \\
 \hline
1.2 & 0.423 & -1.839   \\
 \hline
1.3 & 0.413 & -1.868   \\
 \hline
1.4 & 0.402 & -1.896   \\
 \hline
1.5 & 0.392 & -1.925   \\
 \hline
1.6 & 0.381 & -1.955  \\
 \hline
1.7 & 0.371 & -1.984   \\
 \hline
1.8 & 0.361 & -2.015   \\
 \hline
1.9 & 0.350 & -2.046   \\
 \hline
\end{tabular}
}
\caption{Full numerical values for the functions $\gamma_a^{(n_F)}$ and
  $\gamma_a^{(C_A)}$ for different angularities corresponding to the
  parameter $a$, contributing to the soft anomalous dimensions of
  Eq.~\eqref{eq:gamma2loops}. The values for $\gamma_a^{(n_F)}$ are
  rounded to the nearest 0.001, while for $\gamma_a^{(C_A)}$ the numerical uncertainty is at
  most $\pm 1$ in the last digit.}
\label{tab:gammaNIcoeffs}
\end{table}

As advertised, $\widehat{\gamma}_S[\alpha_s(\mu)]$ now
does not depend on the specific observable, and it is given by the
single logarithmic part of the soft anomalous dimension of the quark
form factor, which coincides with the DGLAP soft anomalous dimension
used for threshold resummation. 
Conversely, the entire observable
dependence is now encoded in $\gamma_{1,1-a}^{(\nu)}$, which is common
to the soft and jet functions and technically simpler to compute in
that it only depends on the soft and collinear limit encoded in the
zero-bin subtraction as discussed in Section~\ref{eq:refactorization}.
The corresponding anomalous dimensions for the jet function can be
immediately derived from the standard consistency relation
\begin{align}
  \label{eq:consistencyrel}
\widehat{\gamma}^{(\mu)}_S[\alpha_s(\mu)] + 2\,
  \widehat{\gamma}^{(\mu)}_J[\alpha_s(\mu)] +
  \gamma_H[\alpha_s(\mu)]=0
\,,
\end{align}
and from Eq.~\eqref{4.13}.
In the next section we will show how to obtain the anomalous
dimensions in standard \scetI{} starting from the results obtained
above using the considerations of Section~\ref{sec:RGRelationship}.

\subsection{Relation to standard \scetI{} anomalous dimensions}
\label{sec:SCETIAnomDim}
While for $a=1$ the results of the previous section directly provide
the standard \scetII{} soft anomalous dimension, they can also be used
to derive the \scetI{} soft anomalous dimension as obtained in pure
dimensional regularization by means of
Eqs.~\eqref{eq:finalAnomDim}. 
Analogous considerations hold for the jet function, therefore we focus
on the soft function first.

The quantities $\widehat{\gamma}_S[\alpha_s(\mu)] $ and
$\gamma_{1, 1-a}^{(\nu)}[\alpha_s(\mu)] $ entering
Eqs.~\eqref{eq:finalAnomDim} are given in Eq.~\eqref{eq:gamma2loops},
and the only missing quantity is the one-loop coefficient
$d_{1, 1-a}^{(1)}$ of the initial condition of the $\Delta_{1,1-a}$
function (cf. Eq.~\eqref{eq:DeltaDef}).
This can be determined either by taking the square root of the ratio
between the initial condition (constant part) of the one-loop soft
function~\eqref{eq:softoneloop} and the corresponding result in pure
dimensional regularization, or equivalently by calculating the
zero-bin subtraction and taking its constant part.
The one-loop soft function in dimensional regularization can be found
in Ref.~\cite{Fleming:2007xt}, and its initial condition in Laplace space reads
\begin{equation}
\label{eq:constdimreg}
\hat{\cal S}^{\rm SCET_I}(\mu=M_S) = 1-\frac{\alpha_s(\mu_S)}{4 \pi}
\frac{\pi^2}{1-a} C_F\,.
\end{equation}
Taking the square root of the ratio of the constant part of
Eq.~\eqref{eq:softoneloop} to the latter equation we
obtain
\begin{equation}
d_{1, 1-a}^{(1)} = \frac{\pi^2}{6}C_F\frac{4+a(3 a-4)}{(2-a)(1-a)}\,.
\end{equation}

We set $\alpha=1$ and $\beta=1-a$ in Eqs.~\eqref{eq:finalAnomDim}, and consider
the series
\begin{equation}
\widehat\gamma^{\rm SCET_I}_{1,1-a;F}[(\alpha_s(\mu))]= \sum_{n=0}
\left(\frac{\alpha_s(\mu)}{4\pi}\right)^{n+1}\widehat{\gamma}^{(n),\,{\rm
  SCET_I}}_{1,1-a;F}\,,
\end{equation}
finding 
\begin{align}
\label{eq:2loopSCETI}
\widehat{\gamma}^{(0),\,{\rm SCET_I}}_\Sobsang &=0\,,\notag\\
\widehat{\gamma}^{(1),\,{\rm SCET_I}}_\Sobsang &=\frac{1}{1-a}\bigg[-2 \,C_F
                           C_A\left(\frac{808}{27}-28\zeta_3\right)+
                            \,2 C_F T_F n_F \frac{224}{27}
                                     +\frac{2}{3}\pi^2 C_F (2 a-3)b_0 \notag\\
& \quad- 32 (2-a)C_F\left(C_A
                                     \gamma_a^{(C_A)}+T_F  n_F
                                     \gamma_a^{(n_F)}\right) \bigg]\,,
\end{align}
where an analytic approximation of $\gamma_a^{(C_A)}$ and
$\gamma_a^{(n_F)}$ is given in Eq.~\eqref{eq:gammaIntegrals-taylor}
or, alternatively, their numerical value is reported in
Table~\ref{tab:gammaNIcoeffs}.

\begin{figure}
 \includegraphics[width=0.5\linewidth]{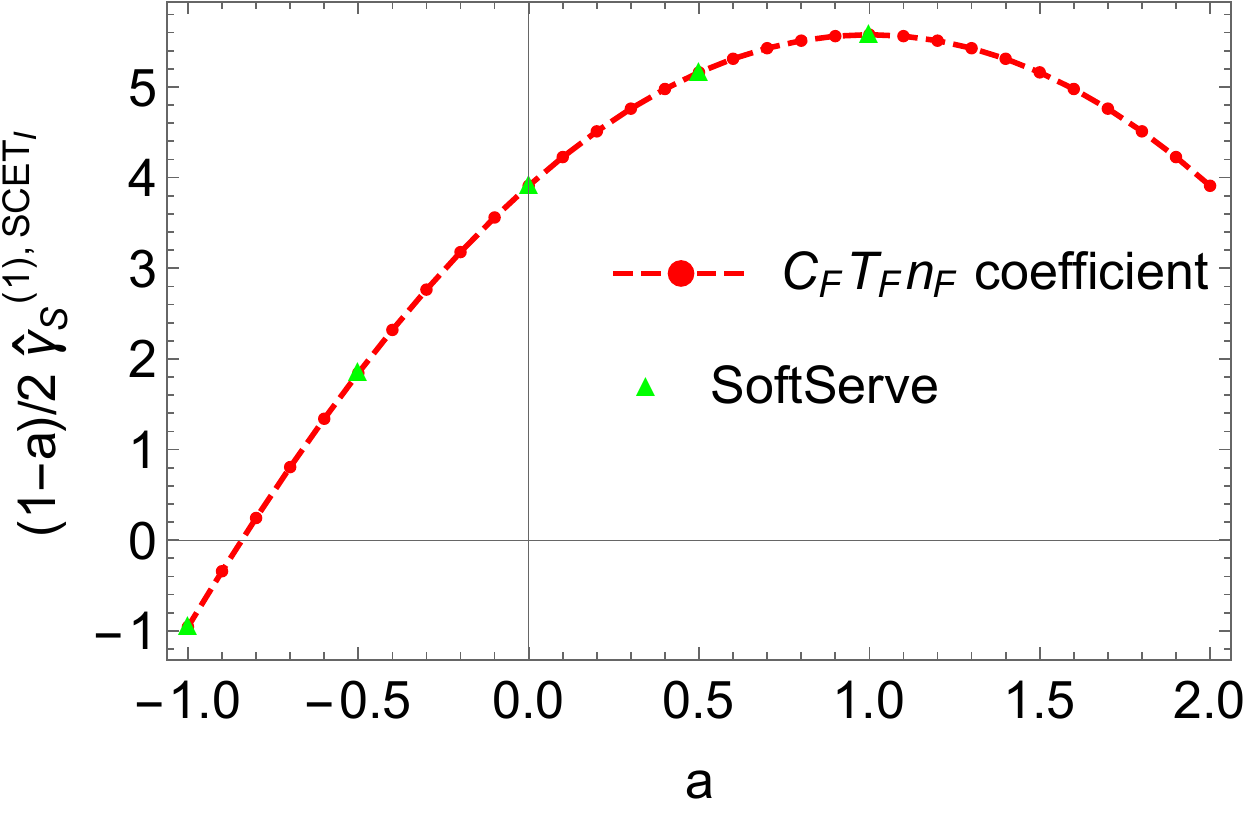}
\hspace{0.5cm}
\includegraphics[width=0.5\linewidth]{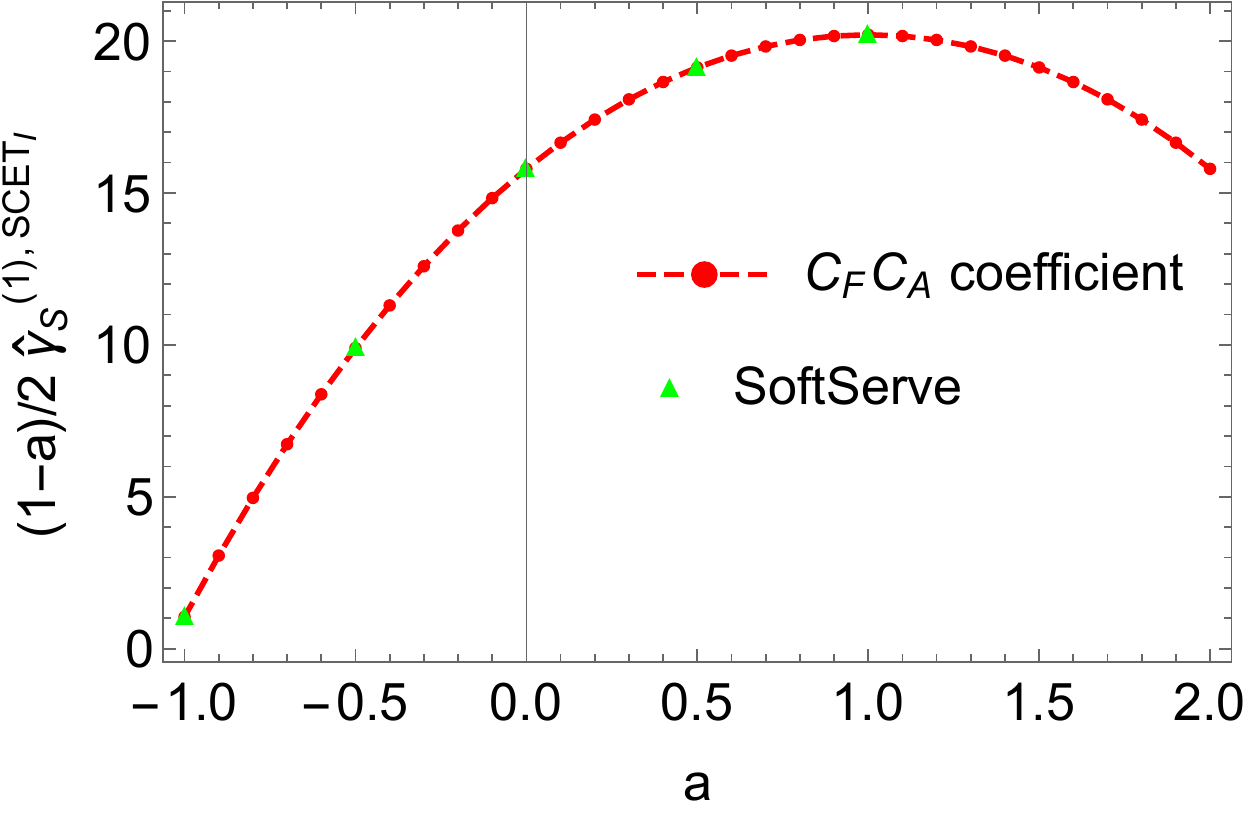}
\caption{Soft anomalous dimension for angularities (multiplied by
  $(1-a)/2$) corresponding to the parameter $a$: coefficient of
  $T_F C_F n_F$ (left) and $C_A C_F$ (right) color factors.}
 \label{fig:gammas}
\end{figure}

To check this result, we compare the second of
Eqs.~\eqref{eq:2loopSCETI} to the result of Ref.~\cite{Bell:2018vaa},
where the soft function was computed numerically.\footnote{We are
  grateful to G. Bell for sharing with us the numerical results of
  Ref.~\cite{Bell:2018vaa} for selected angularities.}
In order to compare to Figure 1 of Ref.~\cite{Bell:2018vaa}, we
consider the quantity
\begin{align}
\frac{1-a}{2}\,\widehat{\gamma}^{(1),\,{\rm SCET_I}}_\Sobsang \,,
\end{align}
and we show the result in Fig.~\ref{fig:gammas}, where we have used
Table~\ref{tab:gammaNIcoeffs} for the constants $\gamma_a^{(C_A)}$ and
$\gamma_a^{(n_F)}$. The result of Eq.~\eqref{eq:2loopSCETI} is given by
the red dashed line, while the green triangles are the numerical
result of Ref.~\cite{Bell:2018vaa} for selected values of the
parameter $a$. The two results are in perfect agreement.
As an additional check, we compare the NNLL resummed cross
section~\eqref{eq:factorization} to the analytic formulae of
Refs.~\cite{Banfi:2014sua,Banfi:2018mcq}, reproducing the results
given there.
As a final check, for $a=0$ the result of Eq.~\eqref{eq:2loopSCETI}
reproduces the soft anomalous dimension for thrust derived in
Refs.~\cite{Becher:2008cf,Kelley:2011ng,Monni:2011gb,Hornig:2011iu}.
Analogous considerations can be used to derive the two loop jet
anomalous dimension, that can be obtained by combining
Eq.~\eqref{eq:finalAnomDim} and the consistency
relation~\eqref{eq:consistencyrel}. 
Alternatively, it can be directly extracted from the soft anomalous
dimension and the hard anomalous dimension (extracted from
Refs.~\cite{Matsuura:1987wt,Matsuura:1988sm,Gehrmann:2005pd,Moch:2005id,Becher:2006mr}),
by imposing that the cross section is independent of the unphysical
scales $\mu$ and $\nu$. We obtain
\begin{align}
\widehat{\gamma}^{(1),\,{\rm SCET_I}}_\Jobsang &=
  C_F^2\left(3-4\pi^2+48\zeta_3\right)+\frac{C_F \, C_A}{1-a}
  \left(\frac{1769}{27}-80\zeta_3-\frac{961}{27} a
  -\frac{11}{9}\pi^2(5 a -6) + 52 \zeta_3 a\right)\notag\\
&\!\!\!-\frac{C_F\,
  n_F}{1-a}\left(\frac{242}{27}+\frac43\pi^2-\frac{130}{27}a-\frac{10}{9}\pi^2
  a\right) + 16 \, \frac{2-a}{1-a}C_F\left(C_A \gamma_a^{(C_A)}+T_F n_F \gamma_a^{(n_F)}\right)\,.
\end{align}

\subsection{Study of the $\nu$-regularization scheme dependence of
  $\gamma^{(\nu)}$}
\label{sec:SchemeDepCalc}

We finally wish to discuss the dependence of the soft and jet
anomalous dimensions on the specific regularization scheme used to
single out the UV divergences in the real radiation. 
In order to verify the validity of Eq.~\eqref{eq:schemeinvariant}, we
perform the calculation of the two loop soft anomalous dimensions
discussed in the previous section using a different UV regularization
scheme for the real radiation integrals.
As an alternative to the exponential regulator, we simply impose a cutoff in the light cone
component of the momentum of each real particle, that is the
constraint
\begin{equation}
\Theta(\nu - \max\{k^+,k^-\}),~\forall~{\rm real}~k\,.
\end{equation}
The integrals in this scheme are similar to the full QCD case with
soft amplitudes. 
Following the same procedure outlined in the previous section, we
obtain
\begin{align}
\label{eq:gamma2loops_cutoff}
\widehat{\gamma}^{(1)}_S[\alpha_s(\mu)] &= \,2 \,C_F
                           C_A\left(\frac{808}{27}-28\zeta_3\right)-
                           2 \,C_F n_F
                           T_F\frac{224}{27} -
  \frac{2}{3}\pi^2 C_F b_0\,,\notag\\
\gamma_{1, 1-a}^{(\nu,\,1)}[\alpha_s(\mu)] &= - \,C_F
                           C_A\left(\frac{808}{27}-28\zeta_3\right)+
                            \,C_F n_F T_F\frac{224}{27}
                                     -\frac{4}{3}\pi^2 C_F \frac{b_0}{(2-a)^2} \notag\\
& \quad- 16 C_F\left(C_A
                                     \gamma_a^{(C_A)}+ T_F n_F
                                     \gamma_a^{(n_F)}\right) \,,
\end{align}
where $\gamma_a^{(C_A)}$ and $\gamma_a^{(n_F)}$ are the same as
before.
We see that $\widehat{\gamma}^{(1)}_S[\alpha_s(\mu)]$ is independent
of the choice of the regulator as expected, while
$\gamma_{1, 1-a}^{(\nu,\,1)}[\alpha_s(\mu)]$ is regulator dependent
and differs from Eq.~\eqref{eq:gamma2loops}.
In order to connect the two
results, we need the one loop coefficient $d_{1,1-a}^{(1)}$ that we can
extract from the finite part of the renormalized one loop soft
function in the light cone cutoff scheme, which reads
\begin{align}
\label{eq:softoneloop-cutoff}
\hat{\cal S}(\mu,\nu) = 1&+\frac{\alpha_s(\mu)}{\pi}
C_F\left[\frac{1}{2-a}\left(2(1-a)\ln^2\frac{\mu}{\nu}+4
    \ln\frac{\mu}{\nu}\ln\frac{\mu u}{Q u_0}-2 \ln^2\frac{\mu u}{Q
      u_0}-\frac{\pi^2}{12}(6-a)\right)\right]\,.
\end{align}
The coefficient $d_{1, 1-a}^{(1)}$ is then obtained as the square root
of the ratio of the constant term of the above soft function to the
result in pure dimensional regularization~\eqref{eq:constdimreg},
obtaining
\begin{equation}
d_{1, 1-a}^{(1)} = \frac{\pi^2}{6}C_F\frac{a(4-a)}{(2-a)(1-a)}\,.
\end{equation}
One can then verify that the quantity~\eqref{eq:schemeinvariant}
evaluated at two loops, namely
\begin{align}
\label{eq:schemeinvariant2}
 \,\frac{2-a}{1-a}\,\gamma_{1, 1-a}^{(\nu,\,1)}[\alpha_s(\mu)]-2\beta(\alpha_s)\frac{\df
                            \ln\Delta_{1,1-a}}{\df\alpha_s} \to {\rm \nu~scheme~invariant~in~SCET_I}\,,
\end{align}
is identical in the two schemes.
This observation can be very useful in performing perturbative
calculations for the anomalous dimensions. Specifically, one can carry
out the computation semi-analytically in a scheme that is very
suitable for a numerical evaluation (such as the light-cone cutoff
scheme), and later convert the result into a scheme with better
analytic properties such as boost invariance, as in the case of the
exponential regulator.

One last comment concerns the constant terms of the two loop soft
function in the extra UV regulator. These are unconstrained by
theoretical arguments and only the combination of soft and jet
functions is independent of the particular UV regularization scheme
adopted in real radiation integrals.
%
\section{Conclusions and Outlook}
\label{sec:Conclusions}
In this article we have studied the observable dependence of anomalous
dimensions in \scetI{} problems, and showed that the introduction of
an extra UV regulator in real radiation integrals can be used to
disentangle this dependence in perturbative calculations.
The system of RGEs of the theory with the additional regulator shares
many analogies with that of \scetII{} problems in the formalism of the
rapidity renormalization group.
This connection highlights some
similarities between the two theories. 
Notably, the whole observable
dependence is encoded in a single anomalous dimension ruling the
evolution in the new UV regularization scale $\nu$ (corresponding to
the rapidity regularization scale in the \scetII{} case), and in the
definition of the initial and final scales of the RGE evolution.
Unlike in the \scetII{} case, however, the dependence of the new soft
and jet functions on the extra UV regulator can be completely
refactorized and shown to cancel in their combination, without leaving
behind a factorization (collinear) anomaly like in \scetII{}.
The explicit cancellation of the $\nu$ dependence makes it natural to
identify the source of the observable dependence in the anomalous
dimensions with the eikonalized jet function that defines the zero-bin
subtraction, which becomes non-trivial in the presence of the extra UV
regulator.

We derived an all-order relation between the anomalous
dimensions of the version of \scetI{} with the extra UV regulator, and
the standard \scetI{} regulated in pure dimensional regularization.
We verified this relation explicitly at 2-loop order for the family of recoil-free
angularities in $e^+e^-$ defined with respect to the winner-take-all axis.
In this context, we carried out a computation of the two loop soft
anomalous dimension and show how to derive the standard \scetI{} soft
anomalous dimension from it. 
This results in new analytic expressions for the perturbative
expansion of this quantity up to two-loop order.
Comparing to previous numerical results from the literature we find
perfect agreement.
We also calculate the new jet functions at one-loop, while the two
loop jet anomalous dimension can be extracted exclusively from
consistency relations, hence providing all necessary ingredients to
carry out the resummation for these observables up to NNLL.

An interesting observation is that the calculation is carried out in
the same framework and regularization scheme for \scetI{} and
\scetII{} theories, hence keeping track of the analogies and
differences between the two limits.
Previous work in the literature which explored the transition between
the \scetI{} and \scetII{} regimes for angularities is that of
Refs.~\cite{Larkoski:2014uqa,Bell:2018vaa}. These papers study the
anomalous dimension in the \scetII{} case ($a=1$ in our notation) as a
limiting case of the \scetI{} anomalous dimension by exploiting the
fact that the factorization theorem is continuous at the transition
point.
In this article we took an orthogonal point of view and formulated
the resummation in \scetI{} in a way that resembles that of the
\scetII{} case, which provides a useful viewpoint on the connection
between the two effective theories.

Although we used angularities to illustrate the structure of the anomalous 
dimensions in the presence of the extra UV regulator, the considerations 
apply more broadly to any \scetI{} observable defined through the particles' final state momenta. 
In future work it will be interesting to explore further the structure
of the zero-bin subtraction for \scetI{} in the presence of the extra
UV regulator, mainly in the context of multi-leg processes where our
observation suggests that the observable dependence in the anomalous
dimensions arises from a quantity that is diagonal in color space.
Moreover, a proof of the cancellation of the $\nu$ dependence between the soft and collinear sectors
at the operator level would be highly desirable.
Finally, we stressed that the introduction of the extra UV regulator
makes real radiation integrals UV finite, and therefore makes the
effective theory suitable for numerical calculations.
A practical advantage of this observation is that the complicated
observable dependence can be separated out from the renormalization
procedure. As a result, the observable dependence of the anomalous
dimensions is to a large extent isolated into finite integrals which
can be also evaluated numerically.
An alternative avenue to exploit this fact is via the numerical
resummation algorithm presented in Ref.~\cite{Bauer:2018svx,Bauer:2019bsp}.


\section*{Acknowledgments}
We would like to thank Guido Bell for providing us with with the
numerical value of the soft anomalous dimension for selected
angularities used to cross check our calculation, Claude Duhr for
discussions about the analytic computation of a class of integrals
appearing in the two loop soft function, and Jonathan Gaunt and Robert
Szafron for discussions on the rapidity renormalization group. We also
thank Thomas Becher, Jonathan Gaunt, Michael Luke, Duff Neill and
Robert Szafron for constructive comments on the manuscript.
We thank Wouter Waalewijn for kindly pointing out an incorrect
statement in the first version of this paper.
This work was supported by the Director, Office of Science, Office of
High Energy Physics of the U.S. Department of Energy under the
Contract No. DE-AC02-05CH11231 (CWB) and by DOE grant DE-SC0009919
(AVM).  PM would like to thank Lawrence Berkeley National Laboratory
for the kind hospitality during the initial stages of this work.

\appendix
\section{Double soft squared amplitude}
\label{app:doublesoft}
In terms of the variables introduced in
Section~\ref{sec:ExplicitCalc}, the double soft tree-level squared matrix
element reads 
\begin{equation}
  \label{eq:tildeM2}
  M_{s,0}^2(k_a,k_b) =(4\pi\alpha_s \mu^{2\epsilon})^2\frac{8 C_F}{m^2(m^2+k_t^2)} C_{ab}(k_a,k_b)
\,,
\end{equation}
where 
\begin{equation}
\label{eq:Corr}
  C_{ab}(k_a,k_b) = C_A(2 \mathcal{S}_{\rm s.o.}+\mathcal{H}_g) + n_f \mathcal{H}_q \,.
 \end{equation}
 The contribution due to two final-state quarks in Eq.~\eqref{eq:Corr}
 has been multiplied by two, to compensate for the overall $1/2$
 factor in Eq.~\eqref{eq:doublereal}.  The three functions
 $\mathcal{S}_{\rm s.o.}$, $\mathcal{H}_g$ and $\mathcal{H}_q$ are the
 $4-2\epsilon$-dimensional counterparts of the homonymous terms
 defined in Ref.~\cite{Dokshitzer:1997iz} and they are taken from
 Ref.~\cite{Banfi:2018mcq}. They depend only on the dimensionless
 variables $z$, $ \phi$ and $\mu^2\equiv m^2/k_t^2$. It is also useful
 to introduce the rescaled momenta $\vec u_i= \vec q_i/k_t$, such that
 \begin{equation}
   \label{eq:xqi}
   u_a^2 = 1+2\sqrt{\frac{1-z}{z}}\mu\cos \phi + \frac{1-z}{z}\mu^2\,,\qquad 
   u_b^2 = 1-2\sqrt{\frac{z}{1-z}}\mu\cos \phi + \frac{z}{1-z}\mu^2\,.
 \end{equation}
In terms of these variables, we have
 \begin{subequations}
\begin{align}
  \label{eq:2S}
  2\mathcal{S}_{\rm s.o.} & = \frac{1}{z(1-z)}\left[\frac{1-(1-z)\mu^2/z}{u_a^2}+\frac{1-z\mu^2/(1-z)}{u_b^2}\right] \\
  \label{eq:Hg}
  \mathcal{H}_g & =-4+(1-\epsilon)\frac{z(1-z)}{1+\mu^2}\left(2\cos  \phi+\frac{(1-2z)\mu}{\sqrt{z(1-z)}}\right)^2\nonumber \\ & +\frac{1}{2(1-z)}\left[1-\frac{1-(1-z)\mu^2/z}{u_a^2}\right] 
+\frac{1}{2z}\left[1-\frac{1-z\mu^2/(1-z)}{u_b^2}\right]
\\
  \label{eq:Hq}
  \mathcal{H}_q& =1-\frac{z(1-z)}{1+\mu^2}\left(2\cos  \phi+\frac{(1-2z)\mu}{\sqrt{z(1-z)}}\right)^2\,.
\end{align}
 \end{subequations}

\addcontentsline{toc}{section}{References}
\bibliographystyle{JHEP}
\bibliography{Rapidity}

\end{document}